\documentclass[a4paper,11pt]{article}
\pdfoutput=1

\usepackage{jheppub}

\usepackage[T1]{fontenc}
\graphicspath{{img/}}

\usepackage{mathrsfs}                   
\usepackage[dvipsnames,table]{xcolor}   
\usepackage{booktabs}                   
\usepackage{multirow}                   
\usepackage{lipsum}                     
\usepackage{longtable}                  
\usepackage{pifont}                     
\usepackage{slashed}
\usepackage{youngtab}
\usepackage{stackengine}
\usepackage{tikz}
\usetikzlibrary{positioning}
\usepackage{subcaption}
\usepackage{amssymb}

\newcommand*\hacketyhack[2][]{%
\begin{tabular}{@{}S[#1]@{}}
 #2
\end{tabular}}

\newcommand{\dagf}[1]{{#1}^{\dagger}}
\newcommand{\ie}{i.e.\ }
\newcommand{\eg}{e.g.\ }

\newcommand{\idlr}[1]{i\!\overleftrightarrow{D}_{\!\!#1}}

\newcommand{\midmidrule}{\arrayrulecolor{black!30}\midrule\arrayrulecolor{black}}

\usepackage[detect-all]{siunitx}
\DeclareSIUnit\year{yr}

\newcommand{\tabnum}[1]{
  \num[
    scientific-notation = true,
    round-mode = figures,
    round-precision = 1,
    exponent-product = \cdot
  ]{#1}
}

\newcommand{\tabnumT}[1]{
  \num[
    scientific-notation = true,
    round-mode = figures,
    round-precision = 2,
    exponent-product = \cdot
  ]{#1}
}

\title{
Model-independent estimates for loop-induced baryon-number-violating nucleon decays
}

\author[a,b]{John Gargalionis,}
\author[a,b]{Juan Herrero-García,}
\author[c]{and Michael A. Schmidt}

\affiliation[a]{Departament de Física Teòrica, Universitat de València, 46100 Burjassot, Spain}

\affiliation[b]{Instituto de Física Corpuscular (CSIC-Universitat de València),
Parc Científic UV, C/Catedrático José Beltrán, 2, E-46980 Paterna, Spain}

\affiliation[c]{Sydney Consortium for Particle Physics and Cosmology,
  School of Physics, University of New South Wales,
  Sydney, New South Wales 2052, Australia}

\preprint{CPPC-2024-02}

\emailAdd{john.gargalionis@ific.uv.es}
\emailAdd{juan.herrero@ific.uv.es}
\emailAdd{m.schmidt@unsw.edu.au}

\abstract{Baryon number is an accidental symmetry of the Standard Model (SM) Lagrangian that so far has been measured to be exactly preserved, although it is expected to be violated at higher energies. In this work we compute order-of-magnitude estimates for the matching contributions of generic ultraviolet models to effective operators that generate nucleon decay processes. This is done in a systematic and automated way using operators constructed from SM fields up to dimension nine and working in a framework that has proved useful in the study of lepton-number violation. For each of the operators we derive estimates for the rates of different nucleon-decay channels. These allow us to establish model-independent lower bounds on the underlying new-physics scale and identify potential correlations between the various decay modes. The results are most relevant for families of models that generate the considered operator. This analysis is especially timely given the expected future sensitivities in numerous experiments such as Hyper-K, DUNE, JUNO and THEIA.}

\keywords{Baryon Number Violation, Nucleon Decay, Effective Field Theory, Standard Model Effective Field Theory, Loop Level, Simplified Models}

\makeatletter
\gdef\@fpheader{}
\makeatother 
\begin{document}
\maketitle

\flushbottom

\newpage
\section{Introduction}\label{sec:intro}

Lepton number ($L$) and baryon number ($B$) are extremely good accidental global symmetries of the Standard Model (SM) Lagrangian. However, they are expected to be violated at the highest energies,\footnote{In fact, any global symmetry is expected to be violated by quantum-gravity effects relevant at Planck-scale energies.} i.e.\ via higher-dimensional operators. Indeed, $L$ is violated in two units at the lowest dimension ($d=5$) through the Weinberg operator~\cite{Weinberg:1979sa}, and this provides a simple explanation for the origin of Majorana neutrino masses after electroweak symmetry breaking.\footnote{Neutrino oscillations require the violation of individual lepton flavors, $L_{e,\mu,\tau}$. For neutrinos to be Dirac fermions, one needs to \emph{impose} total lepton number conservation, $L=L_e+L_\mu+L_\tau$, possibly as a gauge symmetry (more precisely $B-L$ conservation, to avoid anomalies within the SM), for which there is no \textit{a priori} motivation. Therefore, in this EFT approach Majorana neutrino masses are a more natural option. Of course, to actually observe the explicit violation of $L$ in two units, a process such as neutrino-less double beta decay should be observed.} Although proton decay has not been detected so far, $B$ is violated in one unit by four-fermion operators at the next-to-lowest dimension ($d = 6$), making the experimental programme in search of $\Delta B = -1$ processes well grounded, and further motivated by their appearance in Grand Unified Theories (GUTs)~\cite{Georgi:1974sy,Fritzsch:1974nn} and $R$-parity-violating supersymmetry~\cite{Farrar:1978xj}. More broadly, the violation of baryon and lepton number is a necessary ingredient to generate the baryon asymmetry of the Universe~\cite{Sakharov:1967dj}.

Experimental searches set extremely stringent limits, $\sim 10^{24}$ times larger than the age of the Universe, from the non-observation of baryon-number-violating (BNV) nucleon decays. These are much stronger than, for example, bounds on decaying dark matter from indirect detection searches. The next generation of neutrino-oscillation experiments like Hyper-Kamiokande (Hyper-K)~\cite{Hyper-Kamiokande:2018ofw}, Deep Underground Neutrino Experiment (DUNE)~\cite{DUNE:2016evb, DUNE:2020ypp} and Jiangmen Underground Neutrino Observatory (JUNO)~\cite{JUNO:2015zny}, as well as the optical neutrino detector THEIA~\cite{Theia:2019non},
are expected to improve their sensitivity to these decays by one order of magnitude in certain modes.\footnote{For Hyper-K, see Figure~1 in  Ref.~\cite{Hyper-Kamiokande:2022smq} and Table 7 in Ref.~\cite{Dev:2022jbf}. DUNE and JUNO are expected to be comparable with Hyper-K in channels involving kaons.} It is therefore tempting to think that proton decay could be the next big discovery in fundamental physics.

In the last decades there have been many studies of proton decay (see Refs.~\cite{Nath:2006ut,FileviezPerez:2022ypk,Dev:2022jbf,Ohlsson:2023ddi} for reviews), including automated tools to compute decay rates in some GUTs~\cite{Antusch:2020ztu}. There is also a literature of model-independent studies of BNV processes using Effective Field Theory (EFT)~\cite{Abbott:1980zj,Claudson:1981gh,Heeck:2019kgr,He:2021mrt,He:2021sbl,Fridell:2023tpb,us:tree}. In addition to exclusive nucleon decay modes, inclusive nucleon decays provide an interesting avenue to search for BNV~\cite{Heeck:2019kgr}. There is also a variety of dedicated phenomenological studies of leptoquarks in this context~\cite{Dorsner:2012nq,Murgui:2021bdy,Dorsner:2022twk,Kovalenko:2002eh,Arnold:2012sd,Assad:2017iib,Helo:2019yqp,Davighi:2022qgb,Baldes:2011mh}.

In this work we apply an EFT approach to the study of loop-level BNV nucleon decay. In Ref.~\cite{us:tree} we studied nucleon decays induced at tree level in the SMEFT at dimensions 6 and 7, taking into account RGE effects. Here we focus on operators of dimension 8 and 9, but use them to estimate \textit{loop-level} contributions from generic UV models to the operators of dimension 6 and 7. The strategy we follow is in line with a framework that has already proved useful in the investigation of Majorana neutrino masses~\cite{Babu:2001ex,deGouvea:2007qla}. In Refs.~\cite{Babu:2001ex,deGouvea:2007qla}, the goal is to use high-dimensional gauge- and Lorentz-invariant combinations of SM multiplets to estimate the contributions of UV models to neutrino masses. Our approach here is to map that framework onto the case of $|\Delta B| = 1$ physics, and estimate UV contributions to $\Delta B = -1$ nucleon decays in a model-independent way.

For Majorana neutrino masses, the dominant contribution to neutrino masses originates from matching the UV model to the Weinberg operator~\cite{Weinberg:1979sa} $LLHH$ in SMEFT at loop level.\footnote{Weinberg-like operators of the form $LLHH (H^\dagger H)^n$ may dominate for low scales $\Lambda \lesssim 4 \pi \langle H \rangle$.} This can then be matched onto the dimension-3 Majorana mass term $\nu\nu$ in the Low Energy Effective Field Theory (LEFT) through the usual tree-level relation.  We denote this contribution as the \emph{loop-level matching contribution}. Higher-dimensional $|\Delta L|=2$ operators at $d = 7,9,11,\ldots$\footnote{The relation between the dimension of an operator $d$  and its $B$ and $L$ charges has been derived in Refs.~\cite{deGouvea:2014lva,Kobach:2016ami}.} contribute subdominantly, see e.g.~the discussion in Ref.~\cite{Cai:2017jrq}, but they can be used to estimate the loop-level matching contribution by drawing diagrams relating them to the dimension-5 Weinberg operator in an $\mathrm{SU}(2)$-covariant formalism following the procedure in Refs.~\cite{deGouvea:2007qla,Angel:2012ug,delAguila:2012nu,Gargalionis:2020xvt}. In this approach, useful matching estimates have been obtained using operators we dub \textit{field strings},\footnote{Our use of the term \textit{field string} mirrors the concept of \emph{types of operators} presented in Ref.~\cite{Fonseca:2019yya}.} or \textit{field-string operators}, which specify the operator's field content but not the complete Lorentz and gauge structure. These field strings function as accounting devices that summarise the loop effects of large classes of UV models, namely those that generate those field strings at tree level, up to order one factors.

The analogous procedure mapped onto the study of $|\Delta B| = 1$ phenomena is necessarily more involved. First, in place of the unique (up to flavour) $|\Delta L| = 2$ Weinberg operator and the dimension-3 neutrino mass operator in the LEFT, there are four dimension-6 $\Delta (B-L)=0$ and six dimension-7 $\Delta (B-L)=2$ SMEFT operators, which match onto 26 dimension-6 $\Delta B = 1$ operators with light quarks ($u,d,s$) per lepton generation in the basis of Ref.~\cite{Jenkins:2017jig}. Furthermore, each operator can generate a different pattern of decays, and the actual computation of the rates necessarily involves inputs from the lattice. These considerations suggest the utility of an automated approach for deriving the matching estimates. Our algorithm matches each of the field-string operators in our listing first onto the aforementioned 10 SMEFT operators with the smallest number of loops necessary to generate BNV nucleon decay which are subsequently matched to the 26 BNV LEFT operators. From these, the nucleon decay rates are calculated using known lattice inputs or else estimated using dimensional analysis when lattice results are unavailable.

Armed with the matching estimates, we are able to derive model-independent lower limits on the energy scales characterising the UV models which generate the field strings in our catalogue at tree level. We would like to stress that these limits will be more constraining than the limits obtained from the direct contribution of the effective SMEFT operator to nucleon decay, similar to the above discussion of higher-dimensional $|\Delta L|=2$ operators, because the procedure estimates the contribution of the more relevant lower-dimensional BNV SMEFT operator, which is generated via loop-level matching in the underlying UV model. The large hierarchy of the QCD scale $\Lambda_{\rm QCD}$ to the BNV scale $\Lambda$, $\Lambda_{\rm QCD}\ll \Lambda$ ensures that the loop suppression does not overcome the power counting in SMEFT. An example UV model illustrating this in a BNV context is presented in Ref.~\cite{Dorsner:2022twk}, and a systematic decomposition of the one-loop completions of the $\Delta B$ dimension-6 operators in the SMEFT is given in Ref.~\cite{Helo:2019yqp}. The limits provide reliable estimates of the constraints from BNV nucleon decay for the underlying UV models which generate the discussed field strings at tree level. Moreover, this procedure enables us to outline the most constraining decay processes in these models, provide estimates for the respective branching ratios, and discuss the possible correlations among them. We have made our code and results available online~\cite{Gargalionis_Operator_closure_estimates_2023}.

The remainder of the paper is structured as follows. In Sec.~\ref{sec:left} we discuss the LEFT framework for proton decay. In Sec.~\ref{sec:smeft} we introduce the field strings we work with up to dimension 9 as well as the SMEFT operators used in the derivation of our results. In Sec.~\ref{sec:method} we explain our algorithmic approach and the different assumptions that underlie it. We also provide two detailed examples illustrating our methods. This section may be skipped by readers eager to jump to our results, presented in Sec.~\ref{sec:results}. These include distinctive correlations between decay modes predicted by the operators we study, as well as limits on the new-physics scales imposed by current experimental bounds. We also make some brief comments on UV completions. An important part of the results we present is the list of dominant matching estimates associated with the field-string operators in our listing. These are presented in Appendix~\ref{sec:app-ops}. We give our conclusions in Sec.~\ref{sec:conc} and provide additional technical materials in the remaining appendices.

\section{Low-energy framework for BNV nucleon decay}\label{sec:left}

In this section we discuss the $|\Delta B| = 1$ operators in the LEFT, the effective theory obtained after integrating out the heavy quarks ($t,b,c$), tau lepton ($\tau$),\footnote{BNV tau decays have much less stringent limits and are indirectly bounded via proton-decay searches with a virtual $\tau$ lepton~\cite{Marciano:1994bg,Crivellin:2023ter}.} gauge bosons ($W$, $Z$) and the Higgs doublet ($H$). We will consider only the light quarks ($u,d,s$), electron, muon and neutrinos at low energies.

In Table~\ref{tab:ISprocess} we show the allowed two-body $\Delta B = -1$\footnote{The difference is defined as $\Delta B \equiv B_{\rm final}-B_{\rm initial}$ and similarly for all other quantum numbers.} nucleon decays to pseudoscalar mesons ($P$) and (anti)leptons for protons ($p\to P\ell$) and neutrons ($n\to P\ell$), respectively. Each decay channel is accompanied by the units in which hadronic isospin ($I$), strangeness ($S$), $B$ and $L$ are violated and the current most stringent bound on the decay rate, at $90\%$ confidence level. The $\Delta S = -1$ decays are shown separately in the last two rows of the tables since they are not generated at dimensions 6 or 7 in the LEFT. We also show the prospective future sensitivity from Hyper-K~\cite{Hyper-Kamiokande:2018ofw}, which is expected to provide the most sensitive probe of the corresponding nucleon decays in most cases. The JUNO experiment will provide a sensitivity of \SI{9.6e33}{\year} for an exposure of \SI{200}{\kilo \tonne \year} for $p \to K^+ \nu$~\cite{JUNO:2022qgr}. Note that the same applies for the mode with an antineutrino. In the following, we will not distinguish between neutrinos and antineutrinos in our notation. DUNE foresees sensitivity to $n \to K^+e^-$ at \SI{1.1e34}{\year} for \SI{400}{\kilo \tonne \year} exposure~\cite{DUNE:2020ypp}, comparable to Hyper-K. The THEIA experiment with an \SI{800}{\kilo \tonne \year} exposure could set a slightly better limit at \SI{3.8e34}{\year} for $p \to K^+ \nu$~\cite{Theia:2019non}. We were not able to find sensitivity estimates in the literature for the absent entries in the table.

\begin{table}[tbp!]
  \begin{center}
    \begin{tabular}{lclcrrr}
      \toprule
      Decay mode & {Limit [\SI{e34}{\year}]} & Ref. & Hyper-K [\SI{e34}{\year}] & $\Delta I$ & $\Delta S$ & $\Delta L$ \\
      \midrule
      \rowcolor{lightgray}
      \multicolumn{7}{l}{Proton channels}\\
      $p\to \pi^0 e^+$ & \hacketyhack[table-format=1.3]{2.4} & \cite{Super-Kamiokande:2020wjk} & \hacketyhack[table-format=1.1]{7.8} & $-\frac12$ & $0$ & $-1$ \\
      $p\to \pi^0 \mu^+$ & \hacketyhack[table-format=1.3]{1.6} & \cite{Super-Kamiokande:2020wjk} & \hacketyhack[table-format=1.1]{7.7} & $-\frac12$ & $0$ & $-1$ \\
      $p\to \eta^0 e^+$ & \hacketyhack[table-format=1.3]{1.0} & \cite{Super-Kamiokande:2017gev} & \hacketyhack[table-format=1.1]{4.3} & $-\frac12$ & $0$ & $-1$ \\
      $p\to \eta^0 \mu^+$ & \hacketyhack[table-format=1.3]{0.47} & \cite{Super-Kamiokande:2017gev} & \hacketyhack[table-format=1.1]{4.9} & $-\frac12$ & $0$ & $-1$ \\
      $p\to \pi^+ \nu_r$ & \hacketyhack[table-format=1.3]{0.039} & \cite{Super-Kamiokande:2013rwg} & --- & $\frac12$ & $0$ & $\pm1$ \\
      $p\to K^0 e^+$ & \hacketyhack[table-format=1.3]{0.10} & \cite{Super-Kamiokande:2005lev} & --- & $-1$ & $1$ & $-1$ \\
      $p\to K^0 \mu^+$ & \hacketyhack[table-format=1.3]{0.16} & \cite{Super-Kamiokande:2012zik} & --- & $-1$ & $1$ & $-1$ \\
      $p\to K^+ \nu_r$ & \hacketyhack[table-format=1.3]{0.59} & \cite{Super-Kamiokande:2014otb} & \hacketyhack[table-format=1.1]{3.2} & $0$ & $1$ & $\pm1$ \\
      \midmidrule
      $p\to \bar{K}^0 e^+$ & \hacketyhack[table-format=1.3]{0.10} &  \cite{Super-Kamiokande:2005lev} & --- & $0$ & $-1$ & $-1$ \\
      $p\to \bar{K}^0 \mu^+$ & \hacketyhack[table-format=1.3]{0.16} & \cite{Super-Kamiokande:2012zik} & --- & $0$ & $-1$& $-1$ \\
      \midrule
      \rowcolor{lightgray}
      \multicolumn{7}{l}{Neutron channels}
      \\
      $n\to \pi^0 \nu_r$ & \hacketyhack[table-format=1.4]{0.11} & \cite{Super-Kamiokande:2013rwg} & --- & $\frac12$ & $0$ & $\pm1$ \\
      $n\to \eta^0 \nu_r$ & \hacketyhack[table-format=1.4]{0.016} & \cite{McGrew:1999nd} & --- & $\frac12$ & $0$ & $\pm1$ \\
      $n\to \pi^- e^+$ & \hacketyhack[table-format=1.4]{0.53} & \cite{Super-Kamiokande:2017gev} & 2.0 & $-\frac12$ & $0$ & $-1$ \\
      $n\to \pi^- \mu^+$ & \hacketyhack[table-format=1.4]{0.35} & \cite{Super-Kamiokande:2017gev} & 1.8 & $-\frac12$ & $0$ & $-1$  \\
      $n\to \pi^+ e^-$ & \hacketyhack[table-format=1.4]{0.0065} & \cite{Seidel:1988ut} & --- & $\frac32$ & $0$ & $1$ \\
      $n\to \pi^+ \mu^-$ & \hacketyhack[table-format=1.4]{0.0049} & \cite{Seidel:1988ut} & --- & $\frac32$ & $0$ & $1$ \\
      $n\to K^+ e^-$ & \hacketyhack[table-format=1.4]{0.0032} & \cite{Frejus:1991ben} & 1.0 & $1$ & $1$ & $1$ \\
      $n\to K^+ \mu^-$ & \hacketyhack[table-format=1.4]{0.0057} & \cite{Frejus:1991ben}& --- & $1$ & $1$ & $1$ \\
      $n\to K^0 \nu_r$ & \hacketyhack[table-format=1.4]{0.013} & \cite{Super-Kamiokande:2005lev} & --- & $0$ & $1$ & $\pm1$ \\
      \midmidrule%
      $n\to K^- e^+$ & \hacketyhack[table-format=1.4]{0.0017} & \cite{McGrew:1999nd} & --- & $0$ & $-1$ & $-1$   \\
      $n\to \bar{K}^0 \nu_{r}$ & \hacketyhack[table-format=1.4]{0.013} & \cite{Super-Kamiokande:2005lev} & --- & $1$ & $-1$ & $\pm1$ \\
      \bottomrule
    \end{tabular}
  \end{center}
  \caption{\label{tab:ISprocess} Two-body semi-leptonic proton (upper table) and neutron (bottom table) decay modes. The table shows the change in nuclear isospin $\Delta I$, strangeness $\Delta S$ and lepton number $\Delta L$ for two-body proton decays with $\Delta B=-1$ to a pseudoscalar meson and a lepton ($\Delta L=1$) or antilepton ($\Delta L=-1$). The reference given involves the most constraining measurements of the electronic and muonic decays. The index $r$ here enumerates over lepton flavours. The final state $\nu_r$ denotes both neutrinos and antineutrinos. We show the decays $p \to \bar{K}^0 e^+_r$, $n \to K^- e^+_r$ and $n \to \bar{K}^0 \nu_{r}$ separately in the tables since they are $\Delta S = -1$ processes and therefore are not induced at dimensions 6 or 7 in the LEFT\@. All quoted experimental limits are at $90\%$ confidence level. The Hyper-K sensitivities are taken from the design report~\cite{Hyper-Kamiokande:2018ofw} assuming a \SI{1.9}{\mega \tonne \year} exposure, when available. There are noticeably fewer neutron-decay modes mentioned in the Hyper-K design report.}
\end{table}

\begin{table}[tbp!]
  \begin{center}
    \begin{tabular}{llccrr}
      \toprule
      Name~\cite{Jenkins:2017jig}  & Ref.~\cite{Nath:2006ut} & Operator &  Flavour & $\Delta I$ & $\Delta S$ \\
      \midrule
      \([\mathcal{O}_{udd}^{S,LL}]_{111r}\) & \(O^{\nu}_{LL}\) & \((u d)(d \nu_{r})\)  & \((\mathbf{8}, \mathbf{1})\) & $\frac12$ & 0  \\
      \([\mathcal{O}_{udd}^{S,LL}]_{121r}\) & \(\tilde{O}^{\nu}_{LL1}\) & \((u s)(d \nu_{r})\)  & \((\mathbf{8}, \mathbf{1})\) & $0$ & $1$ \\
      \([\mathcal{O}_{udd}^{S,LL}]_{112r}\) & \(\tilde{O}^{\nu}_{LL2}\) & \((u d)(s \nu_{r})\)  & \((\mathbf{8}, \mathbf{1})\) &$0$& $1$\\
      \midmidrule
      \([\mathcal{O}_{duu}^{S,LL}]_{111r}\) & \(O^{e}_{LL}\) &\((d u)(u e_{r})\)  & \((\mathbf{8}, \mathbf{1})\) &$-\frac12$ &$0$ \\
      \([\mathcal{O}_{duu}^{S,LL}]_{211r}\) & \(\tilde{O}^{e}_{LL}\)& \((s u)(u e_{r})\)  & \((\mathbf{8}, \mathbf{1})\) & $-1$ & $1$\\
      \midmidrule
      \([\mathcal{O}_{duu}^{S,LR}]_{111r}\) & \(O^{e}_{LR}\) & \((d u)(\bar{u}^{\dagger} \bar{e}^{\dagger}_{r})\)  & \((\bar{\mathbf{3}}, \mathbf{3})\) &
      $-\frac12$& $0$\\
      \([\mathcal{O}_{duu}^{S,LR}]_{211r}\) & \(\tilde{O}^{e}_{LR}\) & \((s u)(\bar{u}^{\dagger} \bar{e}^{\dagger}_{r})\)  &\((\bar{\mathbf{3}}, \mathbf{3})\) & $-1$&$1$\\
      \midmidrule
      \([\mathcal{O}_{duu}^{S,RL}]_{111r}\) & \(O^{e}_{RL}\) & \((\bar{d}^{\dagger} \bar{u}^{\dagger})(u e_{r})\) & \((\mathbf{3}, \bar{\mathbf{3}})\) &  $-\frac12$&$0$ \\
      \([\mathcal{O}_{duu}^{S,RL}]_{211r}\) & \(\tilde{O}^{e}_{RL}\) & \((\bar{s}^{\dagger} \bar{u}^{\dagger})(u e_{r})\)  & \((\mathbf{3}, \bar{\mathbf{3}})\) & $-1$ & $1$ \\
      \midmidrule
      \([\mathcal{O}_{dud}^{S,RL}]_{111r}\) & \(O^{\nu}_{RL}\) & \((\bar{d}^{\dagger} \bar{u}^{\dagger})(d \nu_{r})\)  & \((\mathbf{3}, \bar{\mathbf{3}})\) &$\frac12$&$0$\\
      \([\mathcal{O}_{dud}^{S,RL}]_{211r}\) & \(\tilde{O}^{\nu}_{RL1}\) & \((\bar{s}^{\dagger} \bar{u}^{\dagger})(d \nu_{r})\)  & \((\mathbf{3}, \bar{\mathbf{3}})\) & $0$&$1$\\
      \([\mathcal{O}_{dud}^{S,RL}]_{112r}\) & \(\tilde{O}^{\nu}_{RL2}\) & \((\bar{d}^{\dagger} \bar{u}^{\dagger})(s \nu_{r})\)  & \((\mathbf{3}, \bar{\mathbf{3}})\) &$0$&$1$\\
      \midmidrule
      \([\mathcal{O}_{ddu}^{S,RL}]_{[12]1r}\) & --- & \((\bar{d}^{\dagger} \bar{s}^{\dagger})(u \nu_{r})\)  & \((\mathbf{3}, \bar{\mathbf{3}})\) &$0$&$1$\\
      \midmidrule
      \([\mathcal{O}_{duu}^{S,RR}]_{111r}\) & \(O^{e}_{RR}\) & \((\bar{d}^{\dagger} \bar{u}^{\dagger})(\bar{u}^{\dagger} \bar{e}^{\dagger}_{r})\)  & \((\mathbf{1}, \mathbf{8})\) & $-\frac12$&$0$\\
      \([\mathcal{O}_{duu}^{S,RR}]_{211r}\) & \(\tilde{O}^{e}_{RR}\) & \((\bar{s}^{\dagger} \bar{u}^{\dagger})(\bar{u}^{\dagger} \bar{e}^{\dagger}_{r})\)  & \((\mathbf{1}, \mathbf{8})\) &$-1$&$1$\\
      \bottomrule
    \end{tabular}
  \end{center}
  \caption{\label{tab:wet-ops-BL0} The table shows the $\Delta B = \Delta L = -1$ operators in the LEFT basis of Ref.~\cite{Jenkins:2017jig} that give rise to nucleon decays at tree level. We also indicate the correspondence to the nomenclature of Ref.~\cite{Nath:2006ut}. (The operator $\mathcal{O}_{ddu}^{S,RL}$ is not studied there.) Square brackets enclose pairs of indices that are antisymmetric under permutation. The operators are presented in Weyl-spinor notation along with their transformation properties under flavour $\mathrm{SU}(3)_L \times \mathrm{SU}(3)_R$ and the units in which they violate hadronic isospin ($\Delta I$) and strangeness ($\Delta S$). The combination of properties $(\Delta I, \Delta S, \Delta L, \Delta B)$ specifies the decay process each operator induces (see Table~\ref{tab:ISprocess}). The index $r$ represents lepton flavour.}
\end{table}

\begin{table}[tbp!]
  \begin{center}
    \begin{tabular}{lccrr}
      \toprule
      Name~\cite{Jenkins:2017jig} &  Operator  & Flavour & $\Delta I$ & $\Delta S$ \\
      \midrule
      \([\mathcal{O}_{ddd}^{S,LL}]_{[12]r1}\) & \((d s)(\bar{e}_{r} d)\) & \((\mathbf{8}, \mathbf{1})\) & $1$ & $1$ \\
      \midmidrule
      \([\mathcal{O}_{udd}^{S,LR}]_{11r1}\) & \((u d)(\nu^{\dagger}_{r} \bar{d}^{\dagger})\)  & \((\bar{\mathbf{3}}, \mathbf{3})\) &
      $\frac12$ & $0$\\
      \([\mathcal{O}_{udd}^{S,LR}]_{12r1}\) &  \((u s)(\nu^{\dagger}_{r} \bar{d}^{\dagger})\)  & \((\bar{\mathbf{3}}, \mathbf{3})\) &
      $0$ & $1$\\
      \([\mathcal{O}_{udd}^{S,LR}]_{11r2}\) &  \((u d)(\nu^{\dagger}_{r} \bar{s}^{\dagger})\)  & \((\bar{\mathbf{3}}, \mathbf{3})\) &
      $0$ & $1$\\
      \midmidrule
      \([\mathcal{O}_{ddu}^{S,LR}]_{[12]r1}\) &  \((d s)(\nu^{\dagger}_{r} \bar{u}^{\dagger})\)  & \((\bar{\mathbf{3}}, \mathbf{3})\) &
      $0$ & $1$ \\
      \midmidrule
      \([\mathcal{O}_{ddd}^{S,LR}]_{[12]r1}\) &  \((d s)(e^{\dagger}_{r} \bar{d}^{\dagger})\)  & \((\bar{\mathbf{3}}, \mathbf{3})\) &
      $1$ & $1$ \\
      \midmidrule
      \([\mathcal{O}_{ddd}^{S,RL}]_{[12]r1}\) &  \((\bar{d}^{\dagger} \bar{s}^{\dagger})(\bar e_r d)
      \)  & \((\mathbf{3},\bar{\mathbf{3}})\)
      & $1$ & $1$ \\
      \midmidrule
      \([\mathcal{O}_{udd}^{S,RR}]_{11r1}\) &  \((\bar{u}^{\dagger} \bar{d}^{\dagger})(\nu^{\dagger}_{r} \bar{d}^{\dagger})\)  & \((\mathbf{1}, \mathbf{8})\) &
      $\frac12$ & $0$ \\
      \([\mathcal{O}_{udd}^{S,RR}]_{12r1}\) &  \((\bar{u}^{\dagger} \bar{s}^{\dagger})(\nu^{\dagger}_{r} \bar{d}^{\dagger})\)  & \((\mathbf{1}, \mathbf{8})\) &
      $0$ & $1$\\
      \([\mathcal{O}_{udd}^{S,RR}]_{11r2}\) &  \((\bar{u}^{\dagger} \bar{d}^{\dagger})(\nu^{\dagger}_{r} \bar{s}^{\dagger})\)  & \((\mathbf{1}, \mathbf{8})\) &
      $0$ & $1$ \\
      \midmidrule
      \([\mathcal{O}_{ddd}^{S,RR}]_{[12]r1}\) &  \((\bar{d}^{\dagger} \bar{s}^{\dagger})(e^{\dagger}_{r} \bar{d}^{\dagger})\)  & \((\mathbf{1}, \mathbf{8})\) & $1$ & $1$ \\
      \bottomrule
    \end{tabular}
  \end{center}
  \caption{\label{tab:wet-ops-BL2} The table shows the $\Delta B = -\Delta L = -1$ operators in the LEFT basis of Ref.~\cite{Jenkins:2017jig} that give rise to nucleon decays at tree level. (These operators are not studied in Ref.~\cite{Nath:2006ut}, so the nomenclature cannot be compared.) We point the reader to the caption of Table~\ref{tab:wet-ops-BL0} or the main text for the common description of the columns.}
\end{table}

The Lagrangian in the LEFT takes the form
\begin{align}
    \mathcal{L} = \sum_i [C_{i}]_{pqrs} [\mathcal{O}_{i}]_{pqrs} +\mathrm{h.c.}\;,
\end{align} 
where $[C_{i}]_{pqrs}$ denote the dimensionful Wilson Coefficients (WCs) with flavour indices $pqrs$ in the order of the fermion fields in the operator. The lowest-dimensional operators that violate baryon and lepton number are of dimension 6. The subset of these that preserve $B-L$ is presented in Table~\ref{tab:wet-ops-BL0}. Disregarding the flavour structure, there are 9 independent operators\footnote{Two of the BNV dimension-6 operators, $\mathcal{O}_{uud}^{S,XY}$ with $X\neq Y$, $X,Y\in \{L,R\}$, are antisymmetric in the up-type quark flavour indices and do not directly contribute to BNV nucleon decays at tree level, but only to charmed $\Lambda_c^+$ decays and, thus, are omitted.} in this group. The remaining 7 independent $\Delta(B-L)=-2$ operators are listed in Table~\ref{tab:wet-ops-BL2}. The table lists all operators following the convention in Ref.~\cite{Jenkins:2017jig} and their relation to Ref.~\cite{Nath:2006ut}. The explicit forms of the operators are presented in the third column using Weyl-spinor notation~\cite{Dreiner:2008tw}. The transformation properties of the LEFT operators with respect to the flavour group ${\mathrm{SU}(3)}_{L}\times {\mathrm{SU}(3)}_{R}$, where the three light quarks form an $\mathrm{SU}(3)$ triplet $(u,d,s)\sim \mathbf{3}$, are listed in the fourth column. The last two columns present the units in which nuclear isospin and strangeness are violated by the operators. These and the units in which the operators change $B$ and $L$ determine the relevant decay channels in each case through the correspondence given in Table~\ref{tab:ISprocess}.

\begin{table}[tbp!]
  \begin{center}
    \begin{tabular}{llccccc}
      \toprule
      Name & Ref.~\cite{Liao:2020zyx} &  Operator & Flavour & Indices & $\Delta I$ & $\Delta S$ \\
      \midrule
          $[\mathcal{O}^{V,LR}_{udd}]_{pqrs}$ & $\mathcal{O}_{d \nu u d D2}$ & $(u_p \idlr{\mu} d_q)(\bar{d}^{\dagger}_r \bar\sigma^\mu \nu_s)$ & $(\mathbf{6},\mathbf{3})$ & $\begin{matrix} 111s \\ 121s \\ 112s \end{matrix}$ & $\begin{matrix} \frac12 \\ 0 \\ 0 \end{matrix}$ & $\begin{matrix} 0 \\ 1 \\ 1 \end{matrix}$ \\
      \midmidrule
      $[\mathcal{O}^{V,LR}_{ddu}]_{\{pq\}rs}$ & $\mathcal{O}_{u \nu d D 3}$ & $(d_p \idlr{\mu} d_q)(\bar{u}^\dagger_r \bar\sigma^\mu \nu_s)$ &  ($\mathbf{6},\mathbf{3}$) & 
      $\begin{matrix} 111s \\ 121s & \end{matrix}$ & $\begin{matrix} \frac12 \\ 0 \end{matrix}$ & $\begin{matrix} 0 \\ 1 \end{matrix}$ \\
      \midmidrule
      $[\mathcal{O}^{V,RR}_{ddu}]_{\{pq\}rs}$ & $\mathcal{O}_{u \nu d D 4}$ & $(\bar d^{\dagger}_p \idlr{\mu} \bar d^{\dagger}_q)(\bar{u}^\dagger_r \bar\sigma^\mu \nu_s)$ & $(\mathbf{1},\mathbf{10})$ & $\begin{matrix} 111s\\121s\end{matrix}$ & $\begin{matrix} \frac12 \\ 0 \end{matrix}$ & $\begin{matrix} 0 \\ 1 \end{matrix}$ \\
      \midmidrule
      $[\mathcal{O}^{V,LR}_{uud}]_{\{pq\}rs}$ & $\mathcal{O}_{d e u D 1}$ & $(u_p \idlr{\mu} u_q)(\bar{d}^\dagger_r \bar\sigma^\mu e_s)$& $(\mathbf{6},\mathbf{3})$  & $\begin{matrix} 111s \\ 112s \end{matrix}$ & $\begin{matrix} -\frac12 \\ -1 \end{matrix}$ & $\begin{matrix} 0 \\ 1 \end{matrix}$ \\
      \midmidrule
      $[\mathcal{O}^{V,RR}_{uud}]_{\{pq\}rs}$ & $\mathcal{O}_{d e u D 2}$ & $(\bar u^{\dagger}_p \idlr{\mu} \bar u^{\dagger}_q)(\bar{d}^\dagger_r \bar\sigma^\mu e_s)$ & $(\mathbf{1},\mathbf{10})$ & $\begin{matrix} 111s &  \\ 112s & \end{matrix}$ & $\begin{matrix} -\frac12 \\ -1 \end{matrix}$ & $\begin{matrix} 0 \\ 1 \end{matrix}$ \\
      \midmidrule
      $[\mathcal{O}^{V,LR}_{udu}]_{pqrs}$ & $\mathcal{O}_{u e u D 1}$ & $(u_p \idlr{\mu} d_q)(\bar{u}^\dagger_r \bar\sigma^\mu e_s)$ & $(\mathbf{6},\mathbf{3}) $ & $\begin{matrix} 111s  \\ 121r  \end{matrix}$ & $\begin{matrix} -\frac12 \\ -1 \end{matrix}$ & $\begin{matrix} 0 \\ 1 \end{matrix}$ \\
      \midmidrule
      $[\mathcal{O}^{V,LL}_{uud}]_{\{pq\}rs}$ & $\mathcal{O}_{d e u D 3}$ & $(u_p \idlr{\mu} u_q)(d_r \sigma^\mu \bar{e}^{\dagger}_s)$ & $(\mathbf{10},\mathbf{1})$  & $\begin{matrix} 111s  \\ 112s \end{matrix}$ & $\begin{matrix} -\frac12 \\ -1 \end{matrix}$ & $\begin{matrix} 0 \\ 1 \end{matrix}$ \\
      \midmidrule
      $[\mathcal{O}^{V,RL}_{uud}]_{\{pq\}rs}$ & $\mathcal{O}_{d e u D 4}$ & $(\bar u^{\dagger}_p \idlr{\mu} \bar u^{\dagger}_q)(d_r \sigma^\mu \bar{e}^{\dagger}_s)$ & $ (\mathbf{3},\mathbf{6})$& $\begin{matrix} 111s \\ 112s\end{matrix}$ & $\begin{matrix} -\frac12 \\ -1 \end{matrix}$ & $\begin{matrix} 0 \\ 1 \end{matrix}$ \\
      \midmidrule
      $[\mathcal{O}^{V,RL}_{udu}]_{pqrs}$ & $\mathcal{O}_{u e d u D 2}$ & $(\bar u^{\dagger}_p \idlr{\mu} \bar d^{\dagger}_q)(u_r \sigma^\mu \bar{e}^{\dagger}_s)$ & $ (\mathbf{3},\mathbf{6})$ & $\begin{matrix} 111s \\ 121s \end{matrix}$ & $\begin{matrix} -\frac12 \\ -1 \end{matrix}$ & $\begin{matrix} 0 \\ 1 \end{matrix}$ \\
      \bottomrule
    \end{tabular}
  \end{center}
  \caption{\label{tab:d7leftops-BL0} The table shows the $\Delta B = \Delta L = -1$ operators that enter the LEFT at dimension 7. Where appropriate, the flavour structures involving a strange quark are listed. The last two columns show the units in which hadronic isospin ($\Delta I$) and strangeness ($\Delta S$) are violated by each operator. The index $s$ labels lepton flavours.}
\end{table}

\begin{table}[tbp!]
  \begin{center}
    \begin{tabular}{llccccc}
      \toprule
      Name & Ref.~\cite{Liao:2020zyx} &  Operator  & Flavour & Indices & $\Delta I$ & $\Delta S$ \\
      \midrule
          $[\mathcal{O}^{V,RL}_{dud}]_{pqrs}$ & $\mathcal{O}_{d \nu u d D1}^{*}$ & $(\bar d^{\dagger}_p \idlr{\mu} \bar u^{\dagger}_q)(\nu^{\dagger}_r \bar\sigma^\mu d_s)$ &$(\mathbf{3},\mathbf{6})$ &  $\begin{matrix} 11r1 \\ 21r1 \\ 11r2 \end{matrix}$ & $\begin{matrix} \frac12 \\ 0 \\ 0 \end{matrix}$ & $\begin{matrix} 0 \\ 1 \\ 1 \end{matrix}$ \\
      \midmidrule
      $[\mathcal{O}^{V,LL}_{ddu}]_{\{pq\}rs}$ & $\mathcal{O}_{u \nu d D 1}^{*}$ & $(d_p \idlr{\mu} d_q)(\nu^{\dagger}_r \bar\sigma^\mu u_s)$ & $(\mathbf{10},\mathbf{1})$ & $\begin{matrix} 11r1 \\ 12r1 \end{matrix}$ & $\begin{matrix} \frac12 \\ 0 \end{matrix}$ & $\begin{matrix} 0 \\ 1 \end{matrix}$ \\
      \midmidrule
      $[\mathcal{O}^{V,RL}_{ddu}]_{\{pq\}rs}$ & $\mathcal{O}_{u \nu d D 2}^{*}$ & $(\bar{d}^{\dagger}_p \idlr{\mu} \bar{d}^{\dagger}_q)(\nu^{\dagger}_r \bar\sigma^\mu u_s)$ & $(\mathbf{3},\mathbf{6})$ & $\begin{matrix} 11r1 \\ 12r1 \end{matrix}$ & $\begin{matrix} \frac12 \\ 0 \end{matrix}$ & $\begin{matrix} 0 \\ 1 \end{matrix}$ \\
      \midmidrule
      $[\mathcal{O}^{V,LL}_{ddd}]^{\tiny\yng(3)}_{pqrs}$ & $\mathcal{O}_{d e d D 1}^{*}$ & $(d_p \idlr{\mu} d_q)(e^{\dagger}_r \bar\sigma^\mu d_s)$ & $ (\mathbf{10},\mathbf{1})$ & $\begin{matrix} 11r1 \\ 12r1 \end{matrix}$ & $\begin{matrix} \frac32 \\ 1 \end{matrix}$ & $\begin{matrix} 0 \\ 1 \end{matrix}$ \\
      \midmidrule
      $[\mathcal{O}^{V,RL}_{ddd}]_{\{pq\}rs}$ & $\mathcal{O}_{d e d D 2}^{*}$ & $(\bar{d}^{\dagger}_p \idlr{\mu} \bar{d}^{\dagger}_q)(e^{\dagger}_r \bar\sigma^\mu d_s)$ &  $(\mathbf{3},\mathbf{6})$ & $\begin{matrix} 11r1   \\ 12r1 \\ 11r2  \end{matrix}$ & $\begin{matrix} \frac32 \\ 1 \\ 1 \end{matrix}$ & $\begin{matrix} 0 \\ 1 \\ 1 \end{matrix}$ \\
      \midmidrule
      $[\mathcal{O}^{V,LR}_{ddd}]_{\{pq\}rs}$ & $\mathcal{O}_{d e d D 3}^{*}$ & $(d_p \idlr{\mu} d_q)(\bar{e}_r \sigma^\mu \bar{d}^{\dagger}_s)$ & $(\mathbf{6},\mathbf{3})$  &  $\begin{matrix} 11r1 \\ 12r1  \\ 11r2 \end{matrix}$ & $\begin{matrix} \frac32 \\ 1 \\ 1 \end{matrix}$ & $\begin{matrix} 0 \\ 1 \\ 1 \end{matrix}$ \\
      \midmidrule
      $[\mathcal{O}^{V,RR}_{ddd}]^{\tiny\yng(3)}_{pqrs}$ & $\mathcal{O}_{d e d D 4}^{*}$ & $(\bar{d}^{\dagger}_p \idlr{\mu} \bar{d}^{\dagger}_q)(\bar{e}_r \sigma^\mu \bar{d}^{\dagger}_s)$ & $(\mathbf{1},\mathbf{10})$ & $\begin{matrix} 11r1 \\ 12r1 \end{matrix}$ & $\begin{matrix} \frac32 \\ 1 \end{matrix}$ & $\begin{matrix} 0 \\ 1 \end{matrix}$ \\
      \bottomrule
    \end{tabular}
  \end{center}
  \caption{\label{tab:d7leftops-BL2} The table shows the $\Delta B = -\Delta L = -1$ operators that enter the LEFT at dimension 7. Where appropriate the flavour structures involving a strange quark are listed, along with the units in which hadronic isospin ($\Delta I$) and strangeness ($\Delta S$) are violated by each operator. For the operators $\mathcal{O}_{ddd}^{V,LL}$ and $\mathcal{O}_{ddd}^{V,RR}$ the permutation symmetry on the down-quark indices has been shown with tableaux to avoid moving the lepton-flavour index with respect to the other operators in the table. These are also the leading-order operators giving rise to the $\Delta I = \tfrac{3}{2}$ process $n \to \pi^+ e^-$.}
\end{table}

We also consider BNV operators at dimension 7 in the LEFT. These involve a single derivative, and can be important in the nucleon decay estimates coming from the operators we present in Sec.~\ref{sec:smeft}. The dimension-7 LEFT operators that violate baryon number in one unit but preserve $B-L$ are presented in Table~\ref{tab:d7leftops-BL0}, while those that violate $B-L$ in two units are shown in Table~\ref{tab:d7leftops-BL2}; in both cases the operators are presented with their $\Delta I$ and $\Delta S$ properties. We also provide the relation to the operator basis in Ref.~\cite{Liao:2020zyx}. We highlight that the $(B-L)$-violating group includes the lowest-dimensional operators mediating the $\Delta I = \tfrac{3}{2}$ process $n \to \pi^+ e_r^-$. These decays are otherwise generated at dimension 8 by the insertion of two dimension-6 operators: one $(B-L)$-preserving BNV operator and another isospin-violating operator generated by the $W$.

\subsection{Rate calculations}
\label{sec:rate-calcs}

In the following analysis we focus on nucleon decays to a meson and lepton, which generally provide the most stringent limits. Decay modes with multiple leptons are phase-space suppressed and radiative decays with a photon in the final state are suppressed by the fine-structure constant from radiating a photon off the initial or final state~\cite{Fajfer:2023gfi}. Neglecting the final-state lepton mass, the decay of a nucleon $N\in\{p,n\}$ to a meson $M$ and (anti)lepton $\ell$ is given by (see \eg Refs.~\cite{Aoki:2017puj,Yoo:2021gql})
\begin{equation} \label{eq:direct-method}
  \Gamma(N\to M + \ell) =
  \frac{m_N}{32\pi} \left(1-\frac{m_M^2}{m_N^2}\right)^2 \left|\sum_I C_I W_0^I(N\to M)\right|^2 \ ,
\end{equation}
where $C_I$ denotes the LEFT WCs, and $W_0^I$ is the nuclear matrix elements, which we take from Ref.~\cite{Yoo:2021gql}. Note that there is no interference between amplitudes with different lepton chirality in the limit of vanishing final-state lepton mass.

Two-body nucleon decays can also come about from four-fermion operators that enter the LEFT at dimension 7, which are presented in Tables \ref{tab:d7leftops-BL0} and \ref{tab:d7leftops-BL2}. We are not aware of lattice calculations of the nuclear matrix elements associated with these operators, and so we estimate the corresponding nucleon decay rates using naive dimensional analysis:
\begin{equation}
  \Gamma(N\to M + \bar{\ell}) \simeq  \frac{m_N}{32\pi f_\pi^2} \left(1 - \frac{m_M^2}{m_N^2}\right)^2 \left|\Lambda_{\rm QCD}^4 C\right|^2\,.
\end{equation}
Here $f_\pi$ is the pion decay constant, $\Lambda_{\rm QCD}$ denotes the QCD scale, and $C$ the WC of the corresponding operator in Tables~\ref{tab:d7leftops-BL0} and \ref{tab:d7leftops-BL2}.

\section{SMEFT operator classification} 
\label{sec:smeft}

Below we outline the SM effective-operator framework in which we conduct our study, along with details of how the framework relates to the LEFT operators introduced in the previous section. We remind the reader that our goal is to derive order-of-magnitude estimates for the $|\Delta B| = 1$ nucleon decay rates associated with the operators, and this informs much of the following discussion. Here and throughout the paper, we use $\mathcal{O}^*$ to represent the conjugate of an operator $\mathcal{O}$, with flavour indices indexing fields as they appear in $\mathcal{O}$.

As mentioned briefly in the Introduction, we use the term \textit{field string} to identify a string of SM multiplets whose product contains a gauge- and Lorentz-invariant part. This is called a \textit{type of operator} in Ref.~\cite{Fonseca:2019yya} and is also discussed in Ref.~\cite{Gargalionis:2020xvt}.\footnote{Note that the use of \emph{field strings} slightly differs from Ref.~\cite{Gargalionis:2020xvt} where the group theory structure is partly fixed, which is not here.} Where it is clear from context, we also simply call these \textit{operators}.

\subsection{SMEFT field strings}

The field strings in our notation are presented in Table~\ref{tab:bviolating-operators} of Appendix~\ref{sec:app-ops}; each operator is shown with a numerical label and organised by mass dimension. In addition to these properties, we present a summary of some results in the next few columns that are discussed in more detail in Sec.~\ref{sec:results}. The field-string WCs $C_i$ are dimensionless and defined as
\begin{equation}
  \mathcal{L}_{\text{eff}} = \sum_i \frac{[C_{i}]_{pqrs}}{\Lambda^{d_i-4}} [\mathcal{O}_{i}]_{pqrs} + \mathrm{h.c.} \ ,
\end{equation}
where $i$ indexes over the operator labels of Table~\ref{tab:bviolating-operators} and $d_i$ denotes the dimension of operator $\mathcal{O}_i$. 

We work with field strings that violate $B$ in one unit up to dimension 9. There are two classes of such operators: those that preserve the combination $B-L$, which enter at even mass dimension, and those that violate $B-L$ in two units, which enter at odd mass dimension. This latter class can be further split into operators that violate lepton number in one unit or in three units. We concentrate here only on those that violate lepton number in one unit, since our concern is two-body nucleon decays.\footnote{We point the interested reader to Refs.~\cite{Fonseca:2018ehk, Appelquist:2001mj} for studies of $\Delta L = 3$ physics.} A number of important comments about the operators are in order:
\begin{enumerate}
\item We remain agnostic with respect to the Lorentz structure of the operators. Different Lorentz structures will in general be related by Fierz transformations, which introduce $\mathcal{O}(1)$ factors that should not affect our estimates excessively. This motivates our use of field strings.
\item Unlike the case of estimating matching contributions to the Majorana neutrino mass term, we find the $\mathrm{SU}(2)$ isospin structure of the operators not to be important when estimating nucleon decay rates. That is, different isospin structures give rise to $\Delta B = -1$ decay amplitudes differing only by $\mathcal{O}(1)$ group-theory factors. This can be understood by noting that at dimensions 6 and 7 in the SMEFT we can work in a basis in which individual operators are uniquely identifiable on the basis of their field content, and there is no possibility to generate the vanishing structure $\epsilon_{ij} H^{i} H^{j}$, unlike the case of the dimension-5 Weinberg operator.\footnote{In Refs.~\cite{deGouvea:2007qla,Angel:2012ug,Gargalionis:2020xvt} different $\mathrm{SU}(2)$ structures are necessary to distinguish since the number of neutrinos present in the operator is important for estimating the matching onto the dimension-3 term in the LEFT\@. Equivalently, we need to be sure that we are not matching onto the structure $L^{i} L^{j} H^{k} H^{l} \epsilon_{ij}\epsilon_{kl}$. No such ambiguity is present among the dimension-6 and -7 field strings we list. Although it seems that there are two $\mathrm{SU}(2)$ contractions allowed for operator $\mathcal{O}_8$, these correspond to different flavour-index permutations on the SMEFT operator $\mathcal{O}_{\bar{l}dqq\tilde{H}}$.}
\item Each numbered operator in our list has free flavour indices, which we have chosen to leave implicit in the table. These are understood to be labelling the SM fermions in each field string in the order $\{p,q,r,s,t,u\}$ as they appear in each row. For example, $\mathcal{O}_{11}$ carries flavour indices $DL_p Q_q Q_r \bar{d}^\dagger_s H$, while $\mathcal{O}_{25}$ is $\bar{e}^\dagger_p \bar{e}^\dagger_q \bar{e}_r \bar{d}_s \bar{d}_t \bar{d}_u$. It is often the case that statements can already be made about the permutation-symmetry properties of the associated coefficients in flavour space, but since we do not work with a definite basis of operators we do not represent these directly on the field strings themselves. These flavour-index permutation symmetries can play an important role in phenomenological predictions, since they can dictate the hierarchy with which the LEFT operators of Tables~\ref{tab:wet-ops-BL0} and \ref{tab:wet-ops-BL2} are generated, and therefore which decay modes dominate. Where appropriate we comment on these when interpreting our results.
\item All of the operators contain at least three quark or antiquark fields contracted into an $\mathrm{SU}(3)_c$-singlet with the totally antisymmetric three-index tensor.
\item We account for operators that contain at most one derivative. This is motivated by the number of novel UV models these operators might play a role in constraining~\cite{Gargalionis:2020xvt}.\footnote{In Ref.~\cite{Gargalionis:2020xvt} it was found that operators with more than one derivative only rarely implied tree-level UV completions not already generating similar operators with fewer derivatives. We point the interested reader to Sec.~3.2.1 of that study for more details.} It should be understood that the derivative can act  on all the different fields within the operator.
\end{enumerate}

All of these points mean that each numbered field string in our listing should be understood to stand in for a family of operators of the same field content with the derivative acting on different fields, and potentially differing in colour, isospin and Lorentz structure. Using the operators as we have presented them, rather than a complete basis up to dimension 9 as has been presented in Refs.~\cite{Murphy:2020rsh,Li:2020gnx,Li:2020xlh,Liao:2020jmn}, allows us to work with significantly fewer structures without compromising our matching estimates by more than an order of magnitude.

Our listing includes a total of 50 numbered field strings. The field content of the operators is derived using the Hilbert series~\cite{Henning:2015alf}, projecting out the $\Delta B = -1$ component for the pertinent operators by the method described in Ref.~\cite{Lehman:2015via} and removing the spurions accounting for redundancies from field redefinitions involving the classical equations of motion and integration by parts. The restriction to at most one derivative is made by hand after the full set of field-string operators has been written down.

\subsection{SMEFT-LEFT operator matching at tree level}\label{sec:smeft-left-tree-level-matching}

Our general approach is to estimate the matching contributions onto the LEFT by closing off the operators in our listing with $B$- and $L$-conserving SM interactions so as to give rise to a SMEFT operator of the lowest allowed dimension, \ie dimension 6 for even-dimensional operators, and dimension 7 for odd-dimensional ones. (We also consider the direct tree-level contributions at dimensions 8 and 9, but these are typically subdominant on dimensional grounds.) We impose a definite structure onto those field strings, i.e.~identify them with genuine SMEFT operators, and match them onto the LEFT at tree level. The LEFT operator coefficients are then used to calculate the relevant two-body proton and neutron decays.

This matching requires that we use a genuine basis of operators in the SMEFT at dimensions 6 through to 9. At dimension 6 we use the Warsaw basis~\cite{Grzadkowski:2010es} and at dimension 7 the basis of Ref.~\cite{Liao:2020zyx}. We note that some of the dimension-6 and dimension-7 LEFT operators are only generated at higher-order in the SMEFT; examples are $\mathcal{O}^{S,RL}_{ddu}$, $\mathcal{O}^{S,LL}_{ddd}$, and $\mathcal{O}^{S,LR}_{ddu}$. The additional operators up to dimension 9 are either included in the lists provided in Refs.~\cite{Liao:2020jmn,Li:2020xlh} or can be related to them using equations of motion.

We present all of these operators in Table~\ref{tab:tableaux}, ordered by mass dimension. Up to flavour, there are four independent operators with $\Delta B=\Delta L=-1$ at dimension 6, and there are six independent operators with $\Delta B = -\Delta L = -1$ at dimension 7. To account for the full matching onto the LEFT, we also include one dimension-8 operator and five dimension-9 ones. In the table we have suppressed colour indices for notational clarity; it is understood that the three quarks in each bilinear are contracted using a Levi-Civita tensor with indices in the order of the quarks in the operators. We account for the permutation symmetries of the flavour indices for each of the corresponding operator coefficient by implementing in our code the relations found in Refs.~\cite{Grzadkowski:2010es,Liao:2019tep,Liao:2020jmn,Li:2020gnx,Li:2020xlh}. These are shown in Table~\ref{tab:tableaux} for repeated fermion fields only. The SMEFT operator coefficients are normalised such that
\begin{equation}
  \mathcal{L}_{\text{eff}} = \sum_i \frac{[C_{i}]_{pqrs}}{\Lambda^{d-4}}[\mathcal{O}_{i}]_{pqrs} + \mathrm{h.c.} \,
\end{equation}
where $i$ indexes over the operator labels of Table~\ref{tab:tableaux}. The leading-order matching contribution of these operators to the dimension-6 LEFT is presented in Table~\ref{tab:SMEFT-LEFT}, following Refs.~\cite{Jenkins:2017jig,Liao:2020zyx} where appropriate.

\begin{table}[tbp!]
  \begin{center}
    \begin{tabular}{llc}
      \toprule
      Name & Operator & Permutation symmetry  \\
      \midrule
      \multicolumn{3}{l}{Dimension 6} \\
      $\mathcal{O}_{qqql}$ & $(Q_p^{i} Q_q^{j}) (Q_r^{l}L_s^k )\epsilon_{ik} \epsilon_{jl}$ & $\tiny\Yvcentermath1 \yng(3) \oplus \yng(2,1) \oplus \yng(1,1,1)$ \\
      $\mathcal{O}_{qque}$ & $(Q^{i}_{p} Q^{j}_{q}) (\bar u_r^{\dagger}\bar e_s^\dagger) \epsilon_{ij}$ & $\tiny\Yvcentermath1 \yng(2)$ \\
      $\mathcal{O}_{duue}$ & $(\bar d_p^{\dagger} \bar u_q^{\dagger}  )(\bar u_r^{\dagger}\bar e_s^\dagger )$ & $\tiny\Yvcentermath1 \yng(2) \oplus \yng(1,1)$ \\
      $\mathcal{O}_{duql}$ & $(\bar d_p^{\dagger}\bar u_q^{\dagger} ) ( Q_r^{i} L_s^j) \epsilon_{ij}$ & --- \\
      \midmidrule
      \multicolumn{3}{l}{Dimension 7} \\
      $\mathcal{O}_{\bar l dddH}$ & $(L_p^{\dagger} \bar d_q^{\dagger}) (\bar d^{\dagger}_{r} \bar d^{\dagger}_{s}) H$ & $\tiny\Yvcentermath1 \yng(2,1)$ \\
      $\mathcal{O}_{\bar ldqq \tilde H}$ & $(L_p^{\dagger} \bar d_q^{\dagger}) (Q_r Q_s^{i} ) \tilde H^j  \epsilon_{ij}$ & $\tiny\Yvcentermath1 \yng(2) \oplus \yng(1,1)$ \\
      $\mathcal{O}_{\bar e qdd\tilde H}$ & $(\bar e_p Q_q^{i})(\bar d_r^{\dagger} \bar d_s^{\dagger}) \tilde H^j \epsilon_{ij}$ & $\tiny\Yvcentermath1 \yng(1,1)$ \\
      $\mathcal{O}_{\bar ldud\tilde H}$ & $(L_p^{\dagger} \bar d_q^{\dagger})(\bar u^{\dagger}_r \bar d^{\dagger}_s) \tilde H$ & $\tiny\Yvcentermath1 \yng(2) \oplus \yng(1,1)$ \\
      $\mathcal{O}_{\bar lqdDd}$ & $(L^{\dagger}_p \bar \sigma^\mu Q_q) ( \bar d_r^{\dagger} \idlr{\mu} \bar d_s^{\dagger})$ & $\tiny\Yvcentermath1 \yng(2)$ \\
      $\mathcal{O}_{\bar edddD}$ & $(\bar e_p \sigma^\mu \bar d_q^{\dagger}) ( \bar d_r^{\dagger} \idlr{\mu} \bar d_s^{\dagger})$ & $\tiny\Yvcentermath1 \yng(3)$ \\
      \midmidrule
      \multicolumn{3}{l}{Dimension 8} \\
      $\mathcal{O}_{ddqlHH}$ & $(\bar d_{p}^{\dagger} \bar d_{q}^{\dagger} )( Q_r^{i} L_s^j) H^k H^l \epsilon_{ik} \epsilon_{jl}$ &  $\tiny\Yvcentermath1 \yng(1,1)$ \\
      \midmidrule
      \multicolumn{3}{l}{Dimension 9} \\
      $\mathcal{O}_{eqqqHHH}$ & $(\bar e_{p}^{\dagger} Q_{qi}^{\dagger} )( Q_{rj}^{\dagger} Q_{sk}^{\dagger}) H^i H^j H^k$ & $\tiny\Yvcentermath1 \yng(2,1)$ \\
      $\mathcal{O}_{luqqHHH}$ & $(L_p^i \bar u_{q})( Q_{rj}^{\dagger} Q_{sk}^{\dagger}) H^{i'} H^j H^k \epsilon_{ii'}$ & $\tiny\Yvcentermath1 \yng(1,1)$ \\
      $\mathcal{O}_{qqedHHD}$ & $(Q^{i}_p \idlr{\mu} Q^{j}_{q}) (\bar e_r \sigma^\mu \bar d^{\dagger}_s) \tilde H^k \tilde H^l \epsilon_{ik}\epsilon_{jl}$ & $\tiny\Yvcentermath1 \yng(2)$ \\
      $\mathcal{O}_{qqlqHHD}$ & $(Q^{i}_p \idlr{\mu} Q^{j}_{q}) (L^{\dagger}_r \bar\sigma^\mu Q_s) \tilde H^k \tilde H^l \epsilon_{ik} \epsilon_{jl}$ & $\tiny\Yvcentermath1 \yng(2)$ \\
      $\mathcal{O}_{udqlHHD}$ & 
      $(\bar u_p \idlr{\mu} \bar d_{q}) (Q^{\dagger}_{ri} \bar\sigma^\mu L_s^j) H^i H^k \epsilon_{jk}$ & --- \\
      \bottomrule
    \end{tabular}
  \end{center}
  \caption{\label{tab:tableaux} The table shows the SMEFT operators that match onto the $|\Delta B| = 1$ LEFT at tree level broken up by mass dimension. We show the permutation symmetry on repeated-fermion flavour indices as tableaux in the last column. These are adapted from the flavour relations given in Refs.~\cite{Grzadkowski:2010es,Liao:2019tep,Liao:2020jmn,Li:2020xlh,Li:2020gnx}.}
\end{table}

As discussed above, we intend the field strings listed in Table~\ref{tab:bviolating-operators} to stand in for operators of a variety of Lorentz, colour and isospin structures, with derivatives, if present, acting on the fields in all possible ways. Our goal here is to establish a relationship between the coefficients of the field strings of Table~\ref{tab:bviolating-operators}, with which we work in our matching algorithm, and the coefficients of the operators introduced in Table~\ref{tab:tableaux}. Of course, any specific UV model will typically generate several concrete and explicit operators. These can always be written as linear combinations of the operators presented above \emph{with $\mathcal{O}(1)$ coefficients} arising from group-theoretic transformations, e.g.\ Fierz transformations and $\mathrm{SU}(2)$ Schouten identities. Thus, a mapping can be made between the coefficients we work with in our matching algorithm and those of the operator spanning set presented in Table~\ref{tab:tableaux} up to $\mathcal{O}(1)$ factors, which do not alter our results in a meaningful way.

\begin{table}[tbp!]
  \begin{center}
  \begin{tabular}{lc}
      \toprule
      Name & SMEFT matching\\ 
      \midrule
      \([\mathcal{O}_{udd}^{S,LL}]_{pqrs}\) & $V_{q'q}V_{r'r}(C_{qqql,r'q'ps}-C_{qqql,q'r'ps}+C_{qqql,q'pr's})$ \\ 
      \([\mathcal{O}_{duu}^{S,LL}]_{pqrs}\) & $V_{p'p} (C_{qqql,rqp's}-C_{qqql,qrp's}+C_{qqql,qp'rs})$ \\
      \([\mathcal{O}_{duu}^{S,LR}]_{pqrs}\) &  $-V_{p'p} (C_{qque,p'qrs}+C_{qque,qp'rs})$\\
      \([\mathcal{O}_{duu}^{S,RL}]_{pqrs}\) & $C_{duql,pqrs}$\\
      \([\mathcal{O}_{dud}^{S,RL}]_{pqrs}\) & $-V_{r'r}C_{duql,pqr's}$\\
      \([\mathcal{O}_{ddu}^{S,RL}]_{pqrs}\)  &  $(C_{ddqlHH,pqrs}-C_{ddqlHH,qprs})  \frac{v^2}{2\Lambda^2} $\\  
      \([\mathcal{O}_{duu}^{S,RR}]_{pqrs}\) & $C_{duue,pqrs}$\\
      \midmidrule
      \([\mathcal{O}_{ddd}^{S,LL}]_{pqrs}\) & $V_{s's}V_{p'p}V_{q'q}
      (C_{eqqqHHH,rs'p'q'}
      -C_{eqqqHHH,rs'q'p'})
      \frac{v^3}{2\sqrt{2}\Lambda^3}$\\
      \([\mathcal{O}_{udd}^{S,LR}]_{pqrs}\) &  $-V_{q'q}C_{\bar ldqq\tilde H,rspq'} \frac{v}{\sqrt{2}\Lambda}$\\ 
      \([\mathcal{O}_{ddu}^{S,LR}]_{pqrs}\) &  $ V_{p'p}V_{q'q} (C_{luqqHHH,rsp'q'}-C_{luqqHHH,rsq'p'})\frac{v^3}{2\sqrt{2}\Lambda^3}$\\
  \([\mathcal{O}_{ddd}^{S,LR}]_{pqrs}\) & $V_{p'p}V_{q'q} (C_{\bar l dqq\tilde H,rsq'p'} -C_{\bar l dqq\tilde H,rsp'q'} )\frac{v}{2\sqrt{2}\Lambda}$\\ 
      \([\mathcal{O}_{ddd}^{S,RL}]_{pqrs}\) &  $V_{s's} ( C_{\bar eqdd\tilde H,rs'qp}-C_{\bar eqdd\tilde H,rs'pq})  \frac{v}{2\sqrt{2}\Lambda}$ \\
      \([\mathcal{O}_{udd}^{S,RR}]_{pqrs}\) & $C_{\bar l dud\tilde H,rspq}\frac{v}{\sqrt{2}\Lambda}$\\
      \([\mathcal{O}_{ddd}^{S,RR}]_{pqrs}\) & $C_{\bar ldddH,rspq} \frac{v}{\sqrt{2}\Lambda}$ \\
      \midmidrule
     \([\mathcal{O}_{ddu}^{V,RL}]_{pqrs}\) &  $-C_{\bar lqdDd,prsq}$\\
      \([\mathcal{O}_{ddd}^{V,RL}]_{pqrs}\) &  $-V_{s's} C_{\bar l qdDd,rs'pq}$\\
      \([\mathcal{O}_{ddd}^{V,RR}]_{pqrs}\) &  $-C_{\bar e dddD,rspq}$\\
      \([\mathcal{O}_{ddu}^{V,LL}]_{pqrs}\) &  $V_{p'p}V_{q'q} (C_{qqlqHHD,p'q'rs}+C_{qqlqHHD,q'p'rs})  \frac{v^2}{4\Lambda^2}$\\ 
      \([\mathcal{O}_{ddd}^{V,LL}]_{pqrs}\) &  $V_{p'p}V_{q'q}V_{s's} (C_{qqlqHHD,p'q'rs'}+C_{qqlqHHD,q'p'rs'})  \frac{v^2}{4\Lambda^2}$ \\
      \([\mathcal{O}_{ddd}^{V,LR}]_{pqrs}\) &  $V_{p'p}V_{q'q}(C_{qqedHHD,p'q'rs} + C_{qqedHHD,q'p'rs})\frac{v^2}{4\Lambda^2}$\\
      \([\mathcal{O}_{dud}^{V,RL}]_{pqrs}\) & $-V_{s's}C_{udqlHHD,qps'r}^* \frac{v^2}{2\Lambda^2}$ \\
      \bottomrule
    \end{tabular}
  \end{center}
  \caption{The table shows the tree-level matching of the SMEFT operators up to dimension 9 (see Refs.~\cite{Jenkins:2017jig,Liao:2016hru,Liao:2019tep}) onto the dimension-6 scalar and dimension-7 vector LEFT operators listed in Tables \ref{tab:wet-ops-BL0} and \ref{tab:wet-ops-BL2} (see Ref.~\cite{Jenkins:2017jig}).}\label{tab:SMEFT-LEFT}
\end{table}

For the field strings not containing derivatives, the mapping of the coefficients simply follows by identifying each field string (e.g.\ $\mathcal{O}_1$) with the operator of the same field content (e.g.\ $\mathcal{O}_{qqql}$):
\begin{equation} \label{eq:unique-mappings}
  \begin{aligned}
    \relax[\mathcal{O}_1]_{pqrs} & \to [\mathcal{O}_{qqql}]_{qrsp}
    &
    [\mathcal{O}_2]_{pqrs} & \to [\mathcal{O}_{qque}]_{spqr}
    &
    [\mathcal{O}_3]_{pqrs} & \to [\mathcal{O}_{duue}]_{sqrp}
    \\
    [\mathcal{O}_4]_{pqrs} & \to [\mathcal{O}_{duql}]_{srqp}
    &
    [\mathcal{O}_5]_{pqrs} & \to [\mathcal{O}_{\bar l dddH}]_{pqrs}^*
    &
    [\mathcal{O}_8]_{pqrs} & \to [\mathcal{O}_{\bar ldqq\tilde H}]_{psqr}^*
    \\
    [\mathcal{O}_9]_{pqrs} & \to [\mathcal{O}_{\bar eqdd\tilde H}]_{pqrs}^*
    &
    [\mathcal{O}_{10}]_{pqrs} & \to [\mathcal{O}_{\bar ldud\tilde H}]_{prqs}^*
    &
    [\mathcal{O}_{16}]_{pqrs} &\to [\mathcal{O}_{ddqlHH}]_{rsqp} 
    \\ 
    [\mathcal{O}_{26}]_{pqrs} & \to [\mathcal{O}_{eqqqHHH}]_{pqrs}
    &
    [\mathcal{O}_{37}]_{pqrs} &\to [\mathcal{O}_{luqqHHH}]_{psqr}
    \;.
  \end{aligned}
\end{equation}
For the operators with derivatives, integration-by-parts relations and field redefinitions involving the equations of motion also need to be taken into account when mapping onto the genuine SMEFT operators. This means that in this case the mapping is not a bijection. Here, the exact pattern of operators generated in a specific UV model cannot be predicted, since it depends on the position of the derivative in the operator generated. By covering every possibility in our results, we allow anyone to query for the actual operators generated in our online database. For the results presented in the paper, we assume that all of the operators shown are generated. We highlight that all SMEFT operators except those containing derivatives enter these expressions suppressed by SM Yukawa couplings. At dimension 7 the relations are
\begin{equation}
  \label{eq:67mapping}
  \begin{aligned}
    \mathcal{O}_{6,{pqrs}} & \to [\mathcal{O}_{\bar lqdDd}]_{pqrs}^*\\
    \mathcal{O}_{6,pqrs} & \to [y_e]_p [\mathcal{O}_{\bar{e}qdd\tilde{H}}]_{pqrs}^* \\
    \mathcal{O}_{6,pqrs} & \to \begin{cases}
    [y_u]_q [\mathcal{O}_{\bar ldud\tilde{H}}]^*_{prqs} + \sum_x [y_d]_q V^{*}_{xq} [\mathcal{O}_{\bar l dddH}]^*_{prxs} \\
    \sum_x [y_d]_r V^{*}_{xr} [\mathcal{O}_{\bar ldqq\tilde{H}}]_{psqx}^*
    \end{cases} \\
    \mathcal{O}_{7,pqrs} & \to [\mathcal{O}_{\bar edddD}]_{pqrs}^* \\
    \mathcal{O}_{7,pqrs} & \to [y_e]_p [\mathcal{O}_{\bar ldddH}]_{pqrs}^* \\
    \mathcal{O}_{7,pqrs} & \to \sum_x [y_d]_x V^{*}_{qx}[\mathcal{O}_{\bar eqdd\tilde{H}}]_{pxrs}^* \;.
  \end{aligned}
\end{equation}
In the third line, we choose to distinguish the cases where the derivative acts upon the quark doublet (top) and the right-handed down-type quark (bottom) in the field-string operator $\mathcal{O}_6$. At dimension 9, not all relations need to be accounted for, since only a subset give rise to the operators listed in Table~\ref{tab:tableaux}. That is, only the mappings of the operators that give tree-level contributions to the $d=7$ LEFT, i.e.\ that are present in Table~\ref{tab:SMEFT-LEFT}, are relevant:
\begin{equation}
  \begin{aligned}
    \mathcal{O}_{34,pqrs} &\to [\mathcal{O}_{qqedHHD}]_{qrps}^* \\
    \mathcal{O}_{34,pqrs} &\to \sum_x [y_d]_x V^{*}_{sx} [\mathcal{O}_{eqqqHHH}]_{pqrx}^* \\
    \mathcal{O}_{38,pqrs} &\to [\mathcal{O}_{qqlqHHD}]_{qrps}^* \\
    \mathcal{O}_{38,pqrs} &\to \begin{cases}
    [y_{e}]_{p} [\mathcal{O}_{eqqqHHH}]_{pqrs} \\
    [y_{u}]_{q} [\mathcal{O}_{luqqHHH}]_{pqrs} + \mathcal{O}(v^2 / \Lambda^2)
    \end{cases} \\
    \mathcal{O}_{44,pqrs} &\to \begin{cases}
    [\mathcal{O}_{udqlHHD}]_{rsqp} \\
    \sum_x [y_{d}]_{x} V^{*}_{xs} [\mathcal{O}_{luqqHHH}]_{prsq}
    \end{cases} \;.
  \end{aligned} \\
\end{equation}
Here we have again distinguished cases where the derivative is acted upon different fields as above. In the case of $\mathcal{O}_{38}$ the dimension-7 operator $\mathcal{O}_{\bar l d qq \tilde H}$ can also be generated if the derivative is acted upon one of the $Q$ fields, but this contribution is suppressed by $v^2 / \Lambda^2$.

We highlight that some of the SMEFT coefficients derived in this way may vanish identically, because of the permutation symmetry of the flavour indices on the operator coefficients.\footnote{For example, a non-zero $[C_7]_{1111}$ does not imply generation of the operator $[\mathcal{O}_{\bar e q dd \tilde H}]_{1111}$, since the term listed in Table~\ref{tab:tableaux} is antisymmetric in the down-quark flavour indices and therefore the operator vanishes.}

\section{Matching algorithm}\label{sec:method}

We are interested in estimating the loop-level matching to the dimension-6 and dimension-7 BNV SMEFT operators in UV models that simultaneously generate dimension-8 and dimension-9 SMEFT operators at tree level, respectively. This is in general a daunting task, since the matching may occur at a high loop level, and flavour-changing interactions imply that a multiplicity of SMEFT operators is generated. Naturally, the problem lends itself to an automated approach, and in this section we describe the algorithm we use to estimate the matching. Readers uninterested in the details of our algorithm may skip this section and directly jump to the results in Sec.~\ref{sec:results}.

Our procedure is to define a term-rewriting system based on reduction rules that act on the set of field strings presented in Table~\ref{tab:bviolating-operators}. These reduction rules implement renormalisable SM interactions, and they are designed to reduce each operator of dimension 8 or 9 down to SMEFT operators of dimension 6 or 7,\footnote{As mentioned earlier, since the renormalisable SM interactions respect $B$ and $L$, the rules reduce dimension-8 operators in our listing to those of dimension 6, and dimension-9 operators to dimension-7 ones.} respectively (\ie to one of those labelled 1--10 in our notation). These then match onto the LEFT operators of Tables~\ref{tab:wet-ops-BL0} and \ref{tab:wet-ops-BL2} at tree level after applying the identification procedure discussed in the previous section. In addition, there may be direct contributions to the tree-level matching onto the LEFT already at dimension 8 or 9. We consider these contributions separately and add them into the expressions derived from the tree-level matching at dimension 6 and 7, which we generically expect to dominate on dimensional grounds.

The reduction of a SMEFT operator $\mathcal{O}_{d}$ (of dimension $d$) to another $\mathcal{O}_{d-2}$ (of dimension $d-2$) can be represented in the following way:
\begin{equation}
  \label{eq:rule-string}
  \mathcal{O}_{d} \to_{a_1} \mathcal{O}_{a_1} \to_{a_2} \mathcal{O}_{a_1 a_2}
  \to_{a_i} \cdots \to_{a_n} \mathcal{O}_{d-2} \ ,
\end{equation}
where the $``\to_{a_i}"$ are the \emph{reduction rules} and the $\mathcal{O}_{\prod_{i} a_{i}}$ are intermediate states. Eq.~\eqref{eq:rule-string} corresponds to a Feynman diagram that relates $\mathcal{O}_{d}$ to $\mathcal{O}_{d-2}$. Each rule $\to_{a_i}$ is assigned a weight analogous to a Feynman rule, such that the product of the weights of a string of rule applications like Eq.~\eqref{eq:rule-string} is equal to the estimate for the amplitude derived from the corresponding diagram. It may be the case that the same rule could be applied in different ways, or indeed that multiple rules could be applied to the same state. (This corresponds to the existence of many diagrams relating $\mathcal{O}_{d}$ and $\mathcal{O}_{d-2}$.) We keep track of each possible history of the reduction, such that all diagrams allowed by the defined set of reduction rules are recovered. This defines a tree structure that we call a \textit{matching tree}, with $\mathcal{O}_{d}$ as the root node and in general multiple $\mathcal{O}_{d-2}^{(i)}$ (corresponding to operators 1--10) as terminal states.

\subsection{Matching estimates through symbolic rewriting}

We organise the rules and their application so that only the matching contributions at leading-loop order are generated. This is done by prioritising certain families of rules, which we introduce and discuss below.

The simplest kind of rule allows us to loop off pairs of the same multiplet in the situation where one of the two fields is conjugated. In the case where $f$ is a fermion, the spinor structure of the LHS of the rule demands that we introduce a derivative on the RHS, and diagrammatically it is clear that without an additional factor of momentum the loop by itself would vanish~\cite{deGouvea:2007qla}. These $2 \to 0$ rules remove two fields from the operator at the cost of at most one loop factor.

\begin{figure}[t]
  \centering
  \includegraphics[width=0.4\textwidth]{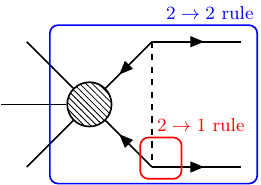}
  \caption{\label{fig:example-yukawa-222} The figure shows an illustration of the composition of two $2 \to 1$ rules into a $2 \to 2$ rule. The overall rule (shown in blue) is $2 \to 2$ since it maintains the number of fields at two through the transformation. The fermion--boson vertex (shown in red) can be interpreted here as a $2 \to 1$ rule or a $1 \to 2$ rule, since it represents the absorption or emission of the boson by the fermion. We follow Ref.~\cite{Dreiner:2008tw} for our arrow conventions.}
\end{figure}

Rules derived directly from a Yukawa interaction $\psi_1 \psi_2 S$ --- with $\psi_i$ distinct fermions and $S$ a scalar --- could be $2 \to 1$ (\eg $\psi_1 \psi_2 \to S^{\dagger}$) or $1 \to 2$ (\eg $S \to \psi_1^{\dagger} \psi_2^{\dagger}$). The $1 \to 2$ rules transform the operators of dimension 8 and 9 into higher-dimensional ones and need to be used together with $2 \to 1$ rules to ensure that all branches in the matching tree converge to one of the operators 1--10. For this reason, we define $1 \to 1$ and $2 \to 2$ rules, which represent the application of a $2 \to 1$ \textit{and} a $1 \to 2$ rule. Diagrammatically, the $1 \to 1$ rules correspond to bubbles --- and are therefore not interesting for our purposes --- and the $2 \to 2$ rules represent loops between fields in the operators. These $2 \to 2$ rules change the field content of the operator at the cost of coupling and loop suppression. They may or may not alter the dimension of the operator. A generalised example of a Yukawa-type $2 \to 2$ rule is pictured in Fig.~\ref{fig:example-yukawa-222}. There can also be $2 \to 2$ rules derived from vector couplings in an analogous way. These can relate different $\mathrm{SU}(2)$ contractions of operators with the same field content, but we find these rules to be unimportant when matching onto BNV operators.\footnote{This is not true in the LNV case, since there it is often necessary to turn an electron into a neutrino; see \eg Refs.~\cite{deGouvea:2007qla,Cai:2014kra,Gargalionis:2020xvt} for more details.}

Reductions involving only $2 \to 0$ and $2 \to 1$ rules correspond to diagrams that match onto a lower dimensional operator with the fewest loops. These tend to dominate the matching expressions, and are therefore prioritised by our algorithm. Our approach is to perform the reduction using only $2 \to 0$ and $2 \to 1$ rules, yielding a collection of constraints on field strings with restricted flavour indices.\footnote{The constraints on these field strings can be generalised to the remaining flavour structures of each field string by a manual application of a $2 \to 2$ rule, adding an extra loop to the matching diagram. We present an example of this procedure in App.~\ref{sec:manual-2to2-rules}.}

Importantly, for the rules to act on $\mathrm{SU}(2)$ irreps consistently, the corresponding weights must also respect the symmetry; that is, the Feynman rules cannot distinguish isospin components explicitly. For this reason we work in the unbroken phase and the weak-eigenstate basis in which the SM Yukawa-coupling matrices $\mathbf{y}_f$ are not diagonal. That is, rotation to the mass basis is performed only at the end of the procedure through
\begin{equation}
  \mathbf{y}_e \to \mathrm{diag}(y_e, y_\mu, y_\tau),\quad  \mathbf{y}_d \to \mathrm{diag}(y_d, y_s, y_b) V^\dagger,\quad \mathbf{y}_u \to \mathrm{diag}(y_u, y_c, y_t) \;,
\end{equation}
where $V$ is the regular Cabibbo-Kobayashi-Maskawa (CKM) (quark mixing) matrix.\footnote{Note that we neglect contributions to the CKM matrix elements from dimension-6 operators.} In this convention, weak-eigenstate and mass-eigenstate flavour indices agree for all fermions except left-handed down-type quarks, which must be rotated according to
\begin{equation}
  \hat{d}_p \to \sum_x V_{px} d_x \;
\end{equation}
when matching onto the LEFT. Here the hat furnishes fields in the weak-eigenstate basis. We add that throughout this paper it is never necessary to distinguish between neutrino flavours, and we work with weak-eigenstate neutrinos that we do not write hatted, \ie $\mathrm{\nu}_i$ with $i \in \{e, \mu, \tau\}$.

The algorithm as discussed above is implemented in \textsc{Mathematica} code and made publicly available~\cite{Gargalionis_Operator_closure_estimates_2023}. The program defines a large set of reduction rules that act on computational representations of the field-string operators of Table~\ref{tab:bviolating-operators}.

Our estimation of the matching necessarily relies on rules, weights and assumptions that greatly simplify an otherwise complicated computation. In the diagrammatic language, we include only a factor of $(16\pi^2)^{-1} \approx 6.3 \cdot 10^{-3}$ for each loop, and take the SM Feynman rules as weights for the reduction rules representing the regular SM interactions in the Feynman gauge. All unknown dimensionless coefficients are set to unity in our numerical calculations. Since we do not work with operators with well-defined Lorentz structure, we cannot ascribe significance to any relative signs between the operator coefficients, and we derive limits assuming only one non-zero coefficient at a time. Finally, we emphasise that we only take into account the leading contribution in the expansion of the SM fermion masses over the cutoff scale.

\subsection{Example of the procedure}\label{sec:example-applications}

Below we illustrate the procedure explicitly for two examples: The dimension-8 operator $\mathcal{O}_{16}$ and the dimension-9 $\mathcal{O}_{49}$. In each case we begin by applying our matching algorithm and deriving the matching estimates onto the field-string operators 1--10 in the $\mathrm{SU}(2)$-covariant formalism. We then apply the identification procedure discussed in Sec.~\ref{sec:smeft-left-tree-level-matching} and match the corresponding SMEFT operators onto the LEFT and change to the mass-eigenstate basis.

\begin{figure}[tbp!]
  \centering
  \includegraphics[width=0.8\textwidth]{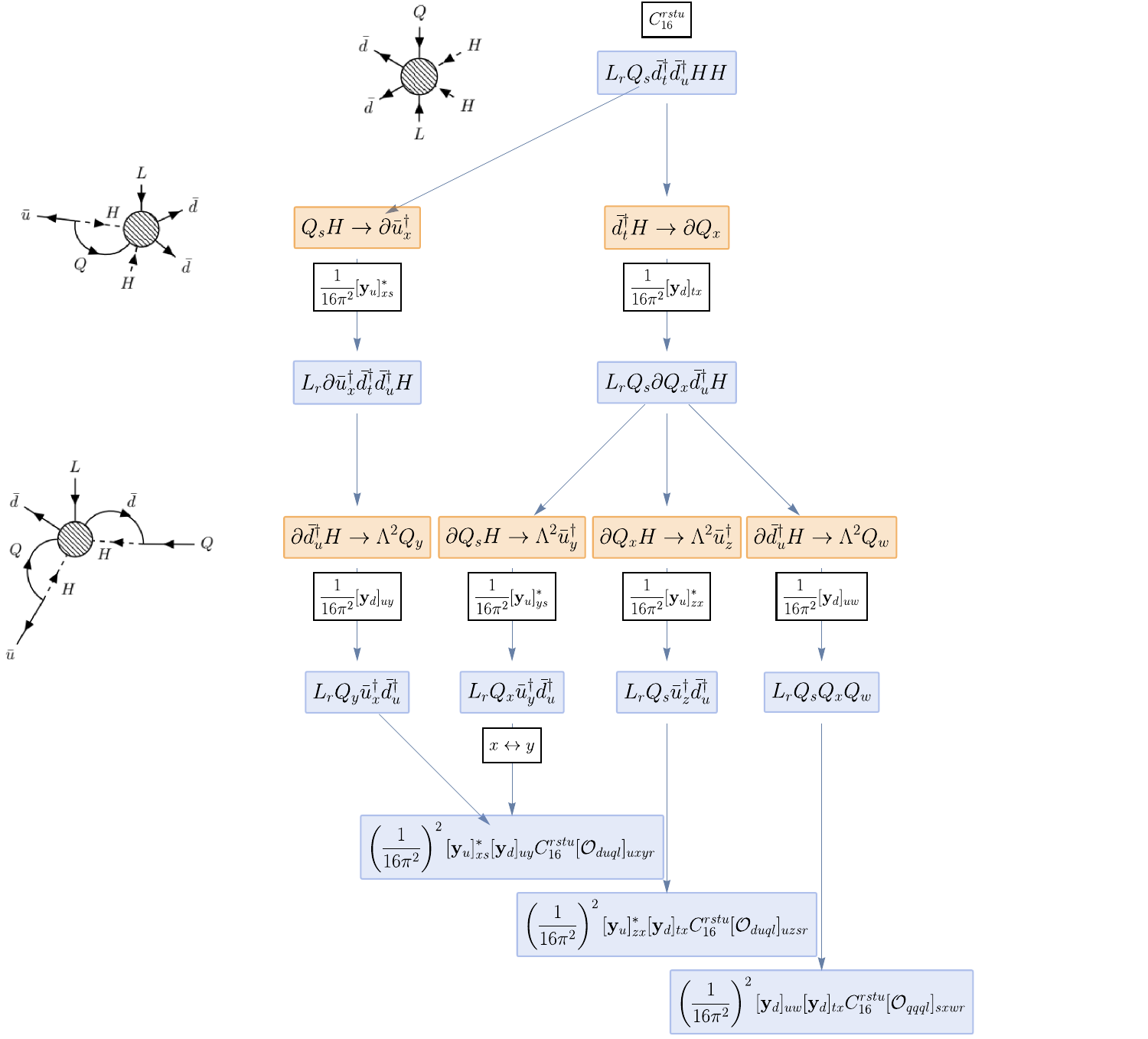}
  \caption{\label{fig:o16-matching-tree} The graph represents the possible reduction paths of the operator $\mathcal{O}_{16}$ onto one of dimension-6 operators in our listing. Blue nodes represent states in the rewrite system, while orange nodes show rule applications. The weight associated with each rule-node is shown immediately below it. The terminal blue nodes represent the matching estimates derived from the product of the weights of each path. Feynman diagrams corresponding to the rules applied are shown only for the left-most branch in the graph. Factors of $\Lambda$ are introduced into the rules to explicitly track changes in operator dimension.}
\end{figure}

The matching tree for the instructive case of $\mathcal{O}_{16}$ is presented in Fig.~\ref{fig:o16-matching-tree}. In our notation, blue nodes in the graph represent states of the rewrite system, while orange nodes show the application of rules. For the left-most branch of the graph only, we show the corresponding diagram next to the appropriate nodes. The weight associated with each rule is shown inside a white box immediately below it. In our convention, the root node also contributes a non-trivial weight representing the field-string operator coefficient. The process begins with the full field-string operator $L_r Q_s \bar{d}^\dagger_t \bar{d}^\dagger_u H H$, which contributes the coefficient ${[C_{16}]}_{rstu}$ to the matching amplitude estimate. There are two reduction rules that can act on this state, derived respectively from the up- and down-type Yukawa terms.\footnote{We note that an additional possibility is for one of the Higgs fields to be replaced by its vacuum expectation value. In this case, a dimension-7 operator in the LEFT would be generated, whose contribution is suppressed compared to the generated dimension-6 operator.} Both rules are $2 \to 1$ in the sense that they remove two fields from the operator $\mathcal{O}_{16}$ at the cost of introducing only one. The presence of the derivative balances Lorentz indices, and can be thought of as arising from the fermion propagator in the loop in the corresponding diagrammatic representation of the rules. Both rules require the introduction of an additional flavour index, which we take from $\{x,y,z,w\}$ here and in the following. (The index $x$ in the left branch is unrelated to the $x$ indexing the flavour of $Q$ in the right branch.) For the application of the up-type Yukawa rule, this diagram is shown next to the appropriate node. These rules contribute similar weights: a loop factor and a Yukawa coupling. Since the down quarks in the root node are equivalent, an additional branch of the tree is present but not shown in which $\bar{d}^{\dagger}_t$ is replaced by $\bar{d}^{\dagger}_u$. Following first the left-most branch of the diagram, we see that the unique rule matching any part of the corresponding state after the $Q$ and $H$ are looped off is again a $2 \to 1$ rule derived from a Yukawa term, in this case the down-type SM Yukawa with the derivative acting on the LHS\@.\footnote{Diagrammatically, the UV propagator that involves both loop momenta introduces a linear dependence on each, which ensures neither loop integral vanishes by even--odd parity~\cite{Angel:2012ug}.} The resulting state has field content $L_r Q_y \bar{u}^{\dagger}_x \bar{d}^{\dagger}_u$ and matches onto the field-string operator $\mathcal{O}_4$. Now we impose the mapping relations discussed in Sec.~\ref{sec:smeft-left-tree-level-matching}. In this case Eq.~\eqref{eq:unique-mappings} shows that $\mathcal{O}_4$ corresponds to the SMEFT operator $\mathcal{O}_{duql}$ whose structure we fix without introducing any new factors to the matching estimate. The estimate can then be constructed by taking the product of the weights associated with each node connecting the root to that representing $\mathcal{O}_{4}$: two loop factors, two Yukawa matrices, and the original operator coefficient $C_{16}$. In this case the estimate is the same as that implied by the left-most branch of the subgraph following from the application of $\bar{d}^\dagger_t H \to \partial Q_x$, but with the dummy indices $x$ and $y$ interchanged. The other branches of this subgraph also involve $2 \to 1$ Yukawa rules and imply the generation of the operators $\mathcal{O}_{duql}$ (with a different coefficient) and $\mathcal{O}_{qqql}$ in a similar way.

\begin{figure}[tbp!]
  \centering
  \includegraphics[width=0.8\textwidth]{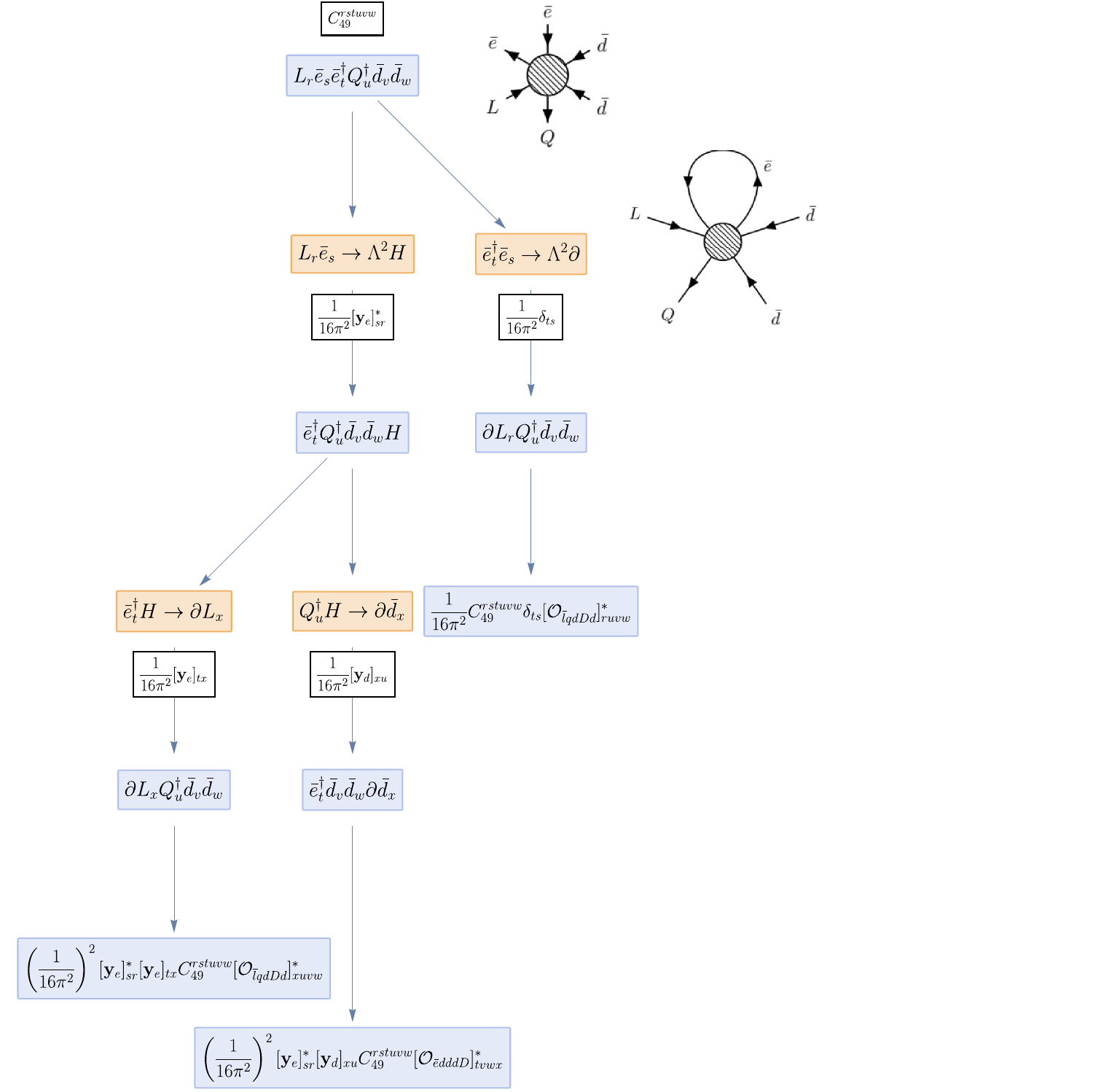}
  \caption{\label{fig:o49-matching-tree} The matching tree for the dimension-9 operator $\mathcal{O}_{49}$. See the caption of Fig.~\ref{fig:o16-matching-tree} for information on the general structure of the graph. In this case, each terminal node represents a sum over dimension-7 operators, among which only the first term is shown in the diagram. We point the reader to the main text for the full matching expressions in this case. Feynman diagrams are here shown only for the right-most branch of the matching tree. Factors of $\Lambda$ explicitly track changes in operator dimension.}
\end{figure}

It is similarly instructive to work through an example of a dimension-9 operator, whose matching trees can display qualitatively different features. For this we use $\mathcal{O}_{49}$ and the matching tree is presented in Fig.~\ref{fig:o49-matching-tree}. In this case we show the Feynman diagrams corresponding to the rule application for the right-most branch only. The algorithm begins as before with the root node contributing the field-string operator coefficient $C_{49}$. In this case, two $2 \to 1$ rules can act on the starting state: an insertion of the electron Yukawa, removing the $L$ and $\bar{e}$ fields and introducing a Higgs, and a direct looping off of the $\bar{e}$ fields in the operator, which is shown diagrammatically. This latter rule application introduces a weight of only a loop factor but also requires that the charged-lepton fields should carry the same flavour. We enforce this by introducing a factor of $\delta_{ts}$ that fixes these flavour indices in the coefficient contributed by the root node. As discussed in the previous section, the fermion loop introduces a derivative into the operator, and this branch matches onto the field-string operator $\mathcal{O}_{6}$ in our listing. Comparing to Eq.~\eqref{eq:67mapping}, we can see that the derivative within $\mathcal{O}_{6}$ can act in four distinct ways, giving rise to a matching estimate in terms of SMEFT operators:
\begin{equation} \label{eq:o6-smeft-op}
  \begin{aligned}
 & \mathcal{L}_{\mathrm{eff}} \supset \frac{1}{16 \pi^2} \frac{C_{49}^{rstuvw}}{\Lambda^3} \delta_{ts} \bigg( {[\mathcal{O}_{\bar{l}qdDd}]}^{*}_{ruvw} + {[y_e]}_{r} {[\mathcal{O}_{\bar{e}qdd\tilde{H}}]}_{ruvw}^{*} + [y_u]_{u} {[\mathcal{O}_{\bar{l}dud \tilde{H}}]}_{rvuw}^{*}\\ &\quad + \sum_{x} [y_d]_{u} V_{xu}^{*} {[\mathcal{O}_{\bar{l}dddH}]}^{*}_{rvxw} + \sum_{x} [y_d]_{v} V^{*}_{xv} {[\mathcal{O}_{\bar{l}dqq\tilde{H}}]}^{*}_{rwux} \bigg) \ ,
  \end{aligned}
\end{equation}
of which only the first expanded term is shown in the matching tree. We emphasise that in any specific UV model, only a subset of these operators will actually be present.\footnote{This caveat arises because we cannot distinguish cases in which the derivative acts on different fields in the field-string operators.} After the application of the electron-Yukawa rule on the initial state, two distinct $2 \to 1$ Yukawa rules can be applied, each of which introduces a derivative into the operator. The first is an application of another rule derived from the electron Yukawa, leading again to the generation of $\mathcal{O}_6$ and the same pattern of SMEFT operators as shown in Eq.~\eqref{eq:o6-smeft-op}, albeit with more Yukawa-suppression. The second generates the field-string operator $\mathcal{O}_7$ after an insertion of the down-quark Yukawa operator. In this case comparison to Eq.~\eqref{eq:67mapping} shows that three SMEFT operators are generated:
\begin{equation} \label{eq:o7-smeft-op}
  \begin{aligned}
 & \mathcal{L}_{\mathrm{eff}} \supset \left(\frac{1}{16 \pi^2}\right)^{2} {[y_e]}^{*}_{sr} {[y_d]}_{xu} \frac{C_{49}^{rstuvw}}{\Lambda^3} \bigg( {[\mathcal{O}_{\bar{e}dddD}]}^{*}_{tvwx} + [y_e]_t{[\mathcal{O}_{\bar{l}dddH}]}^{*}_{tvwx}\\ &\quad + \sum_{y} [y_d]_{y} V_{vy}^{*} {[\mathcal{O}_{\bar{e}qdd\tilde{H}}]}_{tywx}^{*} \bigg) \ ,
  \end{aligned}
\end{equation}
of which, again, only the first expanded term is shown in the matching tree.

It is important to highlight that depending on the pattern of non-zero entries in the coefficients $C_{16}^{rstu}$ (and $C_{49}^{rstuvw}$) some of the operators in Fig.~\ref{fig:o16-matching-tree} (and Fig.~\ref{fig:o49-matching-tree}), including Eqs.~\eqref{eq:o6-smeft-op} and~\eqref{eq:o7-smeft-op}, may vanish identically. We emphasise that this is a general caveat on all of our matching-estimate results, and not special to these operators. Indeed, the phenomenon is more clearly illustrated with the example of $\mathcal{O}_{25} = \bar{e}^{\dagger}_p \bar{e}^{\dagger}_q \bar{e}_r \bar{d}_s \bar{d}_t \bar{d}_u$, which matches onto $\mathcal{O}_{7}$ after looping off two of the $\bar{e}$ fields in the operator, similar to the right-most branch shown in Fig.~\ref{fig:o49-matching-tree}. A scalar Lorentz contraction of $\mathcal{O}_{25}$ is necessarily antisymmetric under the permutation of two of the down-quark flavour indices by Fermi--Dirac statistics. However, $\mathcal{O}_{7}$ is identified in our matching procedure with a sum over SMEFT operators including $\mathcal{O}_{\bar{e}dddD}$, which according to Table~\ref{tab:SMEFT-LEFT} is completely symmetric in down-quark flavour indices. This suggests that the down-flavour-diagonal ($q=r=s$) operators among the ${[\mathcal{O}_{\bar{e}dddD}]}_{pqrs}$ vanish. We note that a similar phenomenon occurs in the $\Delta L = 2$ case, where, for example, $\mathcal{O}_{2}$ of Refs.~\cite{Babu:2001ex,deGouvea:2007qla,Gargalionis:2020xvt} is antisymmetric in $L$ flavour indices, but the dimension-5 Weinberg operator in SMEFT and the dimension-3 neutrino-mass term in the LEFT are necessarily symmetric in the flavour indices.\footnote{A generalisation of this phenomenon in the $\Delta L = 2$ case is discussed in Sec.~4.1 of Ref.~\cite{Gargalionis:2020xvt}.}

\section{Results}\label{sec:results}

In the following we present the dominant matching estimates derived from the algorithm discussed in the previous section for each of the field-string operators in our listing. Using these estimates, we present lower limits on the scale underlying each operator and identify correlations between decay processes. These correlations act as low-energy fingerprints of the high-energy physics underlying each operator, and could help to pin down the origin of a future signal. In the following we omit limits derived from muonic modes, since these provide worse bounds than the corresponding electronic modes.\footnote{As can be seen in Table~\ref{tab:ISprocess}, Hyper-K expects better sensitivity to muonic modes in some cases.} Because of the paucity of sensitivity estimates, we cannot include a full study of the experimental sensitivity of future experiments in our study of correlations amongst two-body decays. Thus, some care should thus be taken when extrapolating our predictions to the next generation of experiments.

The lower limits are derived by substituting the loop-level matching estimates into tree-level expressions for the nucleon decays. We have presented a detailed study of BNV tree-level nucleon decays up to dimension 7 in Ref.~\cite{us:tree}. The expressions we use to derive the lower limits differ from those presented in Ref.~\cite{us:tree} in important ways, which we discuss in more detail below. The tree-level limits we derive serve as templates from which the lower limits on the field-string operators are derived. For this reason, these are presented first, in Sec.~\ref{sec:tree-level-limits}, while the loop-level limits on the field-string operators are presented in Sec.~\ref{sec:loop-level-limits}.

The usual presentational constraints mean that we cannot present all of our multidimensional results in figures on a page. For this reason, we have made all of our results available online~\cite{Gargalionis_Operator_closure_estimates_2023} as two Pandas \texttt{DataFrame} objects~\cite{reback2020pandas,mckinney-proc-scipy-2010} encoding our tree- and loop-level results. These can easily be queried for a fine-grained look at the $\Delta B = -1$ consequences of any specific field string or SMEFT operator. This is particularly relevant in the context of vanishing operators, as discussed in Sec.~\ref{sec:example-applications} with the example of $\mathcal{O}_{25}$. In case such an identically vanishing operator is come across, a simple query on the \texttt{DataFrame} object can list all of the other relevant contributions to nucleon decay.

We also include a short discussion of some UV considerations in Sec.~\ref{sec:UV-completions}. Specifically, we discuss how our results can be used for the purpose of model building by writing down a tree-level UV completion of one of the operators, and also show how our framework can be used in a phenomenological context by constraining an existing model from the literature.

\subsection{Tree-level limits and correlations}\label{sec:tree-level-limits}

Here we provide limits on the UV scale $\Lambda$ underlying each of the operators appearing in Table~\ref{tab:SMEFT-LEFT} in terms of their WCs. These serve as templates from which we calculate the limits on the scale associated with each of the field-string operators of Table~\ref{tab:bviolating-operators} after substituting in the matching estimates.

The tree-level limits are provided in Table~\ref{tab:tree-level-limits} in terms of the absolute value of each SMEFT operator coefficient. The scaling of each limit is with the $1/(d-4)$th power of the coefficient, where $d$ is the operator mass-dimension. Each operator that generates tree-level BNV nucleon decay is included in the table. We highlight that the tree-level matching expressions of Table~\ref{tab:SMEFT-LEFT} include CKM matrices that imply that even SMEFT operators containing third-generation SM fermions can give rise to tree-level BNV nucleon decays. In each case, we show the most-constraining process implied by the operator and the coefficient of the LEFT operator driving the processes.

\begin{figure}[t]
  \centering
  \includegraphics[width=0.56\textwidth]{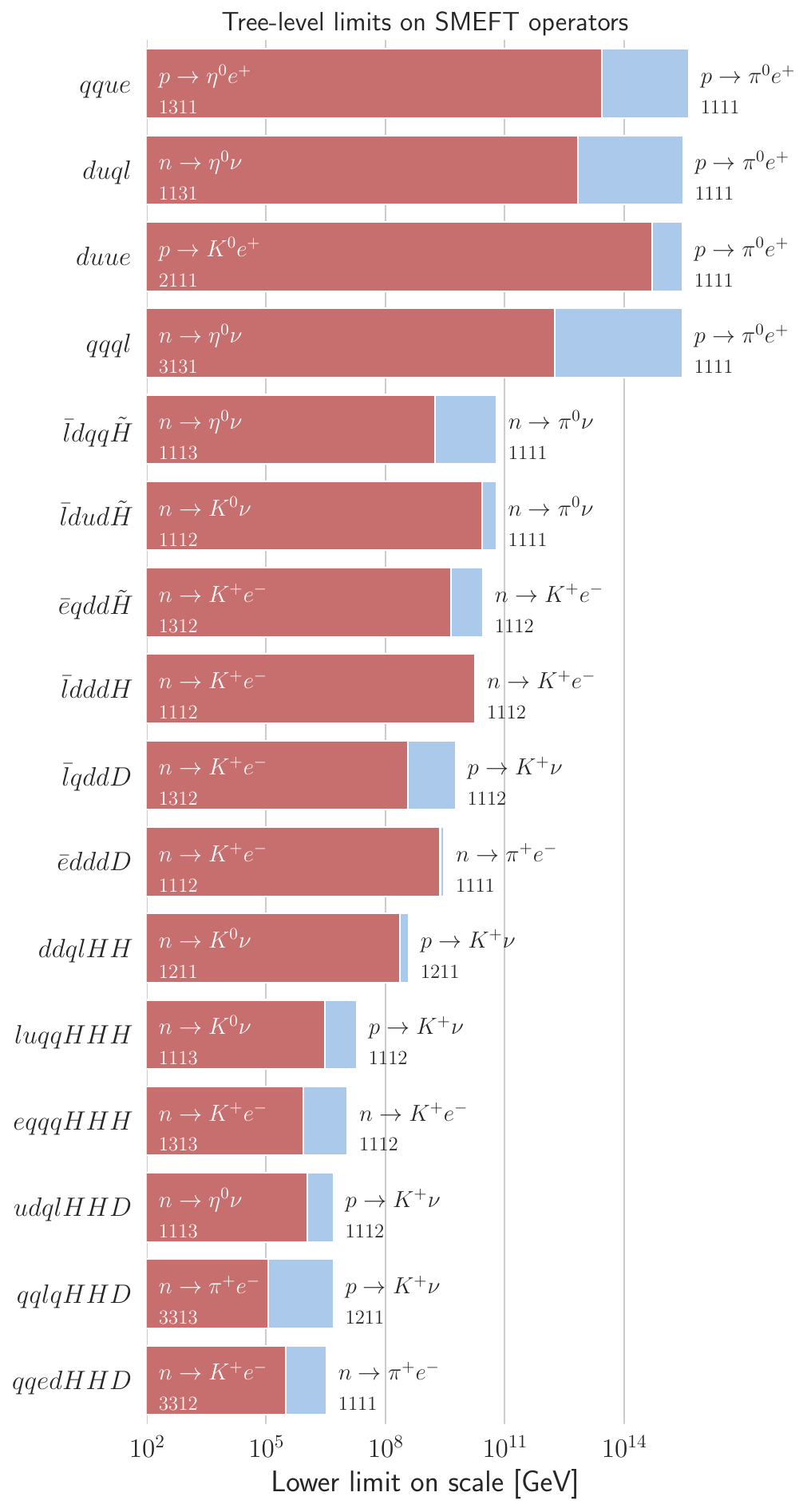}
  \caption{\label{fig:tree-level-limit-barplot} The figure shows the best (blue) and worst (red) tree-level limits on the energy scale underlying the SMEFT operator shown. The limits are calculated assuming only one non-zero operator at a time with unit WC. The decay channel from which the bound is derived and the flavour indices of the operator are shown in black and on the right for the best limit, and in white and on the left for the worst one. We omit limits derived from muon modes and do not distinguish final state neutrinos from antineutrinos.}
\end{figure}

For each labelled operator of Table~\ref{tab:tableaux} we present the lower limits of Table~\ref{tab:tree-level-limits} diagrammatically as a bar plot in Fig.~\ref{fig:tree-level-limit-barplot}, with the operator coefficient of each row of Table~\ref{tab:tree-level-limits} set to unity. We group the limits by the operator label and these are presented in descending order by the limit value. We show the bound on the flavour structure that gives rise to the most-constraining process in blue, while the bound on the operator generating the least-constraining process is shown in red. The associated processes are shown above the flavour indices, and we do not explicitly distinguish antineutrinos from neutrinos. The lepton number of the final state is fixed by the operator dimension. We remind the reader that these bounds are derived at tree level, and indeed worse limits even than those shown in red could be obtained from loop diagrams. In this sense, the red bars of the figure represent the worst tree-level limits in each case, and the intervening region (the visible part of the blue bars in the diagram) represents the range of limits relevant to each operator label. We highlight that the only flavour structure of the operator $\mathcal{O}_{\bar{l}dddH}$ that independently mediates a tree-level nucleon decay is ${[\mathcal{O}_{\bar{l}dddH}]}_{1112}$, and thus there is no intervening visible blue region in that case. For the operator $\mathcal{O}_{ddqlHH}$, the difference between the best and worse limits comes only from the fact that they are derived from different processes. The overall trend in the limit values is clearly driven by the operator dimension.

\begin{figure}[t]
  \centering
  \includegraphics[width=0.9\textwidth]{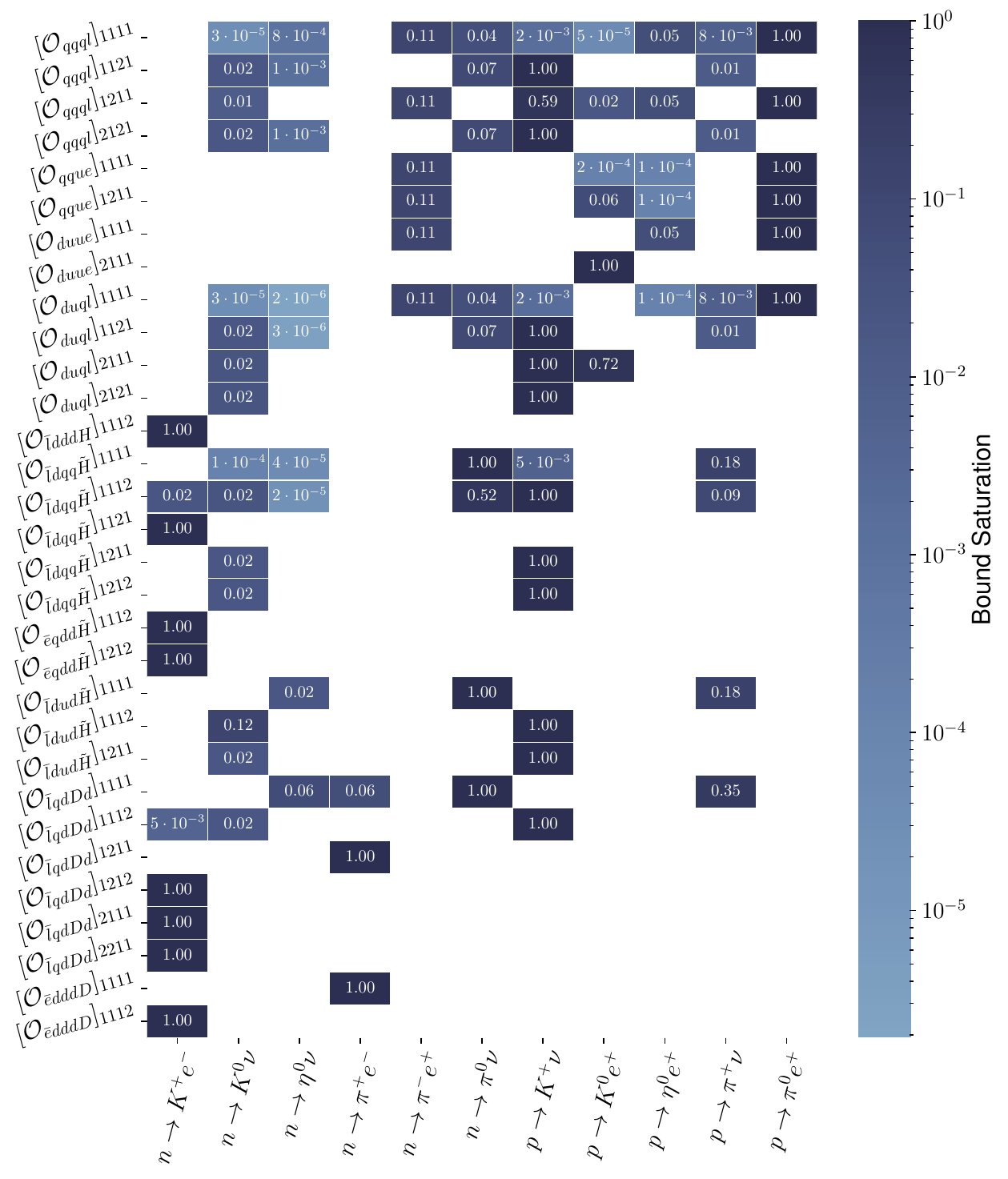}
  \caption{\label{fig:tree-level-correlations} The figure shows the correlations amongst the various decay modes we consider for each SMEFT operator up to dimension 7 without third-generation flavour indices. The colour axis shows the bound saturation, defined in Eq.~\eqref{eq:bound-saturation}, calculated at tree level. All of the dimension-6 operators are best constrained by $p \to K^+ \nu$ or $p \to \pi^0 e^+$, while the most sensitive probes of the dimension-7 operators include $n \to K^+ e^-$, $n \to \pi^+ e^-$, $n \to \pi^0 \nu$ and $p \to K^+ \nu$. This plot is an analogue of Fig.~5 from Ref.~\cite{us:tree} produced using Eq.~\eqref{eq:direct-method}; see the main text for more details on the differences.}
\end{figure}

Our tree-level nucleon-decay results can also be used to study correlations between decay modes, as in Ref.~\cite{us:tree}. We reproduce an analogue of Fig.~5 of Ref.~\cite{us:tree} in our Fig.~\ref{fig:tree-level-correlations}, which shows a heat map of the bound saturation of each decay mode for each operator. The bound saturation is defined as the ratio of the decay rate $\Gamma_{i,\rm th}$ for decay process $i$ to the most constrained decay rate $\Gamma_{\rm max,th}$, both normalised to their current experimental limits:
\begin{align}\label{eq:bound-saturation}
\text{Bound\,Saturation}\equiv \frac{\Gamma_{i,\rm th} / \Gamma_{i,\rm exp}}{\Gamma_{\rm max,th}/\Gamma_{\rm max,exp}}\;.
\end{align}
This quantity is preferable to the branching ratio because it is independent of the UV scale $\Lambda$, and still shows the relative sizes of the contributions of each operator to the various observables we study. In this plot we also include the dimension-7 operators that contain derivatives, that have been omitted in Ref.~\cite{us:tree}. Operators with third-generation flavour indices are presented separately in Appendix~\ref{sec:app-extra-correlations} to reduce clutter. These correlations could be used to help pin down the UV origin of a future BNV signal. For example, an observation of $p \to K^+ \nu$ followed closely by an observation of $p \to K^0 e^+$ could be interpreted as evidence of the operation of $[\mathcal{O}_{duql}]_{2111}$. We do not include the operators of dimension higher than 7 that mediate tree-level nucleon decay in this heat map, since generically we expect their contributions to be suppressed with respect to those at dimensions 6 and 7. Instead, these are also presented in Appendix~\ref{sec:app-extra-correlations}. All of the dimension-6 operators are best constrained by either $p\to K^+\nu$ or $p \to \pi^0 e^+$, while at dimension 7 there is a broader range of saturated decay modes, most of which are neutron decays: $n \to K^+ e^-$, $n \to \pi^+ e^-$, $n \to \pi^0 \nu$ and $p \to K^+ \nu$. As expected, the combination of operators constrained by nucleon decays to the $\eta$ are better constrained by the neutral pion modes. Additionally, $p \to \pi^+ \nu$ is not present among the saturated decay modes since its rate is related to that of the experimentally cleaner (and therefore, better constrained, see Table~\ref{tab:ISprocess}) $n \to \pi^0 \nu$ through a hadronic isospin transformation~\cite{Aoki:2013yxa}: $\Gamma(n \to \pi^0 \nu) = \frac{1}{2} \Gamma(p \to \pi^+ \nu)$. We highlight that some dimension-7 operators predict only the poorly bounded $n \to \pi^+ e^-$ as a sizeable decay channel.

\begin{figure}[t]
  \centering
  \includegraphics[width=0.6\textwidth]{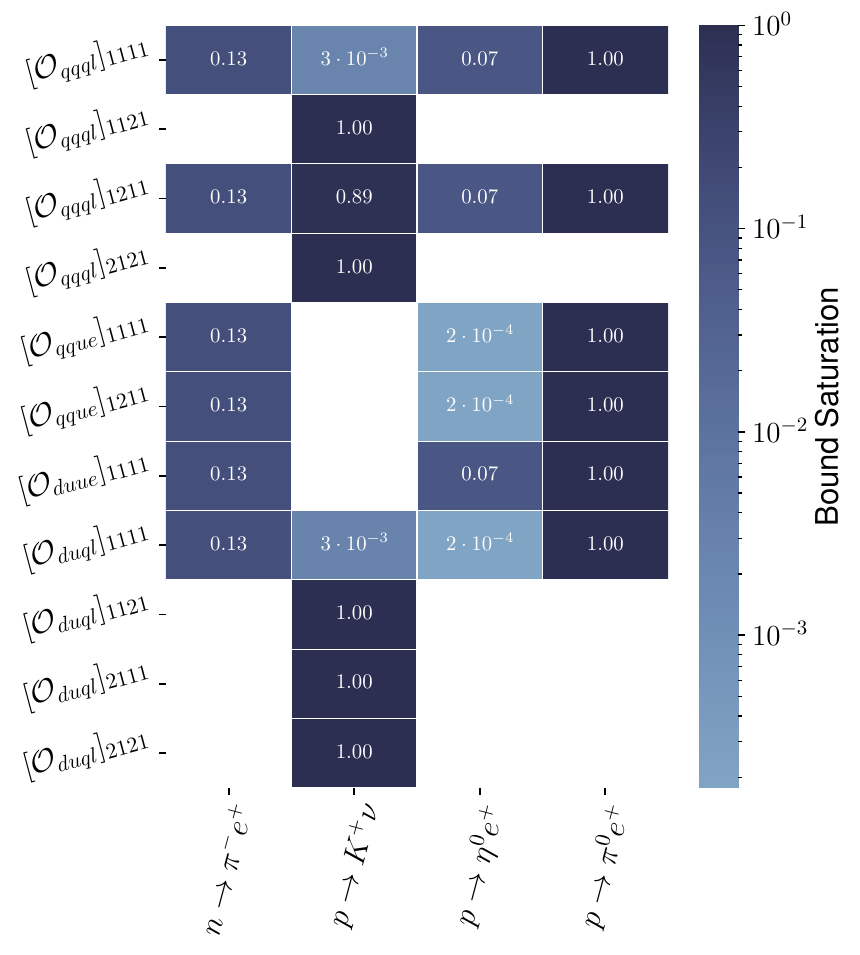}
  \caption{\label{fig:future-dim-6-tree-level-correlations} The figure shows the bound saturation expected at Hyper-K with  \SI{1.9}{\mega \tonne \year} exposure for the dimension-6 operators. The modes shown are only those for which future sensitivities are given in Table~\ref{tab:ISprocess}. Other details are as in Fig.~\ref{fig:tree-level-correlations}.
  }
\end{figure}

In Fig.~\ref{fig:future-dim-6-tree-level-correlations} we also show analogous correlation plots using only the future-sensitivity estimates taken from the Hyper-K design report~\cite{Hyper-Kamiokande:2018ofw}, shown in Table~\ref{tab:ISprocess}. We caution the reader that the full range of two-body decays is not presented in the figure, since we only have sensitivity information for a subset of the allowed modes. Despite this paucity of data, we can still make some statements about interesting correlations that could be used to single out a particular operator in the case where only one of these dominates the nucleon-decay signal. For example, the operator $[\mathcal{O}_{qqql}]_{1211}$ implies roughly equal bound saturation for the modes $p \to \pi^0 e^+$ and $p \to K^+ \nu$ at \SI{1.9}{\mega \tonne \year} exposure, which is a pattern not predicted by other operators.

The decay-rate expressions we use to compute our results differ from those presented in Ref.~\cite{us:tree} in two key ways: 
\begin{enumerate}
    \item We do not include running effects, since the assumptions underlying our estimates significantly limit their accuracy anyway, and 
    \item We use nuclear matrix elements calculated directly on the lattice, rather than Baryon Chiral Perturbation Theory (B$\chi$PT).\footnote{For a comparison between the B$\chi$PT and direct lattice methods, we point the reader to Refs.~\cite{Aoki:2017puj,Yoo:2021gql}.} 
\end{enumerate}     
    We find that the most-constraining process for each operator is the same when comparing to Fig.~5 of Ref.~\cite{us:tree} with the exception of operator $[\mathcal{O}_{qqql}]_{1211}$. Here, using the direct lattice method, we find the $p \to \pi^0 e^+$ rate to be larger than that for $p \to K^+ \nu$, while this situation is inverted when the calculation is performed in B$\chi$PT~\cite{us:tree}. This discrepancy is within the ballpark of difference expected for the two methods~\cite{Aoki:2017puj}.

\subsection{Loop-level limits and correlations}\label{sec:loop-level-limits}

The main results of our study are the matching estimates for each of the field-string operators, which are presented in Table~\ref{tab:bviolating-operators}. There, we show the coefficient of the SMEFT operator (from the basis defined in Table~\ref{tab:SMEFT-LEFT}) that drives the most constrained process generated for unit field-string coefficients. The expressions are shown with an implied sum over primed indices on the RHS, and with bars over conjugated field-string coefficients. The field-string flavour indices are also shown, along with the most constrained process in the last column, and the implied bound on the new-physics scale $\Lambda$ in the penultimate one.

The operators of dimension 6 and 7 mediate tree-level nucleon decay, and thus we do not quote matching estimates for these.\footnote{The tree-level matching expressions onto the LEFT are shown in Table~\ref{tab:SMEFT-LEFT}.} We do however copy over the rest of the information from Table~\ref{tab:tree-level-limits}: the most-constraining process, the flavour indices that give rise to it, and the implied lower limit.

The results show that the matching estimates that we derive range up to two-loop expressions. Many involve suppression from CKM matrix elements and SM Yukawa couplings. The most suppressed leading-loop contribution to nucleon decay comes from field-string operator 26 with flavour indices $1211$. In this case, the non-observation of the decay $n \to K^+ e^-$ allows us to extract a lower bound of $\SI{2e6}{\GeV}$ on the new-physics scale. The process proceeds at dimension-7 in the SMEFT, and the amplitude is further suppressed by two loop factors, and the product of a strange-quark and a down-quark Yukawa.

\begin{figure}[tbp!]
  \centering
  \includegraphics[width=0.43\textwidth]{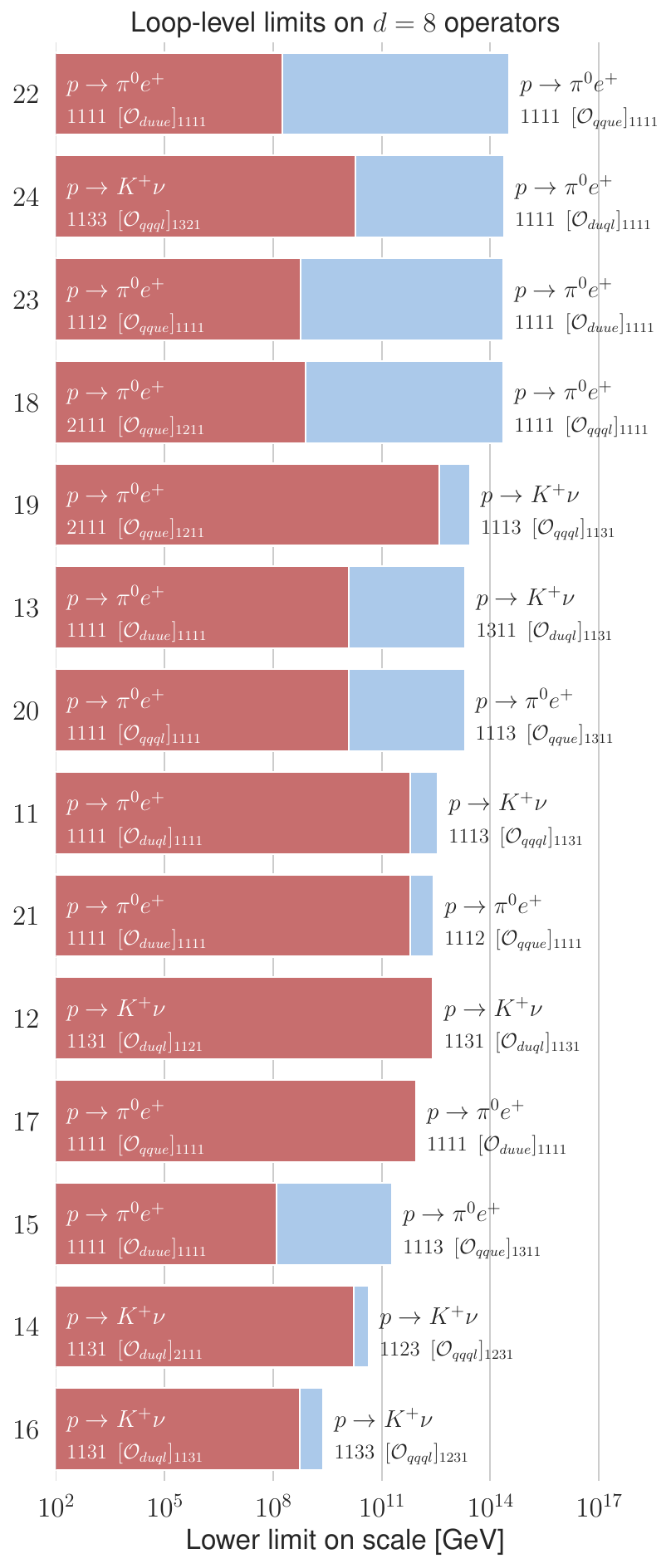}
  \caption{\label{fig:dim-8-loop-level-limit-barplot} The plot shows the lower limits on the scale of the underlying UV model generating the numbered dimension-8 field string. In blue we show the limit derived from the most constrained SMEFT operator, while in red the limit from the next best constrained SMEFT operator, disregarding flavour indices. We also show the decay processes from which the bounds are derived, the field-string flavour indices, and the SMEFT operators driving the matching onto the LEFT\@. The limits are set assuming unit field-string  coefficients, with only one turned on at a time.  Operator $\mathcal{O}_{12}$ generates just one SMEFT operator and thus two different flavour combinations are shown. We note that in some cases a number of flavour-index choices provide the same limit, but in the figure only one of these is shown.}
\end{figure}

\begin{figure}[tp!]
  \centering
  \includegraphics[width=0.56\textwidth]{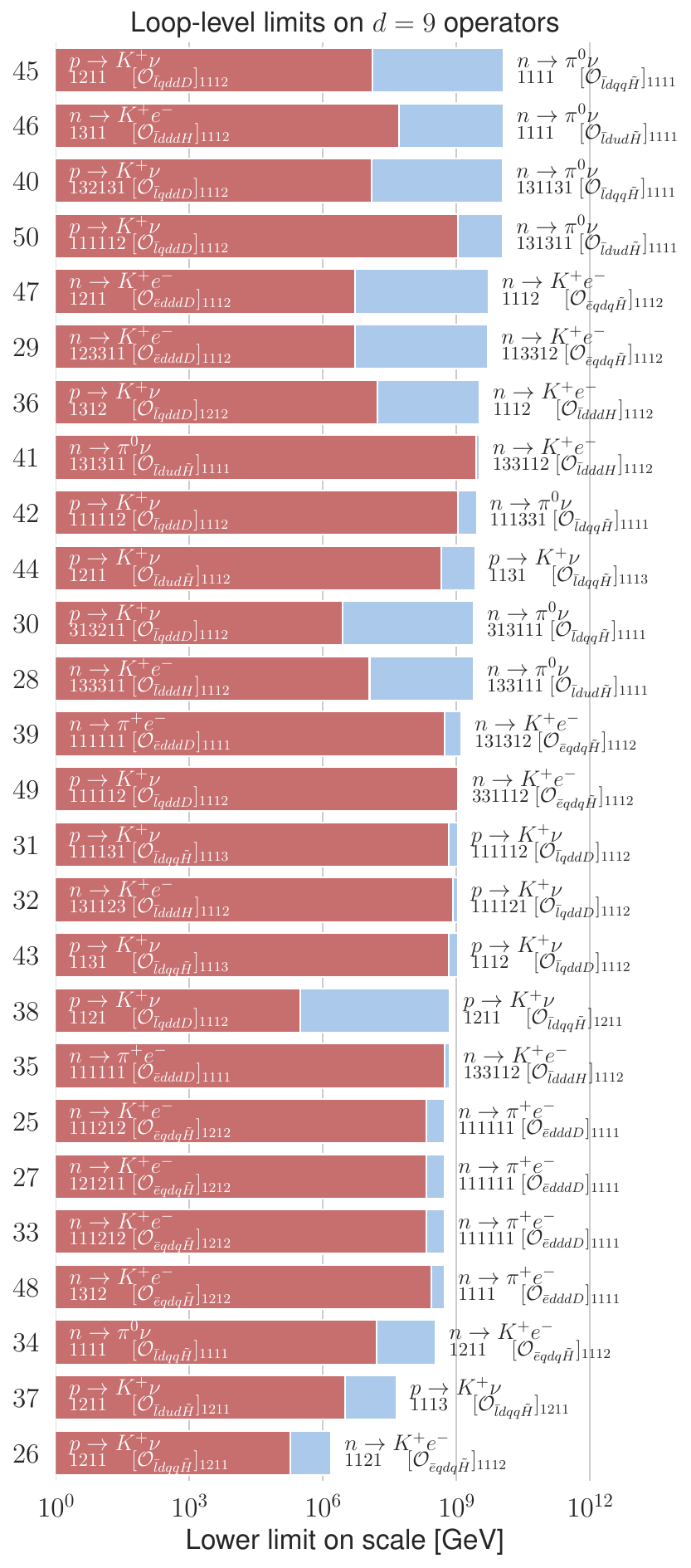}
  \caption{\label{fig:dim-9-loop-level-limit-barplot} Same as Fig.~\ref{fig:dim-8-loop-level-limit-barplot} for dimension-9 field-string operators.}
\end{figure}

In Figs.~\ref{fig:dim-8-loop-level-limit-barplot} and~\ref{fig:dim-9-loop-level-limit-barplot} we present the lower limits on the energy scale of the UV models generating the field-string operators of dimension 8 and 9 at tree level. As before, we consider only one non-zero field-string operator at a time. The limits have been obtained by weighting those on the dimension-6 and -7 SMEFT operators by the estimated loop-level matching contribution. Here, blue bars show the limit from the most constrained flavour structure, assuming unit field-string operator coefficients. To illustrate the diversity of bounds implied by different diagrams, we also show the best limit obtained from the next best constrained SMEFT operator. In each case, we only consider the most constrained flavour combination. For example, in the case of $\mathcal{O}_{22}$ the best limit comes from the matching estimate onto $\mathcal{O}_{qque}$ with flavour $1111$, and this is shown in blue. In red we show the best limit from the closure diagram generating $\mathcal{O}_{duue}$ with flavour $1111$. Although other flavour combinations of $\mathcal{O}_{qqql}$ can provide better limits than $[\mathcal{O}_{duue}]_{1111}$, they most often arise from the same matching diagrams, and thus do not sufficiently illustrate the variety of bounds we derive. Operator $\mathcal{O}_{12}$ is the only one whose matching onto the LEFT is governed by only one SMEFT operator, and so in this case we show two different combinations of flavour indices for the same labelled SMEFT operator. The two limits shown for $\mathcal{O}_{17}$ are coincidentally very close to each other, so much so that the blue bar is not visible. We show the decay process from which each limit is derived in black on the right for the best limit, and on the left in white for the limit shown in red. We also show the flavour indices and the SMEFT operator governing the matching on to the LEFT\@. The limits on the operators are sorted by size. Compared with the tree-level limits shown in Fig.~\ref{fig:tree-level-limit-barplot}, a much richer structure is evident within operators of the same mass dimension. Additionally, because of the presence of SM Yukawa couplings in the matching estimates, it is often the case that the most constrained field-string coefficients feature third-generation flavour indices. Among the most constrained flavour combinations for each numbered field-string operator, the one constrained the least is $[\mathcal{O}_{26}]_{1121}$, for which current limits on $n \to K^+ e^-$ imply a lower limit of roughly $\SI{e6}{\GeV}$ on the scale underlying the dimension-7 operator $[\mathcal{O}_{\bar{e}qdq\tilde{H}}]_{1112}$.

\begin{figure}[t]
  \centering
  \includegraphics[width=0.9\textwidth]{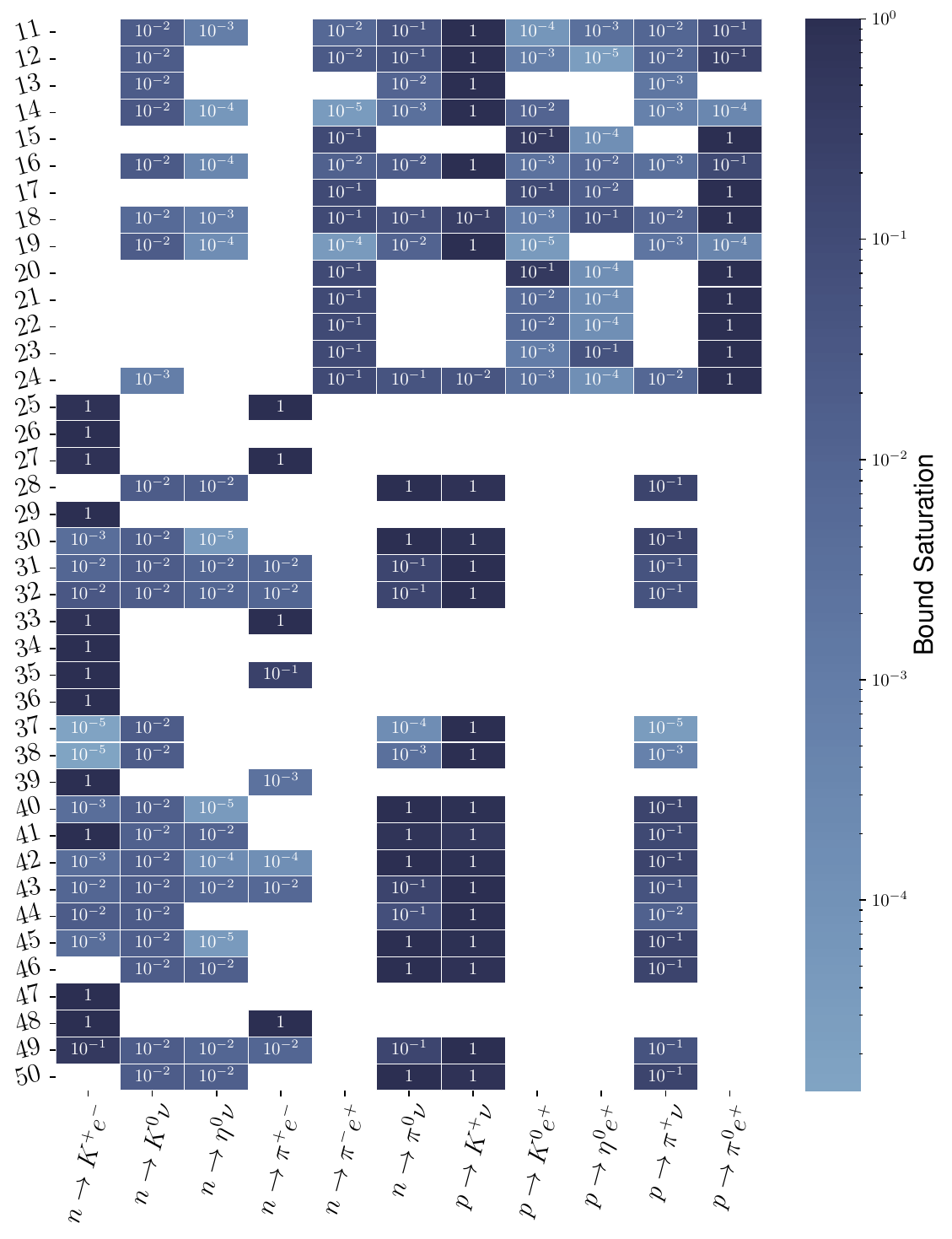}
  \caption{\label{fig:democratic-loop-correlations} The heat map shows the loop-level correlations amongst the decay modes associated with each numbered field string under the democratic flavour assumption, i.e.\ ${[C_i]}_{pqrs\cdots} = 1$. The colour axis shows the order of magnitude of the bound saturation, which we use to emphasise the expected accuracy of our matching estimates. Entries smaller than $10^{-6}$ are left blank in the figure. We point the reader to the caption of Fig.~\ref{fig:tree-level-correlations} for additional details.}
\end{figure}

\begin{figure}[t]
  \centering
  \includegraphics[width=0.9\textwidth]{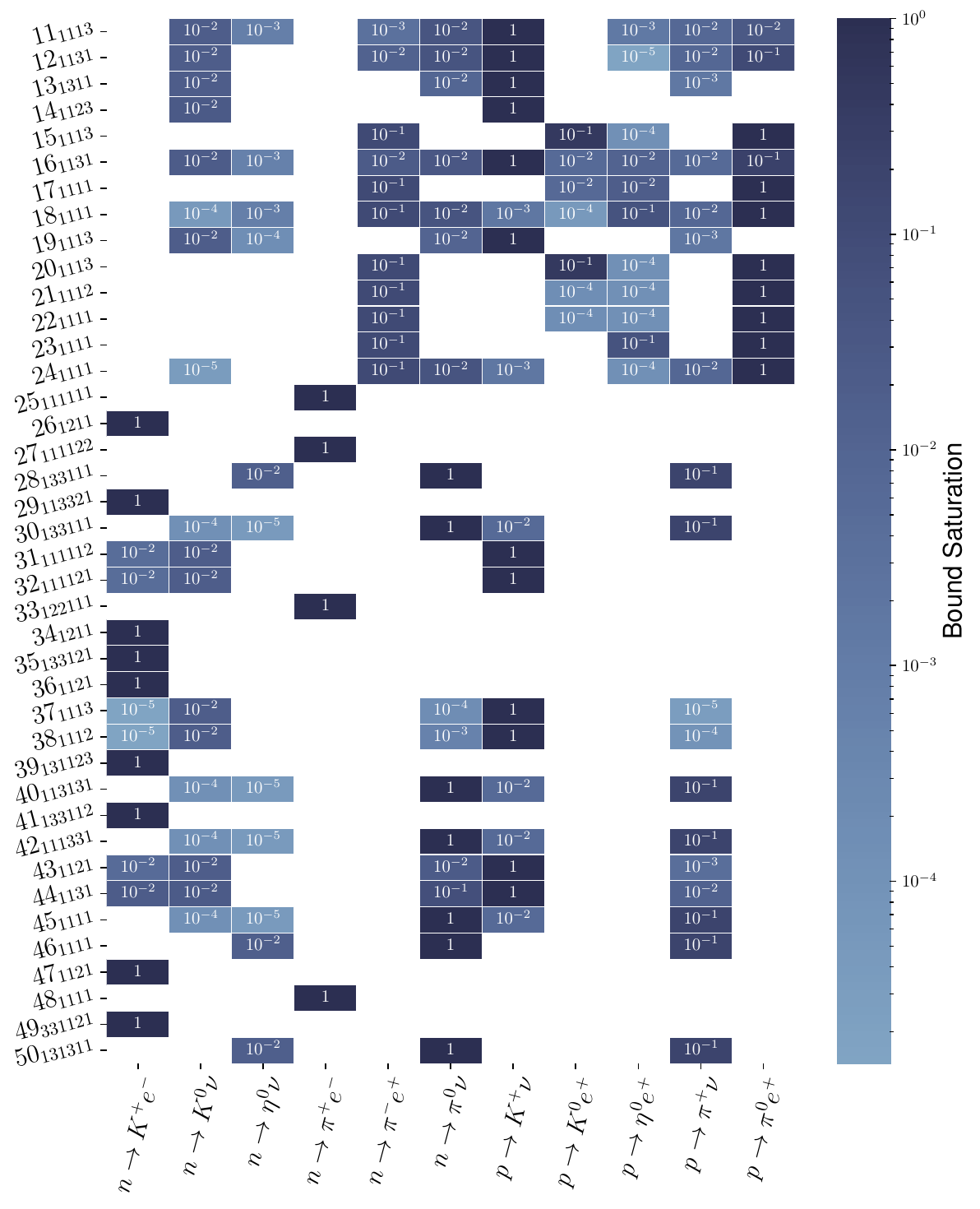}
  \caption{\label{fig:best-loop-correlations} The heat map shows the loop-level correlations amongst the decay modes associated with each numbered field string under the most-constrained flavour assumption. This assumption imposes that all other operator flavour structures vanish except for those shown as subscripts along the $y$-axis. We point the reader to the caption of Fig.~\ref{fig:democratic-loop-correlations} for additional details.}
\end{figure}

The rich structure of the matching estimates is also communicated into non-trivial correlations between decay modes for each field-string operator. Our algorithm identifies just under $\num[group-separator={,}]{60000}$ field-string operators that imply leading-loop contributions\footnote{That is, contributions to nucleon decay identified without the application of $2 \to 2$ rules. Adding in the $2 \to 2$ can be done by hand, and we show an example in Sec~\ref{sec:manual-2to2-rules}.} to BNV nucleon decay, and so we cannot show all of the correlations conveniently within the scope of a work like this.\footnote{If the reader is interested in exploring this space of operators, we encourage them to check our database of results available online~\cite{Gargalionis_Operator_closure_estimates_2023}.} Instead, we present correlation plots in the context of two separate assumptions on the flavour structure of the operators. In Fig.~\ref{fig:democratic-loop-correlations} we show a heat map displaying the correlations among decay modes implied by unit operator coefficients for each flavour index combination: $[C_i]_{pqrs\cdots}=1$. We label this the \textit{democratic} flavour assumption. In Fig.~\ref{fig:best-loop-correlations} we show a similar plot presenting the correlations among decay modes with only the coefficients with the best constrained flavour indices identified in Figs.~\ref{fig:dim-8-loop-level-limit-barplot} and~\ref{fig:dim-9-loop-level-limit-barplot} set to unity, and all others set to vanish. This we label the \textit{most-constrained} flavour assumption. As in the tree-level case, we show these correlations with the bound saturation, defined in Eq.~\eqref{eq:bound-saturation}. We present only the order of magnitude of the calculated bound saturation in each figure to emphasise that our results are only accurate to about this level of precision. Importantly, this means that in these figures there may be more than one decay mode shown to have unit bound saturation. We colour in white the contributions of order $10^{-6}$ and smaller. We note that this necessarily implies that no matching contributions underlying the calculations in these plots involve $2 \to 2$ rules, see App.~\ref{sec:manual-2to2-rules} for justification.

The correlations in the figures are broadly driven by their tree-level analogues, which are presented in Fig.~\ref{fig:tree-level-correlations}. However, it is possible to see new loop-level effects that can make the operator inverse problem more tractable in the event of a positive signal at the next generation of experiments. These appear since at loop level the pattern of decays is dictated by a linear combination of SMEFT operators. For example, looking at the $12_{1131}$ row of Fig.~\ref{fig:best-loop-correlations}, which shows the correlations between decay modes for the most-constrained flavour assumption, we can see that the fingerprint of decay modes is not driven by a single SMEFT operator from Fig.~\ref{fig:tree-level-correlations} (or indeed the additional correlations presented in Appendix~\ref{sec:app-extra-correlations}: Figs.~\ref{fig:tree-level-correlations-gen-3} and~\ref{fig:tree-level-correlations-dim-9}). The combination of SMEFT operators driving the matching onto the LEFT for each numbered field-string operator is easily accessible from our results database.

We remind the reader that field-string operator $\mathcal{O}_{12}$ is the only one generating different flavours of only the \textit{same} SMEFT operator, which happens to be $\mathcal{O}_{duql}$. For this reason, it is simple to follow the origin of the correlations presented in the $\mathcal{O}_{12}$ row of Fig.~\ref{fig:best-loop-correlations} by hand, without the need to query the database directly. From Table~\ref{tab:bviolating-operators} we see that the most-constrained matching estimate comes from a one-loop diagram suppressed by the SM bottom Yukawa and an off-diagonal CKM matrix element. Fixing the flavour indices $1131$ from the figure, we have
\begin{equation}\label{eq:o12-matching}
\mathcal{L}_{\rm eff} \supset \sum_{r} \frac{1}{16 \pi^2} V_{r 3}^* y_b \frac{C_{12}^{1131}}{\Lambda^2} \mathcal{O}_{11r1}^{duql} \; ,
\end{equation}
and the additional SMEFT operators driving the matching follow by expanding the sum. In this case $[\mathcal{O}_{duql}]_{11r1}$ for $r\in \{1,2,3\}$ all give contributions proportional to the bottom Yukawa. Comparing to Table~\ref{tab:tree-level-limits}, we can see that $[\mathcal{O}_{duql}]_{1111}$ is driving the large $p \to \pi^0 e^+$ rate, while $[\mathcal{O}_{duql}]_{1121}$ predicts a leading signal in $p \to K^+ \nu$. This structure explains the most saturated rates in Fig.~\ref{fig:best-loop-correlations} for the most-constraining flavour structure of $\mathcal{O}_{12}$.

\begin{figure}[t]
  \centering
  \includegraphics[width=0.6\textwidth]{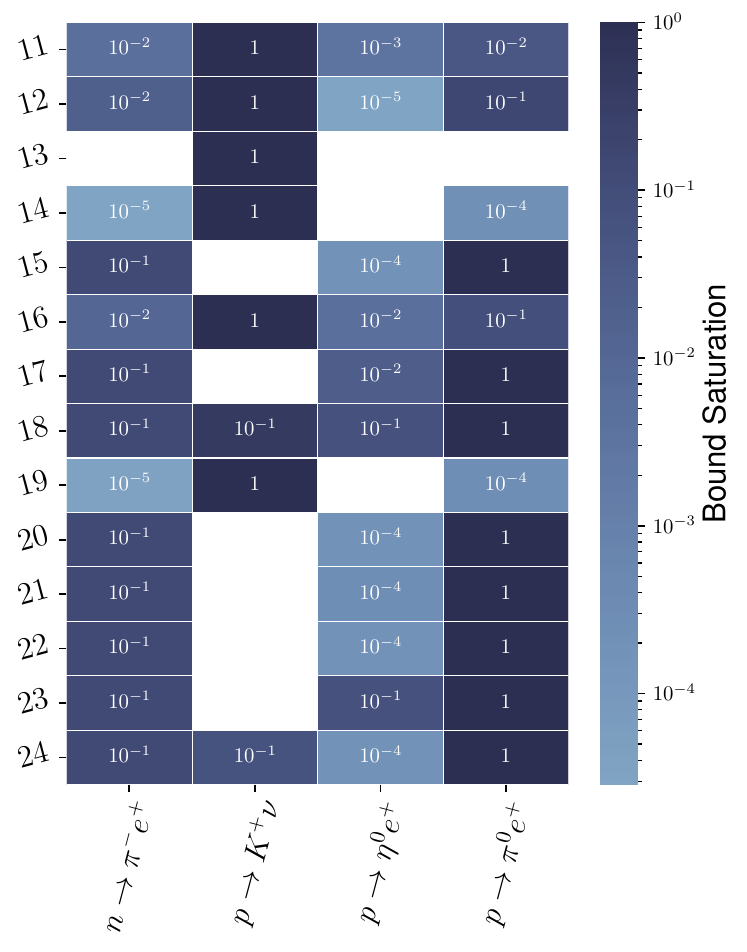}
  \caption{\label{fig:future-democratic-correlations} The heat map shows the loop-level correlations for the Hyper-K sensitivities at \SI{1.9}{\mega \tonne \year} exposure, presented in Table~\ref{tab:ISprocess}. We only show the dimension-8 operators here, since the decay modes for which future sensitivities are given do not provide much insight at dimension 9. We assume ${[C_i]}_{pqrs\cdots} = 1$. Peripheral details are as in Fig.~\ref{fig:democratic-loop-correlations}.}
\end{figure}

Finally, in Fig.~\ref{fig:future-democratic-correlations} we also show the correlations expected at Hyper-K using the future-sensitivity data presented in Table~\ref{tab:ISprocess}, from Ref.~\cite{Hyper-Kamiokande:2018ofw}, assuming \SI{1.9}{\mega \tonne \year} exposure. The correlations are shown under the democratic flavour assumption. We show only the correlations implied by the dimension-8 operators, since at dimension 9 almost all of the field strings imply only visible signals in one of $p\to K^+ \nu$ or $n \to K^+ e^-$. Exceptions are operators $30$--$32$, $34$, $40$, $42$--$45$ and $49$, for which the bound saturations for these decays are within an order of magnitude. Compared to the current bound saturations shown in Fig.~\ref{fig:democratic-loop-correlations}, we see that the ratio of rates for $p \to K^+ \nu$ to $p \to \pi^0 e^+$ is increased by about an order of magnitude for the operator $24$.

\subsection{Comments on ultraviolet completions}\label{sec:UV-completions}

In the following we give two examples of how our results can be used in phenomenological studies for both model engineering and constraint setting. We also present a comparison of some of our results with the literature as validation of our methods.

A powerful advantage of using an effective framework to study BNV nucleon decays is that the process of model discovery can now be automated. Indeed, the analogous framework used in the study of $\Delta L = -2$ phenomena~\cite{Babu:2001ex,deGouvea:2007qla,Gargalionis:2020xvt} has led to a number of UV-model catalogues and systems of model categorisation, see e.g.\ Refs.~\cite{Cai:2014kra,Herrero-Garcia:2019czj,Gargalionis:2020xvt}.\footnote{Connections have also been made to complementary systems of classification~\cite{Bonnet:2012kz,Sierra:2014rxa,Cepedello:2017eqf,Cepedello:2018rfh}.} In this case, the tree-level UV completions of the $\Delta L = -2$ field-string operators correspond to loop-level completions of the dimension-5 Weinberg operator. Thus, listing the tree-level completions of these operators is one way of mapping the space of radiative neutrino mass models.

An important concept in this kind of automated model building is called \textit{filtering} in Ref.~\cite{Gargalionis:2020xvt} and \textit{genuineness} in Refs.~\cite{Bonnet:2012kz,Sierra:2014rxa,Cepedello:2017eqf,Cepedello:2018rfh}. This is the statement that a mechanism of neutrino mass is only useful to consider if it genuinely gives a leading-order contribution to the dimension-3 LEFT term $\nu\nu$. Any radiative diagram correcting the neutrino mass that features a fermionic singlet, for example, would not fit into this category, since in the absence of any additional symmetries it would generate a neutrino mass already at tree level. Such models need to be \textit{filtered} from the list of tree-level completions of the $\Delta L = -2$ operators to identify genuinely interesting radiative models. This concept is particularly relevant in the $\Delta B = -1$ case since here there are 10 operators that generate nucleon decay at the tree level. The tree-level completions of these operators will, in the minimal case, contain one heavy multiplet at dimension 6, and two at dimension 7.\footnote{We are counting a Dirac fermion as one multiplet here.} They are contained in the UV/IR dictionaries of Refs.~\cite{deBlas:2017xtg,Li:2023cwy} and are analogues of the seesaw models in the $\Delta L = -2$ case.

This raises an interesting and pertinent question: Are there any \textit{genuine} contributions to nucleon decays from completions of operators with mass-dimension higher than 7? An interesting finding of Ref.~\cite{Gargalionis:2020xvt} was that, in the $\Delta L = -2$ case, the number of genuine completions grew sharply with increasing operator mass dimension due to combinatorial explosion. Thus, we hypothesise that finding a genuine example by hand likely means that there is an equally vast space of models on the $|\Delta B| = 1$ side to be discovered and categorised. Below we give an example at dimension 8, which only need not contain a tree-level completion of one of the dimension-6 operators, on symmetry grounds. Reading from Ref.~\cite{deBlas:2017xtg} these are: $\omega_1 \sim (\mathbf{3}, \mathbf{1}, -\tfrac{1}{3})_S$, $\omega_4 \sim (\mathbf{3}, \mathbf{1}, -\tfrac{4}{3})_S$, $\zeta \sim (\mathbf{3}, \mathbf{3}, -\tfrac{1}{3})_S$, $\mathcal{Q}_{1} \sim (\mathbf{3}, \mathbf{2}, \tfrac{1}{6})_V$ and $\mathcal{Q}_{5} \sim (\mathbf{3}, \mathbf{2}, -\tfrac{5}{6})_V$. Here we have labelled the multiplets by their SM and Lorentz quantum numbers: ${(\mathrm{SU}(3)_c, \mathrm{SU}(2)_L, \mathrm{U}(1)_Y)}_{R}$, where $R \in \{S, V, F\}$ for the scalar, vector and spinor representations of the Lorentz group. Note that no heavy fermions generate dimension-6 four-fermion operators at tree level. 

We make a connection here with the results of Ref.~\cite{Helo:2019yqp}, where a systematic classification of the one-loop completions of $\Delta B$ SMEFT operators is presented. Indeed, that study identified a family of models, called \textit{Class-II} in their nomenclature, in which the one-loop graph is the dominant contribution to proton decay. The example model we present below happens not to be present in their scheme, since it predicts nucleon decay at two loops.

\subsubsection{Example of a simplified BNV model}\label{sec:example-uv-model}

\begin{figure}[t]
  \centering
  \includegraphics[width=0.42\textwidth]{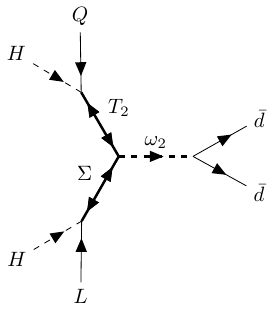}
  \includegraphics[width=0.59\textwidth]{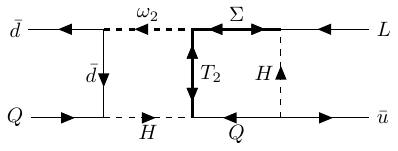}
  \caption{\label{fig:uv-completion-diags} The figure shows two Feynman diagrams relevant to the example UV model discussed in the main text. The exotic field content is shown with bolded propagators. The arrow convention follows Ref.~\cite{Dreiner:2008tw}. (Top) A tree-level completion of operator $\mathcal{O}_{16}$ in our listing. (Bottom) One of the two-loop diagrams generating a dimension-6 $\Delta B = -1$ operator in the example UV model. This diagram follows by pasting the tree-level graph shown in the top panel into the example closure depicted in Fig.~\ref{fig:o16-matching-tree}.}
\end{figure}

The example we present is a completion of the field-string operator $\mathcal{O}_{16} = LQ\bar{d}^\dagger \bar{d}^\dagger H H$, also used as the dimension-8 example in Sec.~\ref{sec:example-applications}. We note that perhaps the most economic model is the two-scalar extension of the SM with one of $\omega_1$ or $\zeta$ coupling to $LQ$ and an exotic fermion coupling to $\bar{d}^\dagger H$; however, both $\omega_1$ and $\zeta$ generate the dimension-6 operator $\mathcal{O}_1 = LQQQ$ at tree level, and this model is thus not genuine in the aforementioned sense.\footnote{Actually, $\omega_1$ generates every dimension-6 $\Delta B = -1$ operator at tree level.} In the top panel of Fig.~\ref{fig:uv-completion-diags} we show the tree-level diagram generating $\mathcal{O}_{16}$ in a genuine model containing a coloured scalar $\omega_{2} \sim (\mathbf{3}, \mathbf{1}, \tfrac{2}{3})_{S}$, a vector-like Dirac fermion $T_{2} \oplus \bar{T}^{\dagger}_{2} \sim (\mathbf{3}, \mathbf{3}, \tfrac{2}{3})_{F}$, and a triplet Majorana fermion $\Sigma \sim (\mathbf{1}, \mathbf{3}, 0)_{F}$. The exotic fields are shown with bold propagators, and we use the convention of Ref.~\cite{Dreiner:2008tw} for the arrows in the diagram. The pertinent part of the interaction Lagrangian of this model is
\begin{equation}
-\mathcal{L}_{\mathrm{int}} = x_{[rs]} \bar{d}_{ar} \bar{d}_{bs} \omega_{2c}^{\dagger} \epsilon^{abc} + y_{r} \bar{T}_2^{I} H^{i} \epsilon_{ij} [\sigma^I]^{j}_{\ k} Q_{r}^k + z_{r} \Sigma^{I} H^{i} \epsilon_{ij} [\sigma^I]^{j}_{\ k}  L_{r}^{k} + f \Sigma^{I} T_{2}^{I} \omega_2^{\dagger} +  \mathrm{h.c.} \; ,
\end{equation}
where $r,s$ are SM flavour indices, $a,b,c$ are colour indices, $i,j,k$ are $\mathrm{SU}(2)_{L}$ fundamental indices, $I$ is an adjoint $\mathrm{SU}(2)_L$ index, and we have emphasised that $x_{rs} = - x_{sr}$ with explicit antisymmetrisation. Colour and spinor indices have been suppressed where there is no ambiguity. Integrating the heavy field content out at tree level using \texttt{MatchingTools}~\cite{Criado:2017khh}, we find that the model generates the operator\footnote{Here we have translated the expression into two-component spinor notation for consistency.} $[Q_{lqd^2H^2}]_{rstu} = (L^{i}_{r} Q^{aj}_{s})(\bar{d}^{\dagger b}_{t} \bar{d}^{\dagger c}_{u}) H^k H^l \epsilon_{ik}\epsilon_{jl} \epsilon_{abc}$ in the dimension-8 basis of Ref.~\cite{Murphy:2020rsh} with coefficient
\begin{equation}\label{eq:lqddhh-matching}
 [C_{lqd^2H^2}]_{rstu} = \frac{3 x^*_{tu} y_s z_r f}{M_{\omega_2}^2 M_{T_2} M_{\Sigma}}  \;,
\end{equation}
where the masses of the exotics $M_i$ are defined in the canonical way, implemented internally within the programme of Ref.~\cite{Criado:2017khh}. The matrix of WCs inherits the antisymmetry in the down-quark indices from the diquark couplings of $\omega_2$, and thus the first-generation flavour structure vanishes identically. We have checked that this is the only $\Delta B$ operator present in the effective Lagrangian up to dimension 8 for the model. We also highlight a connection to the type-III seesaw mechanism for neutrino masses through the exotic Majorana fermion $\Sigma$.

The loop-level nucleon-decay diagrams in this model follow by pasting the tree-level diagram shown in the top panel of Fig.~\ref{fig:uv-completion-diags} into the various closure diagrams associated with the branches of Fig.~\ref{fig:o16-matching-tree}, of which only the left-most diagram is shown. Continuing with the left-most branch as an example, in the bottom panel of Fig.~\ref{fig:uv-completion-diags} we show one of the two-loop diagrams generating the dimension-6 operator $\mathcal{O}_{duql}$, obtained by inserting the tree-level completion diagram into the example two-loop closure shown in Fig.~\ref{fig:o16-matching-tree}. The estimate of the two-loop $\mathcal{O}^{duql}$ coefficient from this diagram is
\begin{equation}
\left( \frac{1}{16\pi^2} \right)^2 \frac{1}{\Lambda^2} \sum_{t^\prime} V_{r t^\prime}^* (y_d)^{t^\prime} (y_u)^{q} x_{t^\prime p} y_{q} z_{s} f \; ,
\end{equation}
where we have taken the particles in the model to share a common mass scale $\Lambda$.

\begin{figure}[t]
  \centering
  \includegraphics[width=0.8\textwidth]{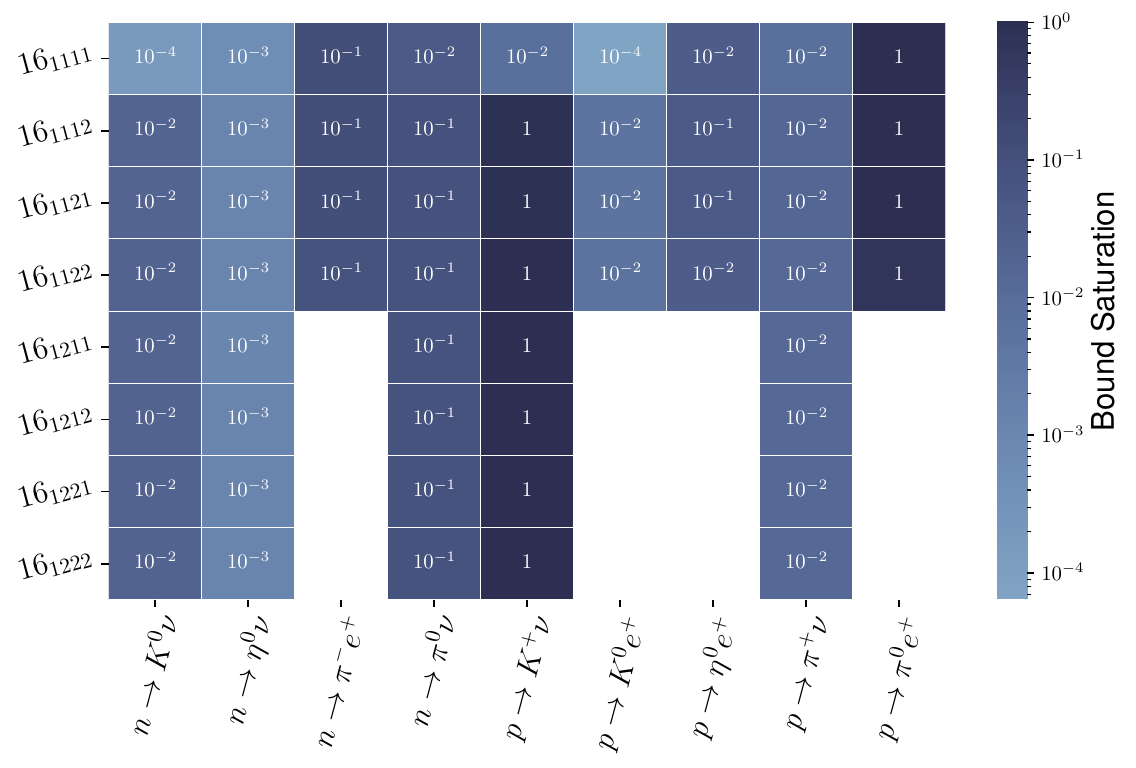}
  \caption{\label{fig:o16-correlations} The plot shows the correlations amongst flavours of $\mathcal{O}_{16}$, excluding third-generation fermions. The peripheral details are as in the caption of Fig.~\ref{fig:best-loop-correlations}. We note that the order-of-magnitude estimates on the bound saturation are not altered in an important way when extrapolating to \SI{1.9}{\mega \tonne \year} exposure at Hyper-K.}
\end{figure}

To study the predictions of the model, we present the correlation plot for flavours of operator $\mathcal{O}_{16}$ involving the first two generations of SM fermions in Fig.~\ref{fig:o16-correlations}.\footnote{We present the query used to generate the data behind this plot as an example in our online repository~\cite{Gargalionis_Operator_closure_estimates_2023}, should the reader wish to perform a similar investigation of any other operators in our listing.} Here we have set the WC to unity in each row, and indicated the order of magnitude of the bound saturation [defined in Eq.~\eqref{eq:bound-saturation}] by the colour and value shown in each box. Since the operator generated in this specific model is antisymmetric in the down-quark flavour indices, the independent flavour structures are $1112$ and $1212$. If both of these are present --- i.e.\ if the generalised Yukawa couplings $y_1$ and $y_2$ are both sizeable --- the model distinctively predicts $p \to K^+ \nu$ and $p \to \pi^0 e^+$ as forerunning signals, with a measurement of $n \to \pi^- e^+$ following with more sensitivity. The presence of $p \to \pi^0 e^+$ and $n \to \pi^- e^+$ can be used to diagnose whether $y_2$ is sizeable or not. We note that the correlations and order-of-magnitude estimates of the bound saturation are not changed significantly when imposing the future-sensitivity estimates for Hyper-K from Table~\ref{tab:ISprocess}.

In this specific model the lower bound provided in Table~\ref{tab:bviolating-operators} cannot be applied, since the corresponding field-string flavour indices $1133$ make $Q_{lqd^2H^2}$ vanish identically. In this case the best limit we can impose is $\Lambda \cdot (x_{23}y_{1}z_{1}f)^{-1/2} > \SI{2e9}{\GeV}$ from $[\mathcal{O}_{16}]_{1123}$ matching onto $[\mathcal{O}_{qqql}]_{1131}$. This can be seen from the flavour-general expression for the estimate given in Table~\ref{tab:bviolating-operators}, or otherwise from our numerical database available online.

We have checked by comparing to the analysis of Ref.~\cite{Helo:2019yqp} that the exotic field content present in this model is not sufficient to generate a $\Delta B$ dimension-6 operator at \textit{one loop}. Since we have also checked that the tree-level effective Lagrangian contains no other dimension-8 operator that violates $B$, we suspect that this model is genuine, in the technical sense discussed above. That is, the leading order contribution to nucleon decay in the model occurs at two loops.

\subsubsection{Application of the bounds in phenomenological studies}
 
Here we note also that our framework can be used to calculate loop-level $\Delta B = -1$ constraints on existing models in a straightforward way. To illustrate this point, we use the example of the model presented in Ref.~\cite{Saad:2017pqj}. There, a Pati--Salam model supplemented by a global $\mathrm{U}(1)_{\mathrm{PQ}}$ is presented. The model's tree-level BNV effects first enter at dimension-9 through the effective Lagrangian of Eq.~(7.105) and are thus suppressed. Using our results, the estimation of the bound from the loop-level nucleon decay is straightforward. Taking as an example the operator generated by the subset of field content $(\bar{\mathbf{6}}, \mathbf{1}, \tfrac{2}{3})_{S}$, $(\mathbf{8}, \mathbf{2}, \tfrac{1}{2})_{S}$ and $(\bar{\mathbf{3}}, \mathbf{2}, -\tfrac{7}{6})_{S}$, which in our
field-string notation is $\mathcal{O}_{29}^{*}$, we find a one-loop contribution to the dimension-7 SMEFT operator $\mathcal{O}_{\bar e qdd\tilde H}$ shown in Tab.~\ref{tab:bviolating-operators} which matches onto the dimension-6 LEFT operator $[\mathcal{O}^{S,RL}_{ddd}]_{1112}$ shown in Table~\ref{tab:SMEFT-LEFT}. The associated neutron decay $n \to K^+ e^-$ imposes a lower bound of $\SI{5e9}{\GeV}$ on the mass scale of the exotic scalar multiplets, assumed to all share a common mass scale for simplicity.

\section{Conclusions}\label{sec:conc}

Proton decay is one of the few currently available experimental probes of GUT-scale energies. Arguably, some of the best motivated scenarios beyond the Standard Model live in this regime, and BNV nucleon decays are characteristic predictions of these theories. More generally, all global symmetries are expected to be violated at the Planck scale, and in particular BNV is an essential ingredient for generating the baryon asymmetry of the Universe.

In this work we have computed estimates for the loop-level matching contributions of generic classes of UV models onto the leading-order operators generating two-body nucleon decays in the SMEFT. Specifically, we match onto dimension-6 and dimension-7 operators, which mediate $(B-L)$-conserving nucleon decays in the first case, and processes that violate $B-L$ in two units in the latter. The contributions from these UV models are organised by the high-dimensional effective operators they generate at tree level. We utilise operators up to dimension 9 for this purpose, and do not consider any Lorentz structure for these. Doing so simplifies our analysis significantly, but inevitably produces order one errors in the loop-level matching coefficients. The estimates themselves are computed in an automated way that simplifies what would otherwise be a complicated calculation. This approach to the computation is sufficient for our purposes, since we are primarily interested in setting order-of-magnitude lower bounds on the scale of the UV models generating these operators. 

Our key findings are summarised in Table~\ref{tab:bviolating-operators}, which presents the matching estimates for each operator that are most constrained by current experimental data. These bounds are represented pictorially in Figs.~\ref{fig:dim-8-loop-level-limit-barplot} and~\ref{fig:dim-9-loop-level-limit-barplot}. In addition, we have studied the correlations and predictions implied by our matching estimates. These results are summarised in Figs.~\ref{fig:democratic-loop-correlations} and~\ref{fig:best-loop-correlations} for different flavour assumptions. Within a given operator, the ratios of the decay rates of different modes are independent of the scale and of some of the uncertainties.  Among other insights, we find out that the poorly measured $n \to \pi^+ e^-$ has impressive power to distinguish between different UV scenarios. Indeed, we wish to highlight the importance of $(B-L)$-violating decay modes (\textit{e.g.}\ $n\to \pi^0 \nu$, in addition to the aforementioned $n \to \pi^+ e^-$) which seem to have received less attention from both experimentalists and phenomenologists, but which can provide the leading-order contributions to nucleon decays in some scenarios. At both dimensions 6 and 7 there are cases in which positive signals in two or three channels could point to particular operators, and therefore signal the direction to possible UV completions, should a positive signal be seen at the next generation of experiments.

We have made all of our results available online as a searchable database~\cite{Gargalionis_Operator_closure_estimates_2023}. We encourage the reader to interact with this should they want to extract bounds or investigate correlations for operators not presented explicitly in the figures of the paper. 

Our analysis sets the stage for a more fine-grained look into the UV origin of the effective operators driving our study. The tree-level completions of these operators necessarily contain exotic multiplets coupling linearly to SM fields. Our results can be used to constrain these multiplets, and classify their contributions to various nucleon decays. Such a study is in line with the analysis of Ref.~\cite{Herrero-Garcia:2019czj}. Additionally, our results can be used to generate the full UV models, analogously to Ref.~\cite{Gargalionis:2020xvt}. This would complement the systematic one-loop analysis presented in Ref.~\cite{Helo:2019yqp} and extend it to include the two-loop contributions present in our framework.

Finally, new applications of the automated procedure developed in this work to obtain order-of-magnitude estimates for other observables would be interesting to pursue. We envisage applications in flavour physics, including in the study lepton-flavour violation and dipole moments.

\acknowledgments

We are grateful to Arnau Bas i Beneito and Arcadi Santamaria for useful discussions that have influenced the development of the project. We acknowledge the use of \texttt{GroupMath}~\cite{Fonseca:2020vke} and \texttt{Sym2Int}~\cite{Fonseca:2017lem,Fonseca:2019yya} for group-theory calculations. All Feynman diagrams were generated using the Ti\textit{k}Z-Feynman package for \LaTeX~\cite{Ellis:2016jkw}. This work is partially supported by the FEDER\slash MCIyU-AEI grant FPA2017-84543-P, by the Spanish ``Agencia Estatal de Investigación'', MICINN\slash AEI (10.13039\slash 501100011033) grants PID2020-113334GB-I00 and PID2020-113644GB-I00. JHG is supported by the ``Generalitat Valenciana'' through the GenT Excellence Program (CIDEGENT\slash 2020\slash 020) and the ``Consolidación Investigadora'' Grant CNS2022-135592 funded also by ``European Union NextGenerationEU/PRTR''. JG is supported by the ``Juan de la Cierva'' programme with reference FJC2021-048111-I, financed by the grant MCIN\slash AEI\slash 10.13039\slash 501100011033 and the European Union ``NextGenerationEU''\slash PRTR, as well as the ``Generalitat Valenciana'' grants PROMETEO\slash 2021\slash 083 and PROMETEO\slash 2019\slash 087. MS acknowledges support by the Australian Research Council through the ARC Discovery Project DP200101470.

\appendix

\section{Operator catalogue, matching results and tree-level limits}\label{sec:app-ops}

In Table~\ref{tab:tree-level-limits} we show the tree-level template limits we use to derive all of our results. These are lower limits on the scale $\Lambda$ underlying each SMEFT operator shown, written in terms of the corresponding WC. They are derived by matching the SMEFT operator onto the LEFT one shown in the table at tree level, and comparing the expected decay rate to the experimental bound.  We note that these limits can be rescaled easily if needed. In the decay processes shown we do not distinguish between $\nu$ and $\bar{\nu}$.

In Table~\ref{tab:bviolating-operators} we present our listing of the $|\Delta B| = 1$ field-string operators, along with the dominant matching estimates for each. Most estimates are proportional to the Yukawa couplings of SM fermions. For example, in our notation $(y_d)^r\equiv m_d^r/v$ denotes the ratio of the down-type quark mass of flavour $r$ over the electroweak vacuum expectation value. The estimates for operators of mass-dimension greater than 7 (field strings 11--50) are written in terms of SMEFT operators that match onto the LEFT at tree-level and therefore give rise to tree-level nucleon decays. The matching estimates represent the rough size of diagrams relating the field strings of mass-dimension greater than 7 in the table to one of the SMEFT operators of Table~\ref{tab:tableaux}, which share field content with the first 10 field strings in our catalogue. For this reason matching estimates for operators 1--10 themselves are not provided. We emphasise that we anticipate the estimates presented in the table to be valid to within an order of magnitude. Each field-string operator 11--50 in general could match onto many SMEFT operators. We present in the table the contribution that provides the currently most stringent experimental limit. The lower bound on the scale of the UV models generating each field string at tree level is also shown, along with the decay channel from which the limit is derived. The flavour indices shown are those of the field-string operator. The corresponding SMEFT flavour indices that label the most constrained operator can be extracted from these and the given matching estimate.

\begin{longtable}{lll}
\caption{The table shows the most stringent lower limits on the UV scale $\Lambda$ underlying each SMEFT operator defined in Tab.~\ref{tab:tableaux} in  Sec.~\ref{sec:smeft-left-tree-level-matching}, which give rise to tree-level BNV nucleon decays, in terms of the dimensionless operator coefficient. These limits are derived by matching the SMEFT operator onto the LEFT operator shown at tree level using Tab.~\ref{tab:SMEFT-LEFT} and comparing the specific process shown to its current experimental bound, provided in Tab.~\ref{tab:ISprocess}.} \label{tab:tree-level-limits} \\
\toprule
Lower limit [GeV] & Process & LEFT coefficient \\
\midrule
\endfirsthead
\toprule
Lower limit [GeV] & Process & LEFT coefficient \\
\midrule
\endhead
\midrule
\multicolumn{3}{r}{Continued on next page} \\
\midrule
\endfoot
\bottomrule
\endlastfoot
$\tabnumT{4211946075676055.5} \cdot \sqrt{{|[C_{qque}]}_{1111}|}$ & $p \to \pi^{0} e^{+}$ & $[C^{S,LR}_{duu}]_{1111}$ \\
$\tabnumT{3016720721021235.5} \cdot \sqrt{\left|{[C_{duql}]_{1111}}\right|}$ & $p \to \pi^{0} e^{+}$ & $[C^{S,RL}_{duu}]_{1111}$ \\
$\tabnumT{2939848956677611.5} \cdot \sqrt{\left|{[C_{duue}]_{1111}}\right|}$ & $p \to \pi^{0} e^{+}$ & $[C^{S,RR}_{duu}]_{1111}$ \\
$\tabnumT{2902403011887124.0} \cdot \sqrt{\left|{[C_{qqql}]_{1111}}\right|}$ & $p \to \pi^{0} e^{+}$ & $[C^{S,LL}_{duu}]_{1111}$ \\
$\tabnumT{2023677780400698.8} \cdot \sqrt{\left|{[C_{qque}]_{1211}}\right|}$ & $p \to \pi^{0} e^{+}$ & $[C^{S,LR}_{duu}]_{1111}$ \\
$\tabnumT{1731467203300215.0} \cdot \sqrt{\left|{[C_{qqql}]_{1121}}\right|}$ & $p \to K^{+} \nu$ & $[C^{S,LL}_{udd}]_{1121}$ \\
$\tabnumT{1394492801995638.5} \cdot \sqrt{\left|{[C_{qqql}]_{1211}}\right|}$ & $p \to \pi^{0} e^{+}$ & $[C^{S,LL}_{duu}]_{1111}$ \\
$\tabnumT{1290865019711307.8} \cdot \sqrt{\left|{[C_{duql}]_{1121}}\right|}$ & $p \to K^{+} \nu$ & $[C^{S,RL}_{dud}]_{1121}$ \\
$\tabnumT{780026072633803.0} \cdot \sqrt{\left|{[C_{duql}]_{2111}}\right|}$ & $p \to K^{+} \nu$ & $[C^{S,RL}_{dud}]_{2111}$ \\
$\tabnumT{588244442399006.1} \cdot \sqrt{\left|{[C_{qqql}]_{2121}}\right|}$ & $p \to K^{+} \nu$ & $[C^{S,LL}_{udd}]_{1121}$ \\
$\tabnumT{510604362164699.5} \cdot \sqrt{\left|{[C_{duue}]_{2111}}\right|}$ & $p \to K^{0} e^{+}$ & $[C^{S,RR}_{duu}]_{2111}$ \\
$\tabnumT{374772469295890.0} \cdot \sqrt{\left|{[C_{duql}]_{2121}}\right|}$ & $p \to K^{+} \nu$ & $[C^{S,RL}_{dud}]_{2111}$ \\
$\tabnumT{358636190973178.56} \cdot \sqrt{\left|{[C_{qqql}]_{1131}}\right|}$ & $p \to K^{+} \nu$ & $[C^{S,LL}_{udd}]_{1121}$ \\
$\tabnumT{267374925062043.1} \cdot \sqrt{\left|{[C_{duql}]_{1131}}\right|}$ & $p \to K^{+} \nu$ & $[C^{S,RL}_{dud}]_{1121}$ \\
$\tabnumT{255969399309761.94} \cdot \sqrt{\left|{[C_{qque}]_{1311}}\right|}$ & $p \to \pi^{0} e^{+}$ & $[C^{S,LR}_{duu}]_{1111}$ \\
$\tabnumT{253594082616048.1} \cdot \sqrt{\left|{[C_{qqql}]_{1311}}\right|}$ & $p \to K^{+} \nu$ & $[C^{S,LL}_{udd}]_{1121}$ \\
$\tabnumT{172310869579086.66} \cdot \sqrt{\left|{[C_{qqql}]_{2311}}\right|}$ & $p \to K^{+} \nu$ & $[C^{S,LL}_{udd}]_{1121}$ \\
$\tabnumT{121842184351522.95} \cdot \sqrt{\left|{[C_{qqql}]_{1231}}\right|}$ & $p \to K^{+} \nu$ & $[C^{S,LL}_{udd}]_{1121}$ \\
$\tabnumT{121842184351522.95} \cdot \sqrt{\left|{[C_{qqql}]_{1321}}\right|}$ & $p \to K^{+} \nu$ & $[C^{S,LL}_{udd}]_{1121}$ \\
$\tabnumT{121842184351522.95} \cdot \sqrt{\left|{[C_{qqql}]_{2131}}\right|}$ & $p \to K^{+} \nu$ & $[C^{S,LL}_{udd}]_{1121}$ \\
$\tabnumT{47403931976023.586} \cdot \sqrt{\left|{[C_{duql}]_{2131}}\right|}$ & $p \to K^{+} \nu$ & $[C^{S,RL}_{dud}]_{2111}$ \\
$\tabnumT{15411480543544.467} \cdot \sqrt{\left|{[C_{qqql}]_{3131}}\right|}$ & $p \to K^{+} \nu$ & $[C^{S,LL}_{udd}]_{1121}$ \\
$\tabnumT{61618974944.33945} \cdot \sqrt[3]{|{[C_{\bar{l}dqq\tilde{H}}]_{1111}}|}$ & $n \to \pi^{0} \nu$ & $[C^{S,LR}_{udd}]_{1111}$ \\
$\tabnumT{61087536504.62788} \cdot \sqrt[3]{|{[C_{\bar{l}dud\tilde{H}}]_{1111}}|}$ & $n \to \pi^{0} \nu$ & $[C^{S,RR}_{udd}]_{1111}$ \\
$\tabnumT{58977287343.01153} \cdot \sqrt[3]{|{[C_{\bar{l}dqq\tilde{H}}]_{1211}}|}$ & $p \to K^{+} \nu$ & $[C^{S,LR}_{udd}]_{1112}$ \\
$\tabnumT{57914434771.0497} \cdot \sqrt[3]{|{[C_{\bar{l}dud\tilde{H}}]_{1211}}|}$ & $p \to K^{+} \nu$ & $[C^{S,RR}_{udd}]_{1112}$ \\
$\tabnumT{42153824656.885826} \cdot \sqrt[3]{|{[C_{\bar{l}dqq\tilde{H}}]_{1112}}|}$ & $p \to K^{+} \nu$ & $[C^{S,LR}_{udd}]_{1211}$ \\
$\tabnumT{37992127018.64059} \cdot \sqrt[3]{|{[C_{\bar{l}dud\tilde{H}}]_{1112}}|}$ & $p \to K^{+} \nu$ & $[C^{S,RR}_{udd}]_{1211}$ \\
$\tabnumT{36179050697.560455} \cdot \sqrt[3]{|{[C_{\bar{l}dqq\tilde{H}}]_{1212}}|}$ & $p \to K^{+} \nu$ & $[C^{S,LR}_{udd}]_{1112}$ \\
$\tabnumT{22100000000.32998} \cdot \sqrt[3]{|{[C_{\bar{e}qdd\tilde{H}}]_{1112}}|}$ & $n \to K^{+} e^{-}$ & $[C^{S,RL}_{ddd}]_{1211}$ \\
$\tabnumT{22185057521.781868} \cdot \sqrt[3]{|{[C_{\bar{l}dqq\tilde{H}}]_{1121}}|}$ & $n \to K^{+} e^{-}$ & $[C^{S,LR}_{ddd}]_{1211}$ \\
$\tabnumT{17953694175.411118} \cdot \sqrt[3]{|{[C_{\bar{l}dddH}]_{1112}}|}$ & $n \to K^{+} e^{-}$ & $[C^{S,RR}_{ddd}]_{1211}$ \\
$\tabnumT{17000000000.113186} \cdot \sqrt[3]{|{[C_{\bar{e}qdd\tilde{H}}]_{1212}}|}$ & $n \to K^{+} e^{-}$ & $[C^{S,RL}_{ddd}]_{1211}$ \\
$\tabnumT{14756959331.16778} \cdot \sqrt[3]{|{[C_{\bar{l}dqq\tilde{H}}]_{1113}}|}$ & $p \to K^{+} \nu$ & $[C^{S,LR}_{udd}]_{1211}$ \\
$\tabnumT{9116363309.860704} \cdot \sqrt[3]{|{[C_{\bar{l}dqq\tilde{H}}]_{1213}}|}$ & $p \to K^{+} \nu$ & $[C^{S,LR}_{udd}]_{1112}$ \\
$\tabnumT{7766412520.650709} \cdot \sqrt[3]{|{[C_{\bar{l}dqq\tilde{H}}]_{1131}}|}$ & $n \to K^{+} e^{-}$ & $[C^{S,LR}_{ddd}]_{1211}$ \\
$\tabnumT{5776430993.75637} \cdot \sqrt[3]{|{[C_{\bar{l}qdDd}]_{1112}}|}$ & $p \to K^{+} \nu$ & $[C^{V,RL}_{ddu}]_{1211}$ \\
$\tabnumT{4764231197.827122} \cdot \sqrt[3]{|{[C_{\bar{l}dqq\tilde{H}}]_{1123}}|}$ & $n \to K^{+} e^{-}$ & $[C^{S,LR}_{ddd}]_{1211}$ \\
$\tabnumT{4764231197.827122} \cdot \sqrt[3]{|{[C_{\bar{l}dqq\tilde{H}}]_{1132}}|}$ & $n \to K^{+} e^{-}$ & $[C^{S,LR}_{ddd}]_{1211}$ \\
$\tabnumT{4740899229.115271} \cdot \sqrt[3]{|{[C_{\bar{l}qdDd}]_{1111}}|}$ & $n \to \pi^{0} \nu$ & $[C^{V,RL}_{ddu}]_{1111}$ \\
$\tabnumT{4600000000.36396} \cdot \sqrt[3]{|{[C_{\bar{e}qdd\tilde{H}}]_{1312}}|}$ & $n \to K^{+} e^{-}$ & $[C^{S,RL}_{ddd}]_{1211}$ \\
$\tabnumT{2957350803.528859} \cdot \sqrt[3]{|{[C_{\bar{e}dddD}]_{1111}}|}$ & $n \to \pi^{+} e^{-}$ & $[C^{V,RR}_{ddd}]_{1111}$ \\
$\tabnumT{2377222135.8358617} \cdot \sqrt[3]{|{[C_{\bar{e}dddD}]_{1112}}|}$ & $n \to K^{+} e^{-}$ & $[C^{V,RR}_{ddd}]_{1211}$ \\
$\tabnumT{2356992665.45016} \cdot \sqrt[3]{|{[C_{\bar{l}qdDd}]_{2111}}|}$ & $n \to K^{+} e^{-}$ & $[C^{V,RL}_{ddd}]_{1112}$ \\
$\tabnumT{1798720495.8833835} \cdot \sqrt[3]{|{[C_{\bar{l}qdDd}]_{1211}}|}$ & $n \to \pi^{+} e^{-}$ & $[C^{V,RL}_{ddd}]_{1111}$ \\
$\tabnumT{1445874521.850213} \cdot \sqrt[3]{|{[C_{\bar{l}qdDd}]_{1212}}|}$ & $n \to K^{+} e^{-}$ & $[C^{V,RL}_{ddd}]_{1211}$ \\
$\tabnumT{1445874521.850213} \cdot \sqrt[3]{|{[C_{\bar{l}qdDd}]_{2211}}|}$ & $n \to K^{+} e^{-}$ & $[C^{V,RL}_{ddd}]_{1112}$ \\
$\tabnumT{453239905.89038223} \cdot \sqrt[3]{|{[C_{\bar{l}qdDd}]_{1311}}|}$ & $n \to \pi^{+} e^{-}$ & $[C^{V,RL}_{ddd}]_{1111}$ \\
$\tabnumT{380953730.4486129} \cdot \sqrt[4]{|{[C_{ddqlHH}]_{1211}}|}$ & $p \to K^{+} \nu$ & $[C^{S,RL}_{ddu}]_{1211}$ \\
$\tabnumT{364330107.8252578} \cdot \sqrt[3]{|{[C_{\bar{l}qdDd}]_{1312}}|}$ & $n \to K^{+} e^{-}$ & $[C^{V,RL}_{ddd}]_{1211}$ \\
$\tabnumT{364330107.8252578} \cdot \sqrt[3]{|{[C_{\bar{l}qdDd}]_{2311}}|}$ & $n \to K^{+} e^{-}$ & $[C^{V,RL}_{ddd}]_{1112}$ \\
$\tabnumT{18980274.37533411} \cdot \sqrt[5]{|{[C_{luqqHHH}]_{1112}}|}$ & $p \to K^{+} \nu$ & $[C^{S,LR}_{ddu}]_{1211}$ \\
$\tabnumT{11018808.034168914} \cdot \sqrt[5]{|{[C_{eqqqHHH}]_{1112}}|}$ & $n \to K^{+} e^{-}$ & $[C^{S,LL}_{ddd}]_{1211}$ \\
$\tabnumT{10111100.419155428} \cdot \sqrt[5]{|{[C_{luqqHHH}]_{1113}}|}$ & $p \to K^{+} \nu$ & $[C^{S,LR}_{ddu}]_{1211}$ \\
$\tabnumT{8218603.738031011} \cdot \sqrt[5]{|{[C_{eqqqHHH}]_{1212}}|}$ & $n \to K^{+} e^{-}$ & $[C^{S,LL}_{ddd}]_{1211}$ \\
$\tabnumT{7541571.415237498} \cdot \sqrt[5]{|{[C_{luqqHHH}]_{1123}}|}$ & $p \to K^{+} \nu$ & $[C^{S,LR}_{ddu}]_{1211}$ \\
$\tabnumT{5869897.996715198} \cdot \sqrt[5]{|{[C_{eqqqHHH}]_{1113}}|}$ & $n \to K^{+} e^{-}$ & $[C^{S,LL}_{ddd}]_{1211}$ \\
$\tabnumT{4882688.273917859} \cdot \sqrt[5]{|{[C_{qqlqHHD}]_{1211}}|}$ & $p \to K^{+} \nu$ & $[C^{V,LL}_{ddu}]_{1211}$ \\
$\tabnumT{4378183.689929657} \cdot \sqrt[5]{|{[C_{eqqqHHH}]_{1123}}|}$ & $n \to K^{+} e^{-}$ & $[C^{S,LL}_{ddd}]_{1211}$ \\
$\tabnumT{4378183.689929657} \cdot \sqrt[5]{|{[C_{eqqqHHH}]_{1213}}|}$ & $n \to K^{+} e^{-}$ & $[C^{S,LL}_{ddd}]_{1211}$ \\
$\tabnumT{4336902.7054122165} \cdot \sqrt[5]{|{[C_{qqlqHHD}]_{1111}}|}$ & $n \to \pi^{0} \nu$ & $[C^{V,LL}_{ddu}]_{1111}$ \\
$\tabnumT{3641853.0911168717} \cdot \sqrt[5]{|{[C_{qqlqHHD}]_{2211}}|}$ & $p \to K^{+} \nu$ & $[C^{V,LL}_{ddu}]_{1211}$ \\
$\tabnumT{3594319.8038545484} \cdot \sqrt[5]{|{[C_{eqqqHHH}]_{1312}}|}$ & $n \to K^{+} e^{-}$ & $[C^{S,LL}_{ddd}]_{1211}$ \\
$\tabnumT{3301100.850904567} \cdot \sqrt[5]{|{[C_{qqedHHD}]_{1111}}|}$ & $n \to \pi^{+} e^{-}$ & $[C^{V,LR}_{ddd}]_{1111}$ \\
$\tabnumT{3265558.0102912844} \cdot \sqrt[5]{|{[C_{eqqqHHH}]_{1223}}|}$ & $n \to K^{+} e^{-}$ & $[C^{S,LL}_{ddd}]_{1211}$ \\
$\tabnumT{2895735.867748272} \cdot \sqrt[5]{|{[C_{qqedHHD}]_{1112}}|}$ & $n \to K^{+} e^{-}$ & $[C^{V,LR}_{ddd}]_{1112}$ \\
$\tabnumT{2895735.867748272} \cdot \sqrt[5]{|{[C_{qqedHHD}]_{1211}}|}$ & $n \to K^{+} e^{-}$ & $[C^{V,LR}_{ddd}]_{1211}$ \\
$\tabnumT{2601087.3434566515} \cdot \sqrt[5]{|{[C_{qqlqHHD}]_{1311}}|}$ & $p \to K^{+} \nu$ & $[C^{V,LL}_{ddu}]_{1211}$ \\
$\tabnumT{2424607.558626088} \cdot \sqrt[5]{|{[C_{qqlqHHD}]_{1112}}|}$ & $n \to \pi^{+} e^{-}$ & $[C^{V,LL}_{ddd}]_{1111}$ \\
$\tabnumT{2126873.2431495483} \cdot \sqrt[5]{|{[C_{qqlqHHD}]_{1212}}|}$ & $n \to K^{+} e^{-}$ & $[C^{V,LL}_{ddd}]_{1211}$ \\
$\tabnumT{1940074.2891234462} \cdot \sqrt[5]{|{[C_{qqlqHHD}]_{2311}}|}$ & $p \to K^{+} \nu$ & $[C^{V,LL}_{ddu}]_{1211}$ \\
$\tabnumT{1914752.535008738} \cdot \sqrt[5]{|{[C_{eqqqHHH}]_{1313}}|}$ & $n \to K^{+} e^{-}$ & $[C^{S,LL}_{ddd}]_{1211}$ \\
$\tabnumT{1586371.9861769439} \cdot \sqrt[5]{|{[C_{qqlqHHD}]_{2212}}|}$ & $n \to K^{+} e^{-}$ & $[C^{V,LL}_{ddd}]_{1211}$ \\
$\tabnumT{1428157.409842204} \cdot \sqrt[5]{|{[C_{eqqqHHH}]_{1323}}|}$ & $n \to K^{+} e^{-}$ & $[C^{S,LL}_{ddd}]_{1211}$ \\
$\tabnumT{1133019.9192613778} \cdot \sqrt[5]{|{[C_{qqlqHHD}]_{1312}}|}$ & $n \to K^{+} e^{-}$ & $[C^{V,LL}_{ddd}]_{1211}$ \\
$\tabnumT{1060376.5849201328} \cdot \sqrt[5]{|{[C_{qqlqHHD}]_{1113}}|}$ & $n \to \pi^{+} e^{-}$ & $[C^{V,LL}_{ddd}]_{1111}$ \\
$\tabnumT{930165.6171553375} \cdot \sqrt[5]{|{[C_{qqlqHHD}]_{1213}}|}$ & $n \to K^{+} e^{-}$ & $[C^{V,LL}_{ddd}]_{1211}$ \\
$\tabnumT{848471.0615840056} \cdot \sqrt[5]{|{[C_{qqlqHHD}]_{3311}}|}$ & $p \to K^{+} \nu$ & $[C^{V,LL}_{ddu}]_{1211}$ \\
$\tabnumT{845086.1213689782} \cdot \sqrt[5]{|{[C_{qqlqHHD}]_{2312}}|}$ & $n \to K^{+} e^{-}$ & $[C^{V,LL}_{ddd}]_{1211}$ \\
$\tabnumT{693783.0838358335} \cdot \sqrt[5]{|{[C_{qqlqHHD}]_{2213}}|}$ & $n \to K^{+} e^{-}$ & $[C^{V,LL}_{ddd}]_{1211}$ \\
$\tabnumT{495514.33111660334} \cdot \sqrt[5]{|{[C_{qqlqHHD}]_{1313}}|}$ & $n \to K^{+} e^{-}$ & $[C^{V,LL}_{ddd}]_{1211}$ \\
$\tabnumT{369589.5165188811} \cdot \sqrt[5]{|{[C_{qqlqHHD}]_{2313}}|}$ & $n \to K^{+} e^{-}$ & $[C^{V,LL}_{ddd}]_{1211}$ \\
$\tabnumT{369589.5165188811} \cdot \sqrt[5]{|{[C_{qqlqHHD}]_{3312}}|}$ & $n \to K^{+} e^{-}$ & $[C^{V,LL}_{ddd}]_{1211}$ \\
$\tabnumT{161636.083313478} \cdot \sqrt[5]{|{[C_{qqlqHHD}]_{3313}}|}$ & $n \to K^{+} e^{-}$ & $[C^{V,LL}_{ddd}]_{1211}$ \\
\end{longtable}

\newpage

\begin{longtable}[c]{llllll}
  \caption{The table displays our listing of the $|\Delta B| = 1$ operators along with our loop-level matching estimates. The operators are grouped by mass dimension. Operators of even dimension in the list carry $(\Delta B, \Delta L)$ numbers of $(-1, -1)$, while odd-dimensional operators have $(+1,-1)$. The matching expressions represent our estimate of the loop-level matching onto the SMEFT of UV models generating the given field string at tree level. The expressions are written in terms of the field-string WCs $C_i$. Conjugated coefficients are shown with bars in the table. There is an implicit sum over all primed indices. The flavour indices label the most strongly constrained field-string operator for each row. We note that these are trivially symmetric on indices of repeated fields, since the order of the fields in each field string is arbitrary. The corresponding flavour indices of the SMEFT operator can be inferred from the matching estimate. The column labelled $\Lambda$ shows the lower limit on the scale of the underlying UV model generating the numbered field string of each row. The nucleon decay process from which this limit is derived is shown in the last column.} \label{tab:bviolating-operators} \\
  \toprule
  \# & Operator & Matching estimate & Flavour & $\Lambda~[\mathrm{GeV}]$ & Process \\
  \midrule
  \endfirsthead
  \toprule
  \# & Operator & Matching estimate & Flavour & $\Lambda~[\mathrm{GeV}]$ & Process \\
  \midrule
  \endhead
  \midrule
  \multicolumn{6}{r}{Continued on next page} \\
  \midrule
  \endfoot
  \bottomrule
  \endlastfoot
  \multicolumn{6}{l}{Dimension 6} \\
  $1$ & $L Q Q Q$ & --- & $1111$ & \tabnum{2902403011887124.0} & $p\to \pi^0 e^+$ \\
  $2$ & ${\bar{e}^{\dagger}} Q Q {\bar{u}^{\dagger}}$ & --- & $1111$ & \tabnum{4211946075676055.5} & $p\to \pi^0 e^+$ \\
  $3$ & ${\bar{e}^{\dagger}} {\bar{u}^{\dagger}} {\bar{u}^{\dagger}} {\bar{d}^{\dagger}}$ & --- & $1111$ & \tabnum{2939848956677611.5} & $p\to \pi^0 e^+$ \\
  $4$ & $L Q {\bar{u}^{\dagger}} {\bar{d}^{\dagger}}$ & --- & $1111$ & \tabnum{3016720721021235.5} & $p\to \pi^0 e^+$ \\
  \midmidrule
  \multicolumn{6}{l}{Dimension 7} \\
  $5$ & $L \bar{d} \bar{d} \bar{d} \dagf{H}$ & --- & $1112$ & \tabnum{17953694175.411118} & $n \to K^+ e^-$ \\
  $6$ & $DL \dagf{Q} \bar{d} \bar{d}$ & --- & $1112$ & \tabnum{5776430993.75637} & $p \to K^+ \nu$ \\
  $7$ & $D{\bar{e}^{\dagger}} \bar{d} \bar{d} \bar{d}$ & --- & $1111$ & \tabnum{2957350803.528859} & $n \to \pi^+ e^-$ \\
  $8$ & $L \dagf{Q} \dagf{Q} \bar{d} H$ & --- & $1111$ & \tabnum{61618974944.33945} & $n \to \pi^0 \nu$ \\
  $9$ & ${\bar{e}^{\dagger}} \dagf{Q} \bar{d} \bar{d} H$ & --- & $1112$ & \tabnum{28191320923.32998} & $n \to K^+ e^-$ \\
  $10$ & $L \bar{u} \bar{d} \bar{d} H$ & --- & $1111$ & \tabnum{61087536504.62788} & $n \to \pi^0 \nu$ \\
  \midmidrule
  \multicolumn{6}{l}{Dimension 8} \\
  $11$ & $DL Q Q {\bar{d}^{\dagger}} H$  & $C_{qqql}^{pqrs} =  \frac{1}{16\pi^2} V_{r u^\prime}^* (y_d)^{u^\prime}  C_{11}^{s p q u^\prime}$ & 1113 & \tabnum{3555355767528.56} & $p \to K^{+} \nu$ \\
$12$ & $DL {\bar{u}^{\dagger}} {\bar{d}^{\dagger}} {\bar{d}^{\dagger}} H$  & $C_{duql}^{pqrs} =  \frac{1}{16\pi^2} V_{r t^\prime}^* (y_d)^{t^\prime}  C_{12}^{s q t^\prime p}$ & 1131 & \tabnum{2650633164857.99} & $p \to K^{+} \nu$ \\
$13$ & $DL {\bar{u}^{\dagger}} {\bar{u}^{\dagger}} {\bar{d}^{\dagger}} \dagf{H}$  & $C_{duql}^{pqrs} = \frac{1}{16\pi^2} (y_u)^{r}  C_{13}^{s r q p}$ & 1311 & \tabnum{20522849131639.1} & $p \to K^{+} \nu$ \\
$14$ & $L Q {\bar{u}^{\dagger}} {\bar{u}^{\dagger}} \dagf{H} \dagf{H}$  & $C_{qqql}^{pqrs} = (\frac{1}{16\pi^2})^2 (y_u)^{q} (y_u)^{r}  C_{14}^{s p q r}$ & 1123 & \tabnum{43470451891.7671} & $p \to K^{+} \nu$ \\
$15$ & ${\bar{e}^{\dagger}} Q Q {\bar{d}^{\dagger}} H H$  & $C_{qque}^{pqrs} =  (\frac{1}{16\pi^2})^2 V_{p u^\prime}^* (y_d)^{u^\prime} (y_u)^{p}  C_{15}^{q r s u^\prime}$ & 1113 & \tabnum{194775342945.279} & $p \to \pi^{0} e^{+}$ \\
$16$ & $L Q {\bar{d}^{\dagger}} {\bar{d}^{\dagger}} H H$  & $C_{qqql}^{pqrs} =   (\frac{1}{16\pi^2})^2 V_{q t^\prime}^* V_{r u^\prime}^* (y_d)^{t^\prime} (y_d)^{u^\prime}  C_{16}^{s p t^\prime u^\prime}$ & 1133 & \tabnum{2448657939.95774} & $p \to K^{+} \nu$ \\
$17$ & $D{\bar{e}^{\dagger}} Q {\bar{u}^{\dagger}} {\bar{u}^{\dagger}} \dagf{H}$  & $C_{duue}^{pqrs} =  \frac{1}{16\pi^2} V_{s^\prime p}^* (y_d)^{p}  C_{17}^{s s^\prime q r}$ & 1111 & \tabnum{885653774981.998} & $p \to \pi^{0} e^{+}$ \\
$18$ & $L Q Q Q H \dagf{H}$  & $C_{qqql}^{pqrs} = \left( \frac{1}{16\pi^2} + \frac{v^2}{\Lambda^2} \right)  C_{18}^{s p q r}$ & 1111 & \tabnum{230965893093320.} & $p \to \pi^{0} e^{+}$ \\
$19$ & $DL Q Q {\bar{u}^{\dagger}} \dagf{H}$  & $C_{qqql}^{pqrs} = \frac{1}{16\pi^2} (y_u)^{r}  C_{19}^{s p q r}$ & 1113 & \tabnum{27527773738619.4} & $p \to K^{+} \nu$ \\
$20$ & $D{\bar{e}^{\dagger}} Q Q Q H$  & $C_{qque}^{pqrs} = \frac{1}{16\pi^2} (y_u)^{p}  C_{20}^{q r s p}$ & 1113 & \tabnum{19647397238658.6} & $p \to \pi^{0} e^{+}$ \\
$21$ & $D{\bar{e}^{\dagger}} Q {\bar{u}^{\dagger}} {\bar{d}^{\dagger}} H$  & $C_{qque}^{pqrs} =  \frac{1}{16\pi^2} V_{s u^\prime}^* (y_d)^{u^\prime}  C_{21}^{q r p u^\prime}$ & 1112 & \tabnum{2718436745633.01} & $p \to \pi^{0} e^{+}$ \\
$22$ & ${\bar{e}^{\dagger}} Q Q {\bar{u}^{\dagger}} H \dagf{H}$  & $C_{qque}^{pqrs} = \left( \frac{1}{16\pi^2} + \frac{v^2}{\Lambda^2} \right)  C_{22}^{q r s p}$ & 1111 & \tabnum{335176018990177.} & $p \to \pi^{0} e^{+}$ \\
$23$ & ${\bar{e}^{\dagger}} {\bar{u}^{\dagger}} {\bar{u}^{\dagger}} {\bar{d}^{\dagger}} H \dagf{H}$  & $C_{duue}^{pqrs} = \left( \frac{1}{16\pi^2} + \frac{v^2}{\Lambda^2} \right)  C_{23}^{s q r p}$ & 1111 & \tabnum{233945746699397.} & $p \to \pi^{0} e^{+}$ \\
$24$ & $L Q {\bar{u}^{\dagger}} {\bar{d}^{\dagger}} H \dagf{H}$  & $C_{duql}^{pqrs} = \left( \frac{1}{16\pi^2} + \frac{v^2}{\Lambda^2} \right)  C_{24}^{s r q p}$ & 1111 & \tabnum{240063007339138.} & $p \to \pi^{0} e^{+}$ \\
  \midmidrule
  \multicolumn{6}{l}{Dimension 9} \\
$25$ & ${\bar{e}^{\dagger}} {\bar{e}^{\dagger}} \bar{e} \bar{d} \bar{d} \bar{d}$  & $C_{\bar{e}dddD}^{pqrs} =  \frac{1}{16\pi^2}  \bar{C}_{25}^{r^\prime p r^\prime q r s}$ & 111111 & \tabnum{547137230.716205} & $n \to \pi^{+} e^{-}$ \\
$26$ & ${\bar{e}^{\dagger}} \dagf{Q} \dagf{Q} \dagf{Q} H H H$  & $C_{\bar{e}qdd\tilde{H}}^{pqrs} =   (\frac{1}{16\pi^2})^2 V_{s^\prime r} V_{t^\prime s} (y_d)^{r} (y_d)^{s}  \bar{C}_{26}^{p s^\prime t^\prime q}$ & 1121 & \tabnum{1542386.84159534} & $n \to K^{+} e^{-}$ \\
$27$ & ${\bar{e}^{\dagger}} \bar{d} \bar{d} \bar{d} \bar{d} {\bar{d}^{\dagger}}$  & $C_{\bar{e}dddD}^{pqrs} =  \frac{1}{16\pi^2}  \bar{C}_{27}^{p q r s w^\prime w^\prime}$ & 111111 & \tabnum{547137230.716205} & $n \to \pi^{+} e^{-}$ \\
$28$ & $L L \bar{e} \bar{u} \bar{d} \bar{d}$  & $C_{\bar{l}dud\tilde{H}}^{pqrs} =  \frac{1}{16\pi^2} (y_e)^{t^\prime}  \bar{C}_{28}^{p t^\prime t^\prime r q s}$ & 133111 & \tabnum{2433707401.04298} & $n \to \pi^{0} \nu$ \\
$29$ & ${\bar{e}^{\dagger}} \dagf{Q} \dagf{Q} {\bar{u}^{\dagger}} \bar{d} \bar{d}$  & $C_{\bar{e}qdd\tilde{H}}^{pqrs} =  \frac{1}{16\pi^2} (y_u)^{u^\prime}  \bar{C}_{29}^{p q u^\prime u^\prime r s}$ & 113312 & \tabnum{5091667984.95834} & $n \to K^{+} e^{-}$ \\
$30$ & $L L \bar{e} \dagf{Q} \dagf{Q} \bar{d}$  & $C_{\bar{l}dqq\tilde{H}}^{pqrs} =  \frac{1}{16\pi^2} (y_e)^{t^\prime}  \bar{C}_{30}^{t^\prime p t^\prime r s q}$ & 313111 & \tabnum{2454879734.02496} & $n \to \pi^{0} \nu$ \\
$31$ & $L L \dagf{L} \dagf{Q} \bar{d} \bar{d}$  & $C_{\bar{l}qdDd}^{pqrs} =  \frac{1}{16\pi^2}  \bar{C}_{31}^{t^\prime p t^\prime q r s}$ & 111112 & \tabnum{1068693120.06403} & $p \to K^{+} \nu$ \\
$32$ & $L \dagf{Q} \bar{d} \bar{d} \bar{d} {\bar{d}^{\dagger}}$  & $C_{\bar{l}qdDd}^{pqrs} =  \frac{1}{16\pi^2}  \bar{C}_{32}^{p q w^\prime r s w^\prime}$ & 111121 & \tabnum{1068693120.06403} & $p \to K^{+} \nu$ \\
$33$ & ${\bar{e}^{\dagger}} \bar{u} {\bar{u}^{\dagger}} \bar{d} \bar{d} \bar{d}$  & $C_{\bar{e}dddD}^{pqrs} =  \frac{1}{16\pi^2}  \bar{C}_{33}^{p t^\prime t^\prime q r s}$ & 111111 & \tabnum{547137230.716205} & $n \to \pi^{+} e^{-}$ \\
$34$ & $D{\bar{e}^{\dagger}} \dagf{Q} \dagf{Q} \bar{d} H H$  & $C_{\bar{e}qdd\tilde{H}}^{pqrs} =  \frac{1}{16\pi^2} V_{s^\prime s} (y_d)^{s}  \bar{C}_{34}^{p s^\prime q r}$ & 1211 & \tabnum{343215851.811040} & $n \to K^{+} e^{-}$ \\
$35$ & $L \dagf{L} {\bar{e}^{\dagger}} \bar{d} \bar{d} \bar{d}$  & $C_{\bar{l}dddH}^{pqrs} =  \frac{1}{16\pi^2} (y_e)^{t^\prime}  \bar{C}_{35}^{p t^\prime t^\prime q r s}$ & 133112 & \tabnum{715269282.261041} & $n \to K^{+} e^{-}$ \\
$36$ & $L \bar{d} \bar{d} \bar{d} \dagf{H} \dagf{H} H$  & $C_{\bar{l}dddH}^{pqrs} = \left( \frac{1}{16\pi^2} + \frac{v^2}{\Lambda^2} \right)  \bar{C}_{36}^{p q r s}$ & 1112 & \tabnum{3321599351.87217} & $n \to K^{+} e^{-}$ \\
$37$ & $L \dagf{Q} \dagf{Q} \bar{u} H H H$  & $C_{\bar{l}dqq\tilde{H}}^{pqrs} =  (\frac{1}{16\pi^2})^2 V_{u^\prime q} (y_d)^{q} (y_u)^{u^\prime}  \bar{C}_{37}^{p r s u^\prime}$ & 1113 & \tabnum{45398473.7604176} & $p \to K^{+} \nu$ \\
$38$ & $DL \dagf{Q} \dagf{Q} \dagf{Q} H H$  & $C_{\bar{l}dqq\tilde{H}}^{pqrs} =  \frac{1}{16\pi^2} V_{s^\prime q} (y_d)^{q}  \bar{C}_{38}^{p s^\prime r s}$ & 1211 & \tabnum{718020271.841351} & $p \to K^{+} \nu$ \\
$39$ & ${\bar{e}^{\dagger}} Q \dagf{Q} \bar{d} \bar{d} \bar{d}$  & $C_{\bar{e}qdd\tilde{H}}^{pqrs} =   \frac{1}{16\pi^2} V_{s^\prime u^\prime} (y_d)^{u^\prime}  \bar{C}_{39}^{p s^\prime q u^\prime r s}$ & 131312 & \tabnum{1300972330.63902} & $n \to K^{+} e^{-}$ \\
$40$ & $L \dagf{Q} \dagf{Q} \dagf{Q} {\bar{u}^{\dagger}} \bar{d}$  & $C_{\bar{l}dqq\tilde{H}}^{pqrs} =  \frac{1}{16\pi^2} (y_u)^{v^\prime}  \bar{C}_{40}^{p v^\prime r s v^\prime q}$ & 131131 & \tabnum{11129076315.4841} & $n \to \pi^{0} \nu$ \\
$41$ & $L Q \bar{u} \bar{d} \bar{d} \bar{d}$  & $C_{\bar{l}dddH}^{pqrs} =  \frac{1}{16\pi^2} (y_u)^{t^\prime}  \bar{C}_{41}^{p t^\prime t^\prime q r s}$ & 133112 & \tabnum{3242638047.83348} & $n \to K^{+} e^{-}$ \\
$42$ & $L \dagf{Q} \dagf{Q} Q \bar{d} \bar{d}$  & $C_{\bar{l}dqq\tilde{H}}^{pqrs} =   \frac{1}{16\pi^2} V_{u^\prime v^\prime} (y_d)^{v^\prime}  \bar{C}_{42}^{p r s u^\prime v^\prime q}$ & 111331 & \tabnum{2843590822.25848} & $n \to \pi^{0} \nu$ \\
$43$ & $DL \dagf{Q} \bar{d} \bar{d} H \dagf{H}$  & $C_{\bar{l}qdDd}^{pqrs} = \left( \frac{1}{16\pi^2} + \frac{v^2}{\Lambda^2} \right)  \bar{C}_{43}^{p q r s}$ & 1112 & \tabnum{1068693120.06403} & $p \to K^{+} \nu$ \\
$44$ & $DL \dagf{Q} \bar{u} \bar{d} H H$  & $C_{\bar{l}dqq\tilde{H}}^{pqrs} = \frac{1}{16\pi^2} (y_u)^{s}  \bar{C}_{44}^{p r s q}$ & 1131 & \tabnum{2665271967.43230} & $p \to K^{+} \nu$ \\
$45$ & $L \dagf{Q} \dagf{Q} \bar{d} H H \dagf{H}$  & $C_{\bar{l}dqq\tilde{H}}^{pqrs} = \left( \frac{1}{16\pi^2} + \frac{v^2}{\Lambda^2} \right)  \bar{C}_{45}^{p r s q}$ & 1111 & \tabnum{11400079852.0040} & $n \to \pi^{0} \nu$ \\
$46$ & $L \bar{u} \bar{d} \bar{d} H H \dagf{H}$  & $C_{\bar{l}dud\tilde{H}}^{pqrs} = \left( \frac{1}{16\pi^2} + \frac{v^2}{\Lambda^2} \right)  \bar{C}_{46}^{p r q s}$ & 1111 & \tabnum{11301758829.0624} & $n \to \pi^{0} \nu$ \\
$47$ & ${\bar{e}^{\dagger}} \dagf{Q} \bar{d} \bar{d} H H \dagf{H}$  & $C_{\bar{e}qdd\tilde{H}}^{pqrs} = \left( \frac{1}{16\pi^2} + \frac{v^2}{\Lambda^2} \right)  \bar{C}_{47}^{p q r s}$ & 1112 & \tabnum{5215654917.17021} & $n \to K^{+} e^{-}$ \\
$48$ & $D{\bar{e}^{\dagger}} \bar{d} \bar{d} \bar{d} H \dagf{H}$  & $C_{\bar{e}dddD}^{pqrs} = \left( \frac{1}{16\pi^2} + \frac{v^2}{\Lambda^2} \right)  \bar{C}_{48}^{p q r s}$ & 1111 & \tabnum{547137230.716205} & $n \to \pi^{+} e^{-}$ \\
$49$ & $L \bar{e} {\bar{e}^{\dagger}} \dagf{Q} \bar{d} \bar{d}$  & $C_{\bar{e}qdd\tilde{H}}^{pqrs} =  \frac{1}{16\pi^2} (y_e)^{s^\prime}  \bar{C}_{49}^{s^\prime s^\prime p q r s}$ & 331112 & \tabnum{1123132971.17634} & $n \to K^{+} e^{-}$ \\
$50$ & $L \dagf{Q} \bar{u} {\bar{u}^{\dagger}} \bar{d} \bar{d}$  & $C_{\bar{l}dud\tilde{H}}^{pqrs} =  \frac{1}{16\pi^2} (y_u)^{u^\prime}  \bar{C}_{50}^{p u^\prime r u^\prime q s}$ & 131311 & \tabnum{11033092587.1297} & $n \to \pi^{0} \nu$ \\
\end{longtable}

\section{Additional correlation plots}\label{sec:app-extra-correlations}

\begin{figure}[t]
  \centering
  \includegraphics[width=0.9\textwidth]{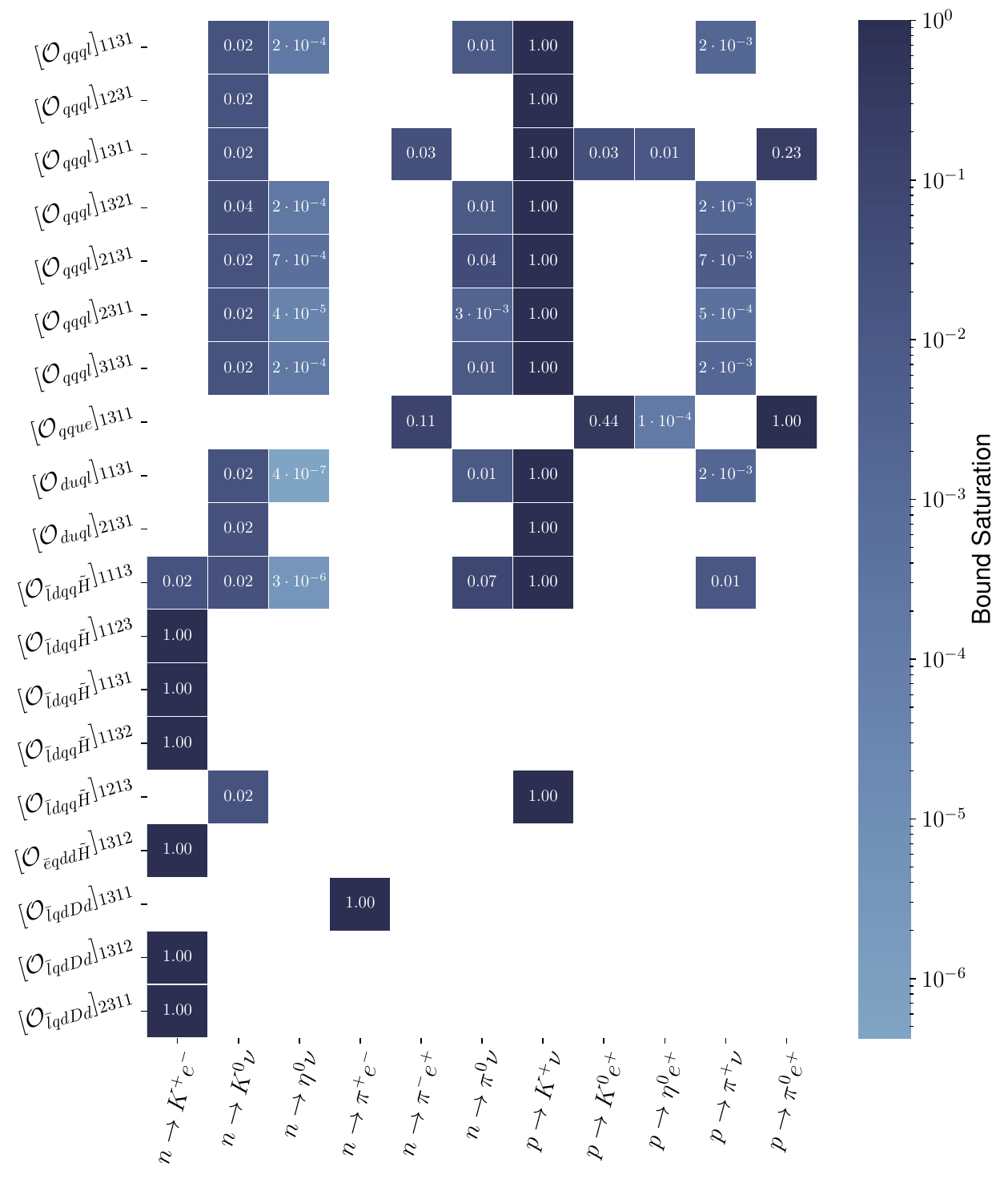}
  \caption{\label{fig:tree-level-correlations-gen-3} The figure shows the tree-level correlations among the SMEFT operators up to dimension-7 including third generation indices. The remaining details are as in the caption of Fig.~\ref{fig:tree-level-correlations}.}
\end{figure}

\begin{figure}[ptb!]
  \centering
  \includegraphics[width=0.9\textwidth]{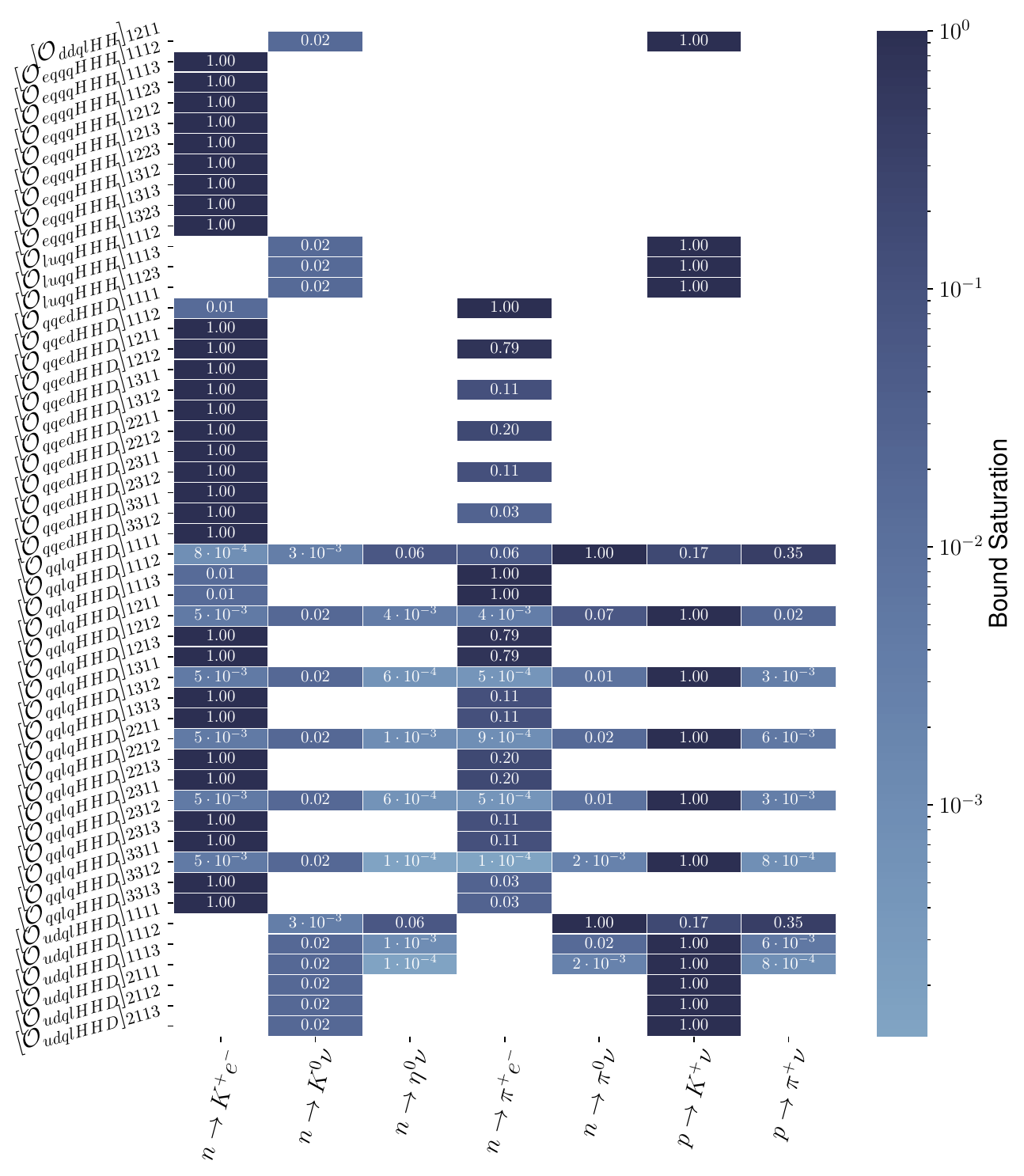}
  \caption{\label{fig:tree-level-correlations-dim-9} The figure shows the tree-level correlations among the SMEFT operators of dimensions 8 and 9. The remaining details are as in the caption of Fig.~\ref{fig:tree-level-correlations}.}
\end{figure}

Here we present the additional correlation plots discussed in Sec.~\ref{sec:tree-level-limits}.  The tree-level correlations including third-generation flavour indices for operators up to dimension 7 are shown in Fig.~\ref{fig:tree-level-correlations-gen-3}. The description of the figure is as for Fig.~\ref{fig:tree-level-correlations}. The analogous heat map for the SMEFT operators of mass-dimension greater than 7 is shown in Fig.~\ref{fig:tree-level-correlations-dim-9}. This includes all flavour structures mediating tree-level nucleon decays.

\section{Direct application of \texorpdfstring{$2 \to 2$}{2-to-2} rules} \label{sec:manual-2to2-rules}

Any operators that are not present in our matching-estimate database generate nucleon decay in such a way that at least one additional loop can be added to the diagram to relate it to one of those contributions that are present. Again taking advantage of the simplicity of $\mathcal{O}_{12}$, we use the matching estimates associated with it to show how this additional loop can be used to extend our results beyond what we dub the \textit{leading-loop contributions} to nucleon decay. 

\begin{figure}[t]
  \centering
  \includegraphics[width=0.4\textwidth]{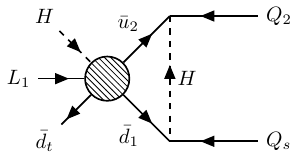}
  \caption{\label{fig:o12-222-rule} The diagram shows a one-loop graph relating ${[\mathcal{O}_{12}]}_{12t1}$ to ${[\mathcal{O}_{11}]}_{12st}$ through a loop with a Higgs and SM fermions. This is equivalent to the application of the $2 \to 2$ rule discussed in the main text. Subscripts on SM fermions here refer to their flavour. We follow Ref.~\cite{Dreiner:2008tw} for our arrow conventions.}
\end{figure}

It is clear that ${[\mathcal{O}_{duql}]_{1211}}$ does not mediate nucleon decay at tree level, since it contains a charm quark.\footnote{This is also the reason why it does not appear in Fig.~\ref{fig:tree-level-correlations}.} Additionally, the most-constrained matching estimate for $\mathcal{O}_{12}$ --- which is the only leading-loop contribution to nucleon decay for this field-string operator --- shows that ${[\mathcal{O}_{duql}]_{1211}}$ is generated from non-zero $[C_{12}]_{12 t 1}$. Thus, none of the $[C_{12}]_{12 t 1}$ are present in the results shown here and published in our online database. However, the flavour structures $[C_{12}]_{12 t 1}$ can all be related to another operator that is present through a $2 \to 2$ rule. In this case one such rule is $\bar{u}_p^{\dagger} \bar{d}_q^{\dagger} \to Q_r Q_s$ with weight $\sum_{pq} (16\pi^2)^{-1} [y_u]_{p} \delta_{p r} [y_d]_{q} V_{s q}^*$. Diagrammatically this corresponds to the closure of the $\bar{u}_2$ and $\bar{d}_1$ lines with a Higgs doublet into two $Q$ fields, and we show this graph in Fig.~\ref{fig:o12-222-rule}. The diagram relates $[C_{12}]_{1 2 t 1}$ to ${[C_{11}]_{1 2 s t}}$, which is present in our database and directly constrained by our results. The coefficient ${[C_{11}]_{1 2 s t}}$ is constrained most strongly by $p\to K^+\nu$. The corresponding constraint on $[C_{12}]_{1 2 t 1}$ can be derived by reweighting the limit on the operators ${[C_{11}]_{1 2 s t}}$ by the aforementioned weight in the appropriate way. Explicitly,
\begin{equation}
[C_{11}]_{12st} = \sum_q \frac{1}{16\pi^2} [y_u]_2 [y_d]_q V_{sq}^* [C_{12}]_{12t1} \equiv \epsilon_{s} [C_{12}]_{12t1} \;,
\end{equation}
and concretely taking $s = t = 1$ for the sake of example, the limit on the scale suppressing $[\mathcal{O}_{12}]_{1211}$ derived from nucleon decay is $\sqrt{\epsilon_1} \cdot \Lambda_{\mathcal{O}_{11}}$, where $\Lambda_{\mathcal{O}_{11}}$ is the bound on the scale underlying the operator $[C_{11}]_{1211}$: $\Lambda_{\mathcal{O}_{11}} = \SI{5.2e11}{\GeV}$, which we retrieve from our results database and comes from the non-observation of $p\to K^+\nu$.

Although this is not the only such diagram that could exist, this kind of manual application of a $2 \to 2$ rule and reweighting of the associated limit provides a way to estimate the bounds on operators not explicitly constrained by our leading-loop results. We note that the large amount of suppression in the rate ($\sim \epsilon_1^2$) justifies not including these contributions in the correlation plots presented in Sec.~\ref{sec:results}, where we have consistently left entries with bound saturation $< 10^{-6}$ blank.

\clearpage
\bibliographystyle{JHEP}
\bibliography{main}

\end{document}